\documentclass{ws-ijmpa}
\usepackage[super,compress]{cite}

\newcommand{\lsim}{\mathrel{\mathop{\kern 0pt \rlap
  {\raise.2ex\hbox{$<$}}}
  \lower.9ex\hbox{\kern-.190em $\sim$}}}
\newcommand{\gsim}{\mathrel{\mathop{\kern 0pt \rlap
  {\raise.2ex\hbox{$>$}}}
  \lower.9ex\hbox{\kern-.190em $\sim$}}}

\begin{document}

\markboth{R. Bernabei et al.}
{Dark Matter investigation with highly radiopure NaI(Tl)}

\title{Dark Matter investigation by DAMA at Gran Sasso}

\author{R. BERNABEI, P. BELLI, S. d'ANGELO, A. DI MARCO and F. MONTECCHIA\thanks{also: 
Dip. di Ingegneria Civile e Ingegneria Informatica,
Univ. ``Tor Vergata''}
}

\address{Dip. di Fisica, Univ. ``Tor Vergata'', I-00133 Rome, Italy\\
and INFN, sez. Roma ``Tor Vergata'', I-00133 Rome, Italy}

\author{F. CAPPELLA, A. d'ANGELO and A. INCICCHITTI}

\address{Dip. di Fisica, Univ. di Roma ``La Sapienza'', I-00185 Rome, Italy\\
and INFN, sez. Roma, I-00185 Rome, Italy}

\author{V. CARACCIOLO, S. CASTELLANO and R. CERULLI}

\address{Laboratori Nazionali del Gran Sasso, INFN, Assergi, Italy}

\author{C.J. DAI, H.L. HE, X.H. MA, X.D. SHENG, R.G. WANG and Z.P. YE\thanks{also: 
University of Jing Gangshan, Jiangxi, China}}

\address{Institute of High Energy Physics,
CAS, 19B Yuquanlu Road, Shijingshan District, Beijing 100049, China}

\maketitle

\vspace{0.5cm}

\begin{abstract}

Experimental observations and theoretical arguments at Galaxy and larger scales have suggested 
that a large fraction of the Universe is composed by Dark Matter particles.
This has motivated the DAMA experimental efforts to investigate the
presence of such particles in the galactic halo by exploiting a model independent signature and
very highly radiopure set-ups deep underground. Few introductory arguments are summarized before
presenting a review of the present model independent positive results obtained by 
the DAMA/NaI and DAMA/LIBRA set-ups at the Gran Sasso National Laboratory of the INFN.
Implications and model dependent comparisons with other different
kinds of results will be shortly addressed. Some arguments 
put forward in literature will be confuted.
\end{abstract}

\ccode{PACS numbers: 95.35.+d; 29.40.Mc}

\vspace{0.5cm}

\section{The Dark component of the Universe}

Since the publication (1687) of the ``$Philosophiae$ $Naturalis$ $Principia$ $Mathematica$'' by  Isaac
Newton, a large effort has been made to explain the motion of the astrophysical objects by means of the universal 
gravitation law. 
Every time some anomalies in the motion of astrophysical objects 
with respect to the theoretical predictions have been observed,
a question arises about the possibility that either it could be a proof of a violation of the 
gravitational law or a presence of objects not yet observed or which cannot 
be observed; for example, in 1846 the anomalous motion of Uranus allowed the prediction of the 
existence of Neptune and then its observation.

The present topic about Dark Matter (DM) 
is conceptually very similar to the old problem of unobserved planets;
in fact, today in large astrophysical systems -- from galactic 
scale to the cosmological one -- many effects are observed, whose explication requires the 
presence of a large quantity of non-visible matter. 

First attempts to derive the total density of matter in the Solar vicinity were made by \"Opik (1915)\cite{Opi15}, Kapteyn (1922)\cite{Kap22}, 
Jeans (1922)\cite{Jea22}, and Oort (1932)\cite{Oort32}, 
and the term ``Dark Matter'' was firstly used in 1922 by Kapteyn.

In 1930 first measurements on the red-shift in galaxies and in clusters of
galaxies were carried out. In particular, in 1933 investigating the Coma cluster 
Zwicky \cite{Zwi33,Zwi33A} observed a discrepancy between the visible matter (estimated by its luminosity, $L$, in solar units) and the total matter of the 
cluster, $M$, (measured in solar units) equal to $M/L\sim$500. In 1936 Smith \cite{Smi36}
obtained an analogous result: $M/L\sim$200, investigating the Virgo cluster. In 1939, other results 
were obtained for the $M/L$ ratio 
by Babcock \cite{Bab39} investigating the 
rotational velocity of the M31 galaxy; in particular, he observed velocity values in the 
peripheral region of this galaxy surprisingly high.
In 1940 Oort \cite{Oort40} confirmed the result by Babcock also in other galaxies measuring a
$M/L$ ratio of about 250. In the '50 
Page \cite{Page,PageA,PageB} performed similar studies on the double galaxies obtaining 
$M/L$ values always of order of 10$^2$, and observing 
$M/L$ values for elliptical galaxies about 5 times larger than those measured for the 
spiral ones. In the same period Kahn and Woltjer \cite{Kah59} pursued a new 
approach to investigate the mass of galaxies systems: they investigated the binary system made of our galaxy and 
M31 by measuring the relative motion. This 
allowed them to derive the mass of the system, obtaining a result that was about 
10 times larger than those available at that time. 
They attributed this ``extra-mass'' to intergalactic material, indeed 
they measured the masses of the two galaxies including their dark haloes,
as shown today by new measurements \cite{Ein10}.
After the second worldwide war German radar installations were kept in 
measurement in the Netherlands; Oort and his collaborators decided to use them 
to study the radio emissions of astrophysical objects. In 1954 van de Hulst 
\cite{van54} calculated that the hydrogen has an emission line  
in the range of the radio-waves frequencies, and pointed out how this radio-emission can allow 
the measurement of the velocity of the inter-galactic hydrogen gas. Studies of such a radio-emission 
were performed firstly on our galaxy \cite{van54} and, then, on the 
M31 galaxy \cite{van57}. The measurements on M31 showed that the hydrogen gas was 
present well beyond the luminous region. Moreover, 
these measurements allowed the investigation of the rotational curve of the Galaxy
up to a distance of 30 kpc from the galactic center, and thus 
the determination of the $M/L$ ratio as a function of the galactic radius.
About ten years later, in 1966, Roberts performed a new study of the radio-emission 
at 21 cm in order to build a mathematical model of the M31 galaxy
\cite{Rob66}; the new data were in perfect agreement with those obtained by 
van de Hulst and collaborators.

All these efforts and progresses allowed the validation -- 
in the middle of the '70 -- of the existence of Dark Matter. 
In particular, a crucial result to credit the presence of Dark Matter 
was obtained by combining the results of
two groups about the rotational velocity of the M31 galaxy as a function of the galactic 
radius. Rubin and Ford \cite{Rub70} in 1970 obtained the rotational curves of spiral galaxies by studying the emission 
in the visible region, while Roberts and
Rots \cite{Rob75} obtained the rotational curves investigating the radio-emissions.
Subsequent studies allowed the investigation of the rotational curves 
for many spiral galaxies, obtaining typically a local $M/L$ ratio in the peripheral region
of order of $\sim 10^2$. In particular, it was shown that all the spiral galaxies 
have flat rotational curves at large distance \cite{Rub78,Rub80,Bos78}. 
All the results confirmed the presence of a dark halo in the galaxies with 
high confidence level.
With time passing other data have been collected \cite{Rub01}; new measurements have 
allowed the investigation of rotational curves up to 80-100 kpc. In the case of the Milky Way,
the rotational curve has been determined up to 
$\simeq$60 kpc \cite{Xue08} by studying $\sim$2500 stars with the
$Sloan$ $Digital$ $Sky$ $Survey$ (SDSS)
\cite{SDSS7,SDSS7A}. The measurements have shown that the velocity  
tends to decrease with respect to 
the $\sim$ 220 km/s value, which roughly holds in the Sun region.
On the other side, the estimate of the DM halo are still uncertain with values varying in the 
$\sim (1-2.5) \times 10^{12} M_\odot$ interval, where $M_\odot$ = $1.99 \times 10^{30}$ kg is 
the solar mass.
As regards the DM density in the center of the Galaxy, it turns out to be almost equal for 
all galaxies\cite{Gil07}: $\sim 0.1 M_\odot$ pc$^{-3}$.

It is worth noting that the studies of the dispersion velocities of the spherical Dwarf galaxies show
$M/L$ ratio larger than that observed in the regions nearby us \cite{GNS,GNSA}.
The dispersion velocities of satellite spiral galaxies suggest 
the existence of a dark halo around spiral galaxies (as our one) extended up to 
$\sim$ 200 kpc, far away from the galactic disk \cite{GSV,GSVA}. 

The behaviour of the rotational velocity of the galaxies as a function of the 
galactic radius gave very strong support to the existence 
of a Dark Matter component in the galactic haloes.

Recent data \cite{Gil89} show no evidence for the presence of a sizeable quantity 
of Dark Matter in the disk 
of the Milky Way (its quantity is of order of 15\% of the total matter). Moreover, the
DM in the central region of the galaxy has with high probability a baryonic nature 
(proto-stars, non-luminous stars, ...) considering that the 
non-baryonic DM cannot form a flatten structure (as the galactic disk) being
dissipation-less. 

Additional information on the mass distribution in the external region of the Galaxy 
comes from the streams of stars and from the gas\cite{Mat74}. Kinematic data available for such
streams support the existence of a dark and massive galactic halo  
\cite{Ein76}. Using the data of the motions of satellite galaxies and of 
globular clusters, various determinations of the mass of the dark halo give: 
$m_{DarkHalo} \simeq 2 \times 10^{12} M_\odot$, in agreement with previous determinations.
In the recent years other streams of stars and gases have been pointed out both in the Milky Way 
and in nearby galaxies as M31. At present the data available for the kinematics of these 
streams are few, but in next years important advancements could be expected from the data of 
the GAIA satellite \cite{GAIA}, whose aim is the measurement of distances and photometric data of millions 
of stars in the Milky Way.

Other information came from the
the $X$-rays emissions by galaxies and galaxies clusters studied by 
the Einstein observatory \cite{San91},
ROSAT, XMM-Newton and Chandra.
Important data also arise from the gravitational $micro$-$lensing$ produced by 
stars, planets and invisible matter systems. As regards big and distant galaxies, 
the main information is given by the rotational curves and the $X$-rays emission of the hot gas surrounding them;
for the nearby Dwarf galaxies 
the main information arises instead from the analysis of the stellar motion.

Recently, an important observation, exploiting the gravitational lensing and the $X$-rays emission, 
has been performed on the Bullet Cluster (1E 0657-558), that is one of the best evidence of
Dark Matter on cosmological scale \cite{Clo06}. The studies have been confirmed by the investigation
on the MACS J0025.4-1222 cluster too \cite{Bra08}.
In addition, it has been detected  a dark-matter filament, connecting 
the two main components of the Abell 222/223 supercluster system, through
its weak gravitational lensing signal\cite{Die12}. 

The modern cosmological interpretation includes Dark Energy as 
a basic component of the matter-energy content of the Universe. 
Direct observational evidence for the presence of Dark Energy comes 
from distant supernovae\cite{SNe1a,SNe1aA,SNe1aB,SNe1aC,SNe1aD} and from Cosmic Microwave Background (CMB) observations.
After the first studies by IRAS and COBE, WMAP and -- recently -- PLANCK missions have allowed
the measurement of the CMB radiation and its power spectrum with a very high precision \cite{Spe03,Spe03A}.
The position of the first maximum of the power spectrum depends on the total matter-energy density of the Universe. 
The spectrum of the CMB fluctuations is presently known down to the scales of 0.1 degree.
In particular, the positions and the relative sizes of the peaks in the power spectrum of CMB strongly support the 
existence of Dark Matter in the Universe.
In particular, the average density of matter-energy content of the Universe in unit of the cosmological critical density
(the amount of matter-energy required to make the Universe spatially flat), $\Omega$, is consistent with 1.
Considering all the observational data coming from the study of the CMB, of the Supernovae Ia, of the 
Baryonic Acoustic Oscillations (BAO)
and of the large-scale structures, the following contributions to $\Omega$ are obtained\cite{PDG,Planck}:
i) $\Omega_r \approx 5 \times 10^{-5}$ for the radiation density;
ii) $\Omega_b \approx$ 0.05 for the baryonic matter;
iii) $\Omega_{dm} \approx$ 0.27 for the non baryonic Dark Matter;
iv) $\Omega_{\Lambda} \approx$ 0.68 for the Dark Energy. 
A value $\Omega_b \approx$ 0.05 for the baryonic matter is also supported by the Big-Bang nucleosynthesis (BBN),
that is based on the predictions of the abundances of the light elements.
This value is much larger than the cosmic density of the luminous matter, $\Omega_{lum} \approx 0.004$ 
\cite{fuku04}, so that most baryons are dark and, probably, in the form of diffuse intergalactic medium \cite{cen99}.

For completeness, we recall that some efforts to find alternative explanations to Dark Matter have been proposed 
such as  $MOdified$ $Gravity$ $Theory$ (MOG) in the 1980s \cite{MOG,MOGA} and $MOdified$ $Newtonian$ $Dynamics$ (MOND) 
theory in 1981 \cite{Mil87,Mil87A,Mil87B}.
They hypothesize that the theory of gravity is incomplete and that a new gravitational theory could explain 
the experimental observations.
MOND modifies the law of motion for very small accelerations, while MOG modifies the Einstein's theory of 
gravitation to account for an hypothetical fifth fundamental force in addition to the gravitational, electromagnetic, 
strong and weak ones.
However, they are unable to account for small and large scale observations, and generally require some amount of Dark 
Matter particles as seeds for the structure formation; moreover, they fail to reproduce accurately the Bullet Cluster.

Therefore, at the present status of knowledge and considering all the available experimental data, the existence of
Dark Matter in the Universe is well recognized, and there are compelling arguments to investigate the presence of
Dark Matter particles also at galactic scale.

\subsection{DM candidates}

The values of the cosmological parameters support that most of the 
matter in the Universe has a non baryonic nature.
The only DM candidate among the known elementary particles is the neutrino;
the density of light neutrinos is strongly constrained by cosmology. In fact, a value above 
the limit $\Omega_{\nu} \approx$ 0.03 gives an unacceptable lack of small-scale structure \cite{cro99,cro99A}.
In addition, a pure light neutrinos scenario is also ruled out by the measurements of the CMB radiation, which does not 
show sufficiently large inhomogeneity.

Therefore, a significant role should be played by non-baryonic relic particles from the Big Bang, outside the 
Standard Model of particle physics; they have to be stable or with a lifetime comparable with the age of the Universe 
to survive up to now in a significant amount. 

In theories extending the Standard Model of particle physics, many candidates as Dark Matter particles have been 
proposed having different nature and interaction types.
It is worth noting that often the acronym {\it WIMP} (Weakly Interacting Massive Particle) is adopted as a synonymous of Dark Matter particle, referring
usually to a particle with spin-independent elastic scattering on target-nuclei. On the contrary, {\it WIMP} identifies a 
class of Dark Matter candidates which can have different phenomenologies and interaction types among them. This is also the case
when considering a given candidate as for example the neutralino; in fact the basic supersymmetric theory has 
a very large number of parameters which are by the fact unknown and, depending on the assumptions, the candidates can have 
well different features and preferred interaction types.
Often constrained SUGRA models (which allow easier calculations for the predictions e.g. at accelerators) are 
presented as SUSY or as the only way to SUSY, which is indeed not the case.

Among the many DM candidates we recall: SUSY particles (as e.g. neutralino 
\cite{neutr,neutrA,neutrB,neutrC,neutrD,neutrE,neutrF,neutrG,neutrH,neutrI,neutrJ} or 
sneutrino in various scenarios \cite{sneu,sneuA,sneuB,sneuC}), inelastic Dark Matter in various scenarios 
\cite{inel,inel_other,inel_otherA,inel_otherB}, 
electron interacting Dark Matter (also including some {\it WIMP} scenarios) \cite{wimpele,em_int}, a heavy neutrino of the 4-th family, 
sterile neutrino \cite{ldm},  Kaluza-Klein particles, self-interacting Dark Matter \cite{vog12}, axion-like (light pseudoscalar 
and scalar candidate) \cite{ijma}, mirror Dark Matter in various scenarios \cite{mirror,mirrorA,mirrorB}, Resonant Dark Matter 
\cite{resonant}, DM from exotic 4th generation quarks \cite{4gen}, Elementary Black holes \cite{Al,AlA,AlB,AlC}, 
Planckian objects \cite{Al,AlA,AlB,AlC}, 
Daemons \cite{Al,AlA,AlB,AlC}, Composite DM \cite{composite,compositeA}, Light scalar {\it WIMP} through Higgs portal \cite{andr10}, Complex 
Scalar Dark Matter \cite{complDM}, specific two Higgs doublet models, exothermic DM \cite{exoth}, Secluded {\it WIMPs} 
\cite{seclDM}, Asymmetric DM \cite{asymDM}, Isospin-Violating Dark Matter \cite{isoDM,isoDMA}, Singlet DM \cite{singletDM,singletDMA}, 
Specific GU \cite{specGU,specGUA}, SuperWIMPs \cite{superDM,superDMA}, WIMPzilla \cite{wimpzilla}, and also further scenarios 
and models as e.g. those given in Ref. \cite{candidates,candidatesA,candidatesB,candidatesC,candidatesD,candidatesE}. 
Moreover, even a suitable particle not yet foreseen by theories could be 
the solution or one of the solutions\footnote{In fact, it is worth noting that, considering the richness in particles
of the visible matter which is less than 1\% of the Universe density,
one could also expect that the particle component of the Dark Matter in the
Universe may also be multicomponent.}.

Depending on the DM candidate, the interaction processes can be various, as e.g:
1) elastic scatterings on target nuclei with either spin-independent or spin-dependent or mixed 
   coupling; moreover, an additional electromagnetic contribution can arise,
   in case of few GeV candidates, from the excitation of bound electrons by the recoiling nuclei \cite{em_int};
2) inelastic scatterings on target nuclei with either spin-independent or spin-dependent or mixed coupling in various
   scenarios \cite{inel,inel_other,inel_otherA,inel_otherB,Wei01,Wei01A};
3) interaction of light DM (LDM) either on electrons or on nuclei with production of a lighter particle \cite{ldm};
4) preferred interaction with electrons \cite{wimpele};
5) conversion of DM particles into electromagnetic radiation \cite{ijma};
6) etc..
Thus, the Dark Matter interaction processes can have well different nature depending on the candidate.
Often, the elastic scattering on target nuclei is the considered process,
but other processes are possible and considered 
in literature, as those aforementioned where also electromagnetic radiation is produced. 
Hence, considering the richness of particle possibilities and the existing uncertainties on related 
astrophysical (e.g. halo model and related parameters, etc.), nuclear (e.g. form factors, spin factors, 
scaling laws, etc.) and particle physics (e.g. particle nature and interaction types, etc.), a widely-sensitive 
model independent approach is mandatory. Indeed, most of the activities in the field are based 
on a particular {\it a priori} assumption on the nature of the DM particle and of its interaction, in order to try to 
overcome the limitation arising from their generally large originally measured counting rate.

\subsection{The density and velocity distribution of Dark Matter in the Galaxy} \label{p:halo}

The expected energy distribution for the interactions of Dark Matter particles 
in a terrestrial detector depends on their density and velocity distribution at Earth's position.
However, the experimental observations regarding the dark halo of our Galaxy 
do not allow us to get information on this crucial key item without introducing a model for the 
Galaxy matter density.
A widely used density distribution of Dark Matter is the isothermal sphere model;
it consists in a spherical infinite system with a flat rotational curve.
Due to its simplicity, the isothermal sphere model is often the used assumption 
in the evaluation of Dark Matter expected rates.
However many of its underlying assumptions (sphericity of the halo, absence of 
rotation, isotropy of the dispersion tensor, flatness of the rotational curve) 
are not strongly constrained by astrophysical observations.
Moreover, the isothermal sphere is strictly unphysical and may only represent the 
behavior of the inner part of physical systems, since it has a total infinite 
mass and needs some cutoff at large radii.
Thus, the use of more realistic halo models is mandatory in the interpretation and comparison  
procedures, since the model dependent results can significantly vary.
An extensive discussion about some of the more credited halo models has 
been reported e.g. in Ref. \refcite{Hep,RNC} and references therein.

Generally, the halo models can be grouped in the following classes:
i) spherically symmetric matter density with isotropic velocity dispersion;
ii) spherically symmetric matter density with non-isotropic velocity dispersion; 
iii) axisymmetric models;
iv) triaxial models;
v) axisymmetric models either with a halo co-rotation or a halo counter-rotation. 
The possible rotation can be considered also for the other kinds of halo.
A parameterization of these classes of halo models has been given in Ref. \refcite{Hep,RNC},
taking into account the available observational data.
Anyhow, information on the galactic dark halo can be obtained only in indirect way 
\cite{Deh98,Gat96} and considering some hypotheses on its form and characteristic.

In particular, the allowed range of values for the Dark Matter local velocity \cite{Hep} can be estimated 
considering the information coming from the rotational curve of our Galaxy: 
$v_0=(220 \pm 50)$ km s$^{-1}$ (90\% C.L.),
that conservatively relies on purely dynamical observations \cite{koc96}. Similar 
estimates of the $v_0$ central value with smaller uncertainty have been obtained 
studying the proper motion of nearby stars in the hypothesis of circular orbit of 
these objects. 
For example, the value $(221 \pm 18)$ km s$^{-1}$ has been
determined mapping the GD-1 stellar stream \cite{Kop10}. As discussed in Ref. \refcite{McM10},
these determinations are strongly dependent on other parameters (the distance from the galactic
center, the solar velocity $v_{\odot}$, the adopted density profile, etc.). These
parameters are themselves affected by strong uncertainties;
thus, depending on their choice the estimate of the rotational speed may vary from
$(200\pm20)$ to $(279\pm33)$ km s$^{-1}$. Similar considerations can also be done for the
escape velocity of the Galaxy, on which many DM models, and in particular the inelastic DM and the 
low mass DM inducing nuclear elastic scatterings, are critically dependent.

For each model -- after fixing the local velocity -- 
the allowed range of local density $\rho_0$ can be evaluated; it ranges 
for the considered models in the values 0.2--1.7 GeV cm$^{-3}$
when taking into account the following physical constraints: i) the amount of flatness of the rotational curve of our 
Galaxy, considering conservatively $0.8 \cdot v_0  \lsim v_{rot}^{100}  \lsim  1.2 \cdot v_0$, 
where $v_{rot}^{100}$ is the value of rotational curve at distance of 100 kpc from the 
galactic center; ii) the maximal non dark halo components in the Galaxy, considering 
conservatively $1 \times 10^{10} M_{\odot}  \lsim  M_{vis}  \lsim  6 \times 10^{10} 
M_{\odot}$ \cite{Deh98,Gat96}. 

Although the large number of 
self-consistent galactic halo models considered in Refs. \refcite{Hep,RNC}, still many other 
possibilities exist and have been proposed in recent years, such as the Einasto profile \cite{Ein09}. 
Moreover, the possible contributions of non-thermalized Dark Matter components 
to the galactic halo, such as the SagDEG stream\cite{sag06} and other kinds of streams as those arising
from caustic halo models\cite{lin04}, could change the local DM speed 
distribution and the local density. These contributions can also play a significant 
role in the model-dependent investigations for the candidate particle. Some analyses in this scenarios have
been proposed \cite{sag06}.

To improve our knowledge of the galactic haloes, low surface brightness (LSB) galaxies 
dominated by Dark Matter have been studied \cite{deB10}; these studies have shown a cored profile 
distribution in all the considered galaxies excluding the presence of the cusp, expected by $\Lambda$CDM 
model. In the light of these experimental results on LSB galaxies, which show cored profiles, and of 
the theoretical results of N-body simulations (Millennium, DEUS FUR, Horizon, Multi Dark, Bolshoi,
Aquarius, Phoenix etc.), which instead expect a cusp in the galaxies central region, new efforts are in progress
with the aim to develop new N-body simulations able to reproduce all the experimental observations.
A review on the state of art is given, for example, in Ref. \refcite{Kuh12}.

In conclusion, the uncertainties still present on the shape of the DM halo and on the density and velocity  
distribution prevent the definition of a ``standard'' halo and illustrate how the comparisons among 
the experiments of direct detection of DM particles (see later) can be consistent even just considering this
particular aspect (also see Ref. \refcite{mao13}).

\section{The Dark Matter particles detection}

In the following, we will briefly discuss and comment about the indirect and the direct detection 
of Dark Matter particles. Firstly let us comment about the possibility to detect Dark Matter
by accelerator experiments. It is worth noting that experiments at accelerators may prove 
-- when they can state a solid model independent result -- 
the existence of some possible DM candidate particles,
but they could never credit by themselves that a certain particle is a/the only solution 
for Dark Matter particle(s). Moreover, DM candidate particles and 
scenarios (even e.g. in the case of the neutralino candidate) exist which cannot be investigated at accelerators.
Thus, a model independent approach, a ultra-low-background suitable target material, 
a very large exposure and the full control of running conditions are mandatory 
to pursue a widely sensitive direct detection of DM particles in the galactic halo.

\subsection{Indirect approaches}
We preliminary briefly remind the current state of indirect Dark Matter searches, whose results are
strongly model-dependent. They are generally performed 
as by-product of experiments located underground, under-water, under-ice or in space having different main scientific 
purpose. In particular, these experiments search for the presence of secondary particles 
produced by some kind of Dark Matter candidates able to 
annihilate in the celestial bodies when some specific assumptions are fulfilled.
Thus, their results are restricted to some candidates and physical scenarios, and require also the modeling of the 
existing -- and largely unknown -- competing background for the secondary particles they are looking for. 

It is worth noting that no quantitative comparison can be directly performed 
between the results obtained in direct and indirect searches 
because it strongly depends on assumptions and on the considered model framework.
In particular, a comparison would always require the calculation and the consideration of all the possible 
configurations for each given particle model (e.g. for neutralino: in the allowed parameters space), as a biunivocal 
correspondence between the observables in the two kinds of experiments does not exist: cross sections in the direct 
detection case and e.g. flux of muons from neutrinos (or of other secondary particles) in the indirect searches.
In fact, the counting rate in direct search is proportional to the direct detection cross sections, while the flux of 
secondary particles is connected also to the annihilation cross section. In principle, these cross sections can be 
correlated, but only when a specific model is adopted and by non directly proportional relations.

As regards under-water and under-ice experiments, results are available from Antares \cite{antares} and Icecube 
\cite{icecube}, both looking for up-going muons ascribed to muon neutrino interactions in the Earth and assuming they 
have been produced in the annihilation of Dark Matter particles in a certain considered scenario; no excess above 
an estimate of up-going muons from atmospheric neutrinos has been presented by both experiments.
Similar approaches have been considered in some kinds of underground detectors (for example, MACRO, Superkamiokande, ...) as well.
As regards the space investigations, an excess of the measured positron fraction above an assumed background model 
was presented by Pamela \cite{pamela,pamelaA} and AMS-02 \cite{AMS02} experiments, but analogous models also 
exist with different secondary production 
giving no very significant deviation \cite{other_sour,other_sourA,other_sourB}.
In addition, since no excess has been observed in the anti-proton spectrum, a similar candidate should be ``leptophilic''; that 
is, e.g. not observable by those direct detection experiments which select just nuclear recoil-like events from the measured 
counting rate, as e.g. CDMS, Edelweiss, Cresst, XENON, etc..\footnote{In fact, to produce results on electron 
recoils, those experiments should e.g. abandon the many data selections they apply.
Thus, since generally their original counting rate is very large, they are by the fact insensitive to signals from
electron recoils. Therefore, such (leptophilic) candidates can hardly be detected by those experiments.}, 
while it can be detected in DAMA experiments which exploit a different methodology (see later). 
Anyhow, additional aspects arise when trying to explain the Pamela and AMS-02 data in a Dark Matter interpretation since e.g. a 
very large boost factor ($\sim 400$) would be required, whose origin cannot be easily justified unless introducing 
a new kind of interaction \cite{sommerfeld}.
Thus, this excess can be due to an inadequacy of the considered model used to describe and propagate all the possible sources
of secondaries; moreover, in literature it has also been shown that some kinds of known sources can account for a similar positron 
fraction \cite{other_sourC,other_sourD,other_sourE}.
Therefore, no constraint on direct detection phenomenology arises from positron fraction; anyhow, if 
those data were interpreted -- under several assumptions -- in terms of some Dark Matter scenario, this would be not in 
conflict with the DAMA model-independent results described later.

Another possible model-dependent positive hint from space and its compatibility with the DAMA results have been discussed in Ref. 
\refcite{hooper10}, considering a particular analysis of data from FERMI: 
close to the Galactic Center the spectral shape of the observed emission is significantly different, peaking at 1-5 GeV,
with respect to a background model which well describes the spectral shape outside the Galactic Center over the disk.
It has been suggested that this may be due to annihilations of light Dark Matter particles.
Another indication for a possible excess above an assumed background model from FERMI-LAT
is in the form of a narrow bump in the $\gamma$-ray spectrum
near 130 GeV \cite{Wen12}; however, a deeper investigation
shows the presence of two bumps at 135 GeV (2.2$\sigma$ C.L.) and near 130 GeV in the Earth limb data
(about 3$\sigma$ C.L.). The FERMI collaboration points out a few other details \cite{frslide}: i) the 135 GeV feature is not as smooth as expected
if it were due to some Dark Matter annihilation; ii) similar bumps are present with similar significance at other energies 
and other places in the sky (as those near 5 GeV at 3.7$\sigma$ C.L.). In addition, 
the inclusion of other astrophysical objects can account for the FERMI data. 

All that shows the intrinsic uncertainties of the DM indirect searches to unambiguously assess the presence of a Dark 
Matter signal, and the absence of direct constraints on direct searches.

\subsection{Direct approaches}

In the direct detection field, different experimental approaches have been considered 
in order to point out the presence of Dark Matter component(s) in our Galaxy. 
Originally it was pursued an approach based on the simple comparison of the measured energy distribution
in a defined energy interval, with one of the possible expectations fixing some particular astrophysical, nuclear
and particle scenario. This approach is the only feasible one either for small scale or poor duty
cycle experiments and allows the exclusion of some free parameters (e.g. particle mass and 
particle-nucleus cross-section) at a given C.L. for the assumed model scenario.  
Although for long time the limits achieved by this approach
have been presented as robust reference points, similar results are quite uncertain
because e.g. they depend on the specific model framework used to estimate
the expected counting rate as well as obviously on the reliability of the adopted 
experimental procedures and parameters.
In fact, it is worth noting that this approach generally assumes {\it a priori} the nature of the DM candidate, 
its interaction type, astrophysical, nuclear and particle physics aspects and experimental features. 
For example, exclusion plots can be modified even by orders of magnitude
by changing (within the allowed intervals) the values 
of the parameters used in a given scenario. 

In addition, important sources of uncertainties can arise from 
the under-estimation of systematics in
experiments adopting relevant data handling.
In fact, more recently, some experiments -- to try to reduce their experimental counting rate --
apply large data selections and several subtraction procedures 
in order to derive a set of nuclear recoil-like candidates, assuming {\it a priori} 
-- among others -- the nature and the interaction type of the DM candidate.
Thus, each exclusion plot should be considered strictly
correlated with the assumed model framework, with the used experimental/theoretical assumptions and
parameters and with detailed information on possible data reduction/selection, on
efficiencies, calibration procedures, etc. 
Moreover, when for example a generic DM candidate inducing just nuclear recoils with purely spin-independent (SI) coupling 
is considered, in order to try to compare exclusion plots as well as allowed regions 
the results are generally scaled to cross sections on nucleon for the different target nuclei; thus some 
scaling law(s) has(have) to be adopted.
A scaling law according to 
$A^2$ is generally assumed (where $A$ is the atomic number of each considered nucleus), while relevant deviations 
are possible.
In particular, for the case of neutralino candidate the two-nucleon currents in the nucleus
can add different contribution for different nuclei (see as an example of some discussion Ref. \refcite{scalaw}).
Also this problem of scaling laws is one of the aspects that have a great relevance in 
the comparison between results obtained by experiments using different target materials.
By the fact, the scaling laws should be derived in Physics from the experimental measurements
themselves and not imposed {\it a priori}.
Just as another example, we mention that when the case of DM candidates with purely or mixed spin dependent (SD) 
interaction scattering on target nuclei is considered, 
a relevant role is played by the nuclear spin factor of each target nucleus \cite{RNC}; in fact, not only large differences
in the measured rate can be expected for
target nuclei sensitive to the SD component of the interaction (such as e.g.
$^{23}$Na and $^{127}$I) with respect to those largely insensitive to such a
coupling (such as e.g. $^{nat}$Ge and $^{nat}$Si), but also when using different
target nuclei although -- in principle --
all sensitive to such a coupling. In fact, very large differences exist in different nuclear models and 
even within the same model; in particular, e.g. in the case of $^{23}$Na and $^{127}$I, having a proton as unpaired nucleon,
and $^{29}$Si, $^{73}$Ge, $^{129}$Xe, $^{131}$Xe, having a neutron as unpaired nucleon, 
the sensitivities are almost complementary, depending 
on the $tg \theta = a_n/a_p$ value (ratio between the neutron and the
proton effective SD coupling strengths). See some values of spin factors reported e.g. in Ref. \refcite{RNC}.
Other arguments are addressed in the following.
The above mentioned uncertainties in the model evaluations and in the experimental procedures 
must be always taken into account when trying to compare exclusion plots and/or allowed regions, when available,
for each kind of scenario. 

Thus, in order to obtain a reliable signature for the presence of DM particles in the galactic halo,
it is necessary to follow a suitable model independent approach.
In particular, the only DM model independent signature presently feasible is the so-called DM annual modulation signature.
For completeness, we also mention a different approach described in the following 
for the direct detection of DM candidates inducing just nuclear recoils; it
is based on the study of the correlation of the nuclear recoil direction with the Earth velocity.
This approach has relevant technical difficulties, when the detection of
the short nuclear recoil track is pursued;
similar activities are up to now at R\&D stage. In particular, we remind DRIFT
(that has realized multi-wire proportional chambers at low pressure \cite{drift}), and
DM-TPC (that uses a
CCD camera to track recoils in a CF$_{4}$ volume \cite{dmtpc}).
A different strategy, which can allow one to overcome the present difficulties
avoiding the necessity to detect directly the nuclear recoil tracks,
has been suggested: the use
of anisotropic scintillators \cite{anis,anisA}. They offer different light responses depending on
the nucleus recoil direction with respect to the scintillator's axes (other efforts on this subject have
been performed in Refs. \cite{anis2,anis2A}); since the response to $\gamma/\beta$ radiation is isotropic instead,
these detectors can offer the possibility to efficiently study possible anisotropy of nuclear recoils \cite{anis,anisA}.
Low background ZnWO$_{4}$ crystal scintillators have recently been
proposed; the features and performances of
such scintillators are very promising \cite{adamo}.

It is also worthwhile to mention the investigation of possible sidereal diurnal effects, which
can be due to the Earth rotation velocity, to the shadow of the Earth (this latter case sensitive just to DM candidates with 
high cross sections and tiny density), and to channeling effects in crystals detectors 
and/or to anisotropic responses of detectors
in case of DM particle inducing nuclear recoils.

Some other arguments on direct search activities will be addressed in Sect. \ref{moddep}.
In the following we will introduce the annual modulation model independent signature, which
is the approach exploited by DAMA.

\subsection{DM model-independent annual modulation signature}

To obtain a reliable signature for the presence of DM particles in the galactic halo,
it is necessary to follow a suitable model independent approach.
In particular, the only DM model independent signature,
feasible and able to test a large range of
cross sections and of DM particle halo densities, is the so-called DM annual modulation
signature originally suggested in the middle of '80 in Ref. \refcite{freese,freeseA}.

In fact, as a consequence of its annual revolution around the
Sun, which is moving in the Galaxy traveling with respect
to the Local Standard of Rest towards the star Vega near
the constellation of Hercules, the Earth should be crossed
by a larger flux of Dark Matter particles around $\sim$ June $2^{nd}$
(when the Earth orbital velocity is summed to the one of the
solar system with respect to the Galaxy) and by a smaller
one around $\sim$ December $2^{nd}$ (when the two velocities are subtracted).
Thus, this signature has a different origin and peculiarities
than the seasons on the Earth and than effects
correlated with seasons (consider the expected value of the
phase as well as the other requirements listed below). 

Experimentally, the expected differential rate
as a function of the energy, $dR/dE_R$ (see Ref. \refcite{RNC} for detailed
discussion), depends on the Dark Matter particles velocity distribution
and on the Earth's velocity in the galactic frame, $\vec{v}_e(t)$.
Projecting $\vec{v}_e(t)$ on the galactic plane, one can write:
$v_e(t) = v_{\odot} + v_{\oplus} cos\gamma cos\omega(t-t_0)$;
here $v_{\odot}$ is the Sun's velocity with respect
to the galactic halo ($v_{\odot} \simeq v_0 + 12$ km/s and $v_0$ is
the local velocity whose value has been previously discussed);
$v_{\oplus} \simeq$ 30 km/s is the Earth's orbital
velocity around the Sun on a plane with inclination
$\gamma$ = 60$^o$ with respect to the galactic plane. Furthermore,
$\omega$= 2$\pi$/T with T=1 year and roughly t$_0$ $\simeq$ June 2$^{nd}$ 
(when the Earth's speed is at maximum). The Earth's velocity can be
conveniently expressed in unit of
 $v_0$: $\eta(t) = v_e(t)/v_0 = \eta_0 +
 \Delta\eta cos\omega(t-t_0)$,
where -- depending on the assumed value of the
local velocity -- $\eta_0$=1.04-1.07 is the yearly average of $\eta$ and
$\Delta\eta$ = 0.05-0.09. Since $\Delta\eta\ll\eta_0$, the expected counting
rate can be expressed by the first order Taylor expansion:
\begin{equation}
\frac{dR}{dE_R}[\eta(t)] = \frac{dR}{dE_R}[\eta_0] +
\frac{\partial}{\partial \eta} \left( \frac{dR}{dE_R} \right)_{\eta =
\eta_0} \Delta \eta \cos\omega(t - t_0) .
\end{equation}
Averaging this expression in a $k$-th energy interval one obtains:
\begin{equation}
        S_k\lbrack\eta(t)\rbrack = S_k\lbrack\eta_0\rbrack
  + \left[\frac{\partial  S_k}{\partial \eta}\right]_{\eta_0}
\Delta\eta cos\omega(t-t_0) =S_{0,k} + S_{m,k}cos\omega(t-t_0),
\label{eq:sm}
\end{equation}
with the contribution from the highest order terms less than 0.1$\%$.

This DM annual modulation signature is very distinctive
since a Dark Matter-induced effect must simultaneously satisfy
all the following requirements: the rate must contain a component
modulated according to a cosine function (1) with one year period (2)
and a phase that peaks roughly around $\simeq$ June 2$^{nd}$ (3);
this modulation must only be found
in a well-defined low energy range, where DM particles 
can induce signal (4); it must apply to those events in
which just one detector of many actually ``fires'', since
the Dark Matter multi-scattering probability is negligible (5); the modulation
amplitude in the region of maximal sensitivity must be $\lsim$7$\%$ for usually adopted halo distributions (6), but it can
be larger in case of some possible scenarios such as e.g. those in Refs. \cite{Wei01,Wei01A,Fre04,Fre04A}.
Thus, no other effect investigated so far in the field of rare processes offers
a so stringent and unambiguous signature.
With the present technology, the annual modulation remains the
main signature to directly point out the presence of Dark Matter particles in the galactic halo.

It is worth noting that, when exploiting such a signature, the experimental observable is not the constant part of 
the signal S$_0$ (hereafter $k$ is omitted) -- which instead the other approaches 
try to extract from their measured counting rate -- but its 
modulation amplitude, S$_m$, as a function of energy.
This has several advantages; in particular,
in this approach the only background of interest is that able to mimic the
signature, i.e. able to account for the whole observed modulation amplitude and to
simultaneously satisfy all the many specific peculiarities of the signature. No background of this sort has been
found or suggested by anyone over more than a decade.
In particular, the DM model-independent annual modulation 
approach does not require any identification of S$_0$ from the total {\it single-hit} counting rate,
in order to establish the presence of DM particles in the galactic halo. The
S$_0$ value can be worked out by a maximum likelihood analysis, which also takes into account the energy 
behaviour of each detector \cite{allDM,allDMA,allDMB,allDMC,allDMD,allDME,allDMF,allDMG,allDMH,allDMI,allDMJ,allDMK,allDML,allDMM}, 
for each considered scenario.
Thus, the DM annual modulation signature allows
one to avoid {\it a priori} assumptions on the nature and interaction type of
the DM particle(s) and to overcome the large uncertainties associated to the exploitation of
heavy data handling procedures, to the modeling of surviving
background in keV region, etc. pursued instead in approaches trying to extract $S_0$ from the measured counting rate 
assuming {\it a priori} the candidates and/or the interaction type, as mentioned above.

Obviously, the implementation of ULB ($Ultra$ $Low$ $Background$) detectors and set-ups,
the implementation of suitably controlled running conditions and the choice of a widely 
sensitive target material are important as well.

\section{The DAMA project}
\label{c1}

The DAMA experiment has been and is working as an observatory for rare processes
by developing and using low radioactive
scintillators. It is installed deep underground in the Gran Sasso National 
Laboratory (LNGS) of INFN. Several low background set-ups are operative and many different kinds of measurements are 
carried out 
\cite{Nim98,allDM,allDMA,allDMB,allDMC,allDMD,allDME,allDMF,allDMG,allDMH,allDMI,allDMJ,allDMK,allDML,allDMM,Sist,RNC,ijmd,ijma,sag06,em_int,chan,wimpele,ldm,allRare,perflibra,modlibra,modlibra2,papep,cncn,am241,DAMALXe,DAMALXeA,DAMALXeB,DAMALXeC,DAMALXeD,DAMALXeE,DAMALXeF,DAMALXeG,DAMALXeH,DAMALXeI,DAMALXeJ,DAMALXeK,DAMALXeL,DAMALXeM,DAMALXeN,DAMARD,DAMARDA,DAMARDB,DAMARDC,DAMARDD,DAMARDE,DAMARDF,DAMARDG,DAMARDH,DAMARDI,DAMARDJ,DAMARDK,DAMARDL,DAMARDM,DAMARDN,DAMARDO,DAMARDP,DAMARDQ,DAMARDR,DAMARDS,DAMAGE,DAMAGEA,DAMAGEB,DAMAGEC,DAMAGED,DAMAGEE,DAMAGEF,DAMAGEG,DAMAGEH,DAMAGEI,DAMAGEJ,DAMAGEK,DAMAGEL,DAMAGEM}.

In particular, dedicated set-ups have been developed to investigate 
the presence of DM particles in the galactic halo by exploiting the DM annual modulation
signature; they are: 
i) DAMA/NaI ($\simeq$ 100 kg of highly radiopure NaI(Tl)) that took data over 7 annual cycles 
and completed its data taking on July 2002 \cite{Nim98,allDM,allDMA,allDMB,allDMC,allDMD,allDME,allDMF,allDMG,allDMH,allDMI,allDMJ,allDMK,allDML,allDMM,Sist,RNC,ijmd,ijma,sag06,em_int,chan,wimpele,ldm,allRare};
ii) the second generation DAMA/LIBRA set-up ($\simeq$ 250 kg
highly radiopure NaI(Tl)) \cite{perflibra,modlibra,modlibra2,papep,cncn,am241,bot11}.

In particular, DAMA/NaI and DAMA/LIBRA use ULB NaI(Tl) detectors to investigate DM.
In fact, highly radiopure NaI(Tl) scintillators offer many competitive features
to effectively investigate the DM annual modulation signature, such as e.g.:
i)   high duty cycle; 
ii)  well known technology; 
iii) large masses feasible; 
iv)  no safety problems; 
v)   the lowest cost with respect to every other considered technique; 
vi)  necessity of a relatively small underground space; 
vii) reachable high radiopurity by suitable material selections and protocols, 
     by chemical/physical purifications, etc.;
viii) feasibility of well controlled operational conditions and monitoring;
ix)  feasibility of routine calibrations down to few keV in the same conditions as the 
     production runs; 
x)   high light response (that is keV threshold really reachable); 
xi)  absence of the necessity of re-purification or cooling down/warming up 
     procedures (implying high reproducibility, high stability, etc.); 
xii) absence of microphonic noise and effective noise rejection at threshold 
     (time decay of NaI(Tl) pulses is hundreds ns, while that of noise pulses is tens ns); 
xiii) possibility to use fragmented set-up which allows effective signal identification and
      background rejection;
xiv) wide sensitivity to both high and low mass DM candidates and to many different 
     interaction types and astrophysical, nuclear and particle Physics scenarios; 
xv)  possibility to effectively investigate the DM annual modulation 
     signature in all the needed aspects; 
xvi) possibility to achieve significant results on several other rare processes; 
xvii) etc.
Long dedicated R\&D projects have been implemented in order to 
select all the involved materials; moreover, purification procedures, protocols, etc. have
been adopted.
DAMA members made such efforts during about 20 years of developments. 

The DAMA/NaI set-up and its performances are described in Refs. \refcite{Nim98,Sist,RNC,ijmd}, while
the DAMA/LIBRA set-up, its main features and radiopurity have been discussed in 
Refs. \refcite{perflibra,scineghe09,modlibra2,tipp11,pmts}. Here we just summarize few information.

The DAMA/NaI experiment \cite{Prop} has been a pioneer experiment running at LNGS until 2002,
investigating as first the DM annual modulation signature 
with suitable exposed mass, sensitivity and control of the running parameters.
It was composed by nine 9.70 kg highly radiopure NaI(Tl) crystal scintillators in suitable radiopure Cu housing
installed in a low-background passive shield. 
Some upgradings have been performed during its operation
\cite{Nim98,allDM,allDMA,allDMB,allDMC,allDMD,allDME,allDMF,allDMG,allDMH,allDMI,allDMJ,allDMK,allDML,allDMM,Sist,RNC,ijmd,ijma,sag06,em_int,chan,wimpele,ldm,allRare}. 

The installation of DAMA/LIBRA started at the dismounting of the former DAMA/NaI in July 2002.
The experimental site as well as many components of the installation itself have been implemented.
All the procedures performed during the dismounting of DAMA/NaI and the installation of DAMA/LIBRA detectors have 
been carried out in high purity (HP) Nitrogen atmosphere.

The NaI(Tl) DAMA/LIBRA apparatus uses 25 NaI(Tl)
highly radiopure detectors with 9.70 kg mass each one
($10.2\times  10.2 \times  25.4$ cm$^3$ volume) placed in five rows by five
columns. The identification number of each detector (as it will be used later) 
is from the left (looking at the set-up from the door of inner barrack): 1 to 5 in the bottom (first) 
row, 6 to 10 in the second row and so on. 
The granularity of the apparatus is an interesting feature
for Dark Matter particle investigations since Dark Matter particles
can just contribute to events where only one of the 25 detectors
fires (single-hit events) and not to whom where more than one
detector fire in coincidence (multiple-hit events).

The new DAMA/LIBRA detectors have been built by joint efforts with 
{\it Saint Gobain Crystals and Detectors} company. 
The constituting materials
have been selected by using several techniques; moreover,
some chemical/physical purifications of selected powders have been
exploited and results tested. In addition, the more effective growing procedure has
been applied and new rules for handling the bare crystals have
been fixed. 
The bare crystals have been grown by Kyropoulos method in
a platinum crucible according to a strict growing and handling
protocol developed and agreed under confidential restriction.

The bare crystals are enveloped in Tetratec-teflon foils and encapsulated in radiopure OFHC Cu 
housing\footnote{The Cu housing has a shape able to compensate the different thermal
expansion of the NaI(Tl) and of the Copper.}; 
each detector is sealed in low radioactivity freshly
electrolyzed copper housing and has two 10 cm long highly
radiopure quartz (Suprasil B) light guides which also act as optical
windows being directly coupled to the bare crystals.
In each detector the light guides are optically coupled to the two end-faces of the bare crystal 
and to two low background photomultipliers (PMT). 
The threshold of each PMT is set at single photoelectron level; the coincidence of the two PMTs on a detector provides 
the trigger of the detector. The PMTs are shielded from the detectors through the light guides
and through a honeycomb-like shaped Cu shielding.

The detectors are housed in a low radioactivity sealed 
copper box installed in the center of the low-radioactivity Cu/Pb/Cd-foils/polyethylene/paraffin shield;
moreover, about 1 m concrete (made from the Gran Sasso rock material) almost fully surrounds (mostly outside the 
barrack) this passive shield, acting as a further neutron moderator.
The copper box is maintained in HP Nitrogen atmosphere in slightly overpressure with respect to the 
external environment; it is part of the 3-level sealing system which excludes the detectors from environmental air. 
The whole installation is air-conditioned to assure a suitable and stable working temperature for
the electronics. The huge heat capacity of the multi-tons passive shield ($\approx 10^6$ cal$/ ^{\circ}$C)
assures further a relevant stability of the detectors' operating temperature (see also later).
In particular, two independent systems of air conditioning are available for redundancy:
one cooled by water refrigerated by a devoted chiller and the other operating with cooling gas.

Following the same strategy as DAMA/NaI, on the top of the shield a glove-box (also continuously maintained in the HP Nitrogen
atmosphere) is directly connected to the inner Cu box, housing the detectors, through Cu pipes. 
The pipes are filled with low radioactivity Cu bars (covered by 10 cm of low radioactive Cu and 15 cm of low
radioactive Pb) which can be removed to allow the insertion of radioactive sources for calibrating the detectors 
in the same running condition, without any contact with external air. The glove-box is also equipped with a compensation chamber.

A hardware/software system to monitor the running conditions is operative and self-controlled 
computer processes automatically control several parameters and manage alarms.
For the electronic chain, the data acquisition system and for all the other details see Refs. \refcite{perflibra,modlibra2,pmts}.

The DAMA/LIBRA set-up allows the recording both of the {\it single-hit} events 
(those events where just one detector of many actually fires) and of the {\it multiple-hit} events 
(those events where more than one detector fire).
 The experiment takes data up to the MeV scale despite the optimization is done for the lowest energy region. 
The linearity and the energy resolution of the detectors at low and high energy have been investigated using 
several sources as discussed e.g. in Ref. \refcite{perflibra,pmts,modlibra,modlibra2}.
In particular, as regards the low energy region, calibrations down to the 3.2 keV X-ray have been carried out.
During the production runs periodical calibrations (every $\simeq$ 10 days) are carried out with $^{241}$Am 
sources, introduced in the proximity of the detectors by source holders inserted in the Cu pipes mentioned above.

The energy threshold, the PMT gain, the electronic line stability are continuously monitored during 
the data taking by the routine calibrations, by the position and energy resolution of internal lines \cite{perflibra}
and by the study of the behaviour of the hardware rate of each detector above single photoelectron level \cite{perflibra} with time.

The main procedures of the DAMA data taking for the investigation of DM particles 
annual modulation signature are:
1) the data taking of each annual cycle starts when $\cos \omega (t-t_0) \simeq 0$ in autumn towards summer, 
after June 2$^{nd}$ when the maximum is expected;
2) the routine calibrations with radioactive sources for energy calibrations and for
the evaluation of the acceptance windows efficiency \cite{perflibra} are performed about each 10 days 
(collecting typically $\simeq 10^4 - 10^5$ events per keV), moreover regularly intrinsic 
calibration are carried out, etc. \cite{perflibra};
3) the on-line monitoring of all the running parameters is continuously carried out with automatic alarm to operator 
if any would go out of allowed range.

\begin{table}[ht]
\begin{center}
\tbl{DAMA/LIBRA--phase1 annual cycles. Here $\alpha=\langle cos^2\omega (t-t_0) \rangle$ 
is the mean value of the squared cosine and $\beta=\langle cos \omega (t-t_0) \rangle$ 
is the mean value of the cosine (the averages are taken over the live time of the data taking
and $t_0=152.5$ day, i.e. June 2$^{nd}$); 
thus, $(\alpha - \beta^2)$ indicates the variance of the cosine
(i.e. it is 0.5 for a full year of data taking). 
The information on the cumulative exposure, when including the former DAMA/NaI,
is reported.}
{\resizebox{\textwidth}{!}{
\begin{tabular}{|l|c|c|c|c|}
\hline
\hline
 & Period & Mass & Exposure        & $(\alpha - \beta^2)$ \\
 &        & (kg) & (kg$\times$day) & \\
\hline
 & & & & \\
 DAMA/LIBRA-1 & Sept. 9, 2003 - July 21, 2004 & 232.8 &  51405  & 0.562 \\ 
   & &        &       & \\ 
 DAMA/LIBRA-2 & July 21, 2004 - Oct. 28, 2005 & 232.8 &  52597  & 0.467 \\ 
   & &        &       & \\ 
 DAMA/LIBRA-3 & Oct. 28, 2005 - July 18, 2006 & 232.8 &  39445  & 0.591 \\ 
   & &        &       & \\ 
 DAMA/LIBRA-4 & July 19, 2006 - July 17, 2007 & 232.8 &  49377  & 0.541 \\ 
   & &        &       & \\ 
 DAMA/LIBRA-5 & July 17, 2007 - Aug. 29, 2008 & 232.8 &  66105  & 0.468 \\ 
   & &        &       & \\ 
 DAMA/LIBRA-6 & Nov. 12, 2008 - Sept. 1, 2009 & 242.5 &  58768  & 0.519 \\ 
   & &        &       & \\ 
 DAMA/LIBRA-7 & to be released soon           & 242.5 &         &       \\
   & &        &       & \\ 
\hline
 DAMA/LIBRA-1 to -6 & Sept. 9, 2003 - Sept. 1, 2009 &  & 317697 & 0.519 \\ 
   & &        &    $\simeq$ 0.87 ton $\times$ yr     & \\ 
 DAMA/NaI + &   &  &  &  \\ 
 DAMA/LIBRA-1 to -6   & &        &    1.17 ton $\times$ yr    & \\ 
\hline
\hline
\end{tabular}
\label{tb:years}}}
\vspace{-0.4cm}
\end{center}
\end{table}

DAMA/LIBRA starts the data taking in 2003. A first upgrade of this set-up was performed in September 2008. 
One detector was recovered by replacing a broken PMT and a new optimization 
of some PMTs and HVs was done;
the transient digitizers were replaced with new ones 
(the U1063A Acqiris 8-bit 1GS/s DC270 High-Speed cPCI Digitizers)
having better performances and a new DAQ with optical read-out was installed. 
The DAMA/LIBRA--phase1 concluded its data taking in this configuration on 2010;
the data released so far correspond to 6 annual cycles. The data of the
seventh annual cycle of this phase1 will be released soon.
In particular, the measured light response was 5.5-7.5 photoelectrons/keV depending on the detector
and the software energy threshold was 2 keV electron equivalent (hereafter keV). 
Information about the annual cycles of DAMA/LIBRA--phase1 is given in Table \ref{tb:years}. 

A further and more important upgrade has been performed at the end of 2010 when all the 
PMTs have been replaced with new ones having higher quantum efficiency; details on the developments and
on the reached performances in the operative conditions are reported in Ref. \refcite{pmts}.
Since January 2011 DAMA/LIBRA--phase2 is running in order: 
1) to increase the experimental sensitivity lowering the software
energy threshold of the experiment; 
2) to improve the corollary investigation on the
nature of the Dark Matter particle and related astrophysical, nuclear and 
particle physics arguments; 
3) to investigate other signal features; 
4) to improve the sensitivity in the investigation of rare processes other than 
Dark Matter as performed by the former DAMA/NaI apparatus in the past \cite{allRare} and 
by itself so far \cite{papep,cncn,am241}. 
This requires long and heavy full time dedicated 
work for reliable collection and analysis of very large exposures, 
as DAMA collaboration has always done.

\section{The DAMA results}
\label{dama_res}

The DAMA/LIBRA data released so far correspond to
6 annual cycles for an exposure of 0.87 ton$\times$yr (see Table \ref{tb:years}) \cite{modlibra,modlibra2}.
Considering these data together with those previously collected by DAMA/NaI
over 7 annual cycles (0.29 ton$\times$yr), the total exposure collected
over 13 independent annual cycles is 1.17 ton$\times$yr; this
is orders of magnitude larger than the exposures typically collected in the field.

The only treatment, which is performed on the raw data, is to remove noise pulses
(mainly PMT noise, Cherenkov light in the light guides and in the PMT windows, and afterglows)
near the energy threshold in the {\it single-hit} events;
for a description of the used procedure and details see e.g. Ref. \refcite{perflibra}.

As regard calibrations, e.g. during the 6 DAMA/LIBRA annual cycles, about $7.2 \times 10^7$ events have 
been collected for energy calibrations
and about $3 \times 10^6$ events/keV for the evaluation of the acceptance windows 
efficiency for noise rejection near the energy threshold \cite{modlibra2}.
These periodical calibrations and, in particular, those related with the acceptance windows efficiency
mainly affect the duty cycle of the experiment.

\begin{figure}[!ht]
\resizebox{\columnwidth}{!}{\includegraphics{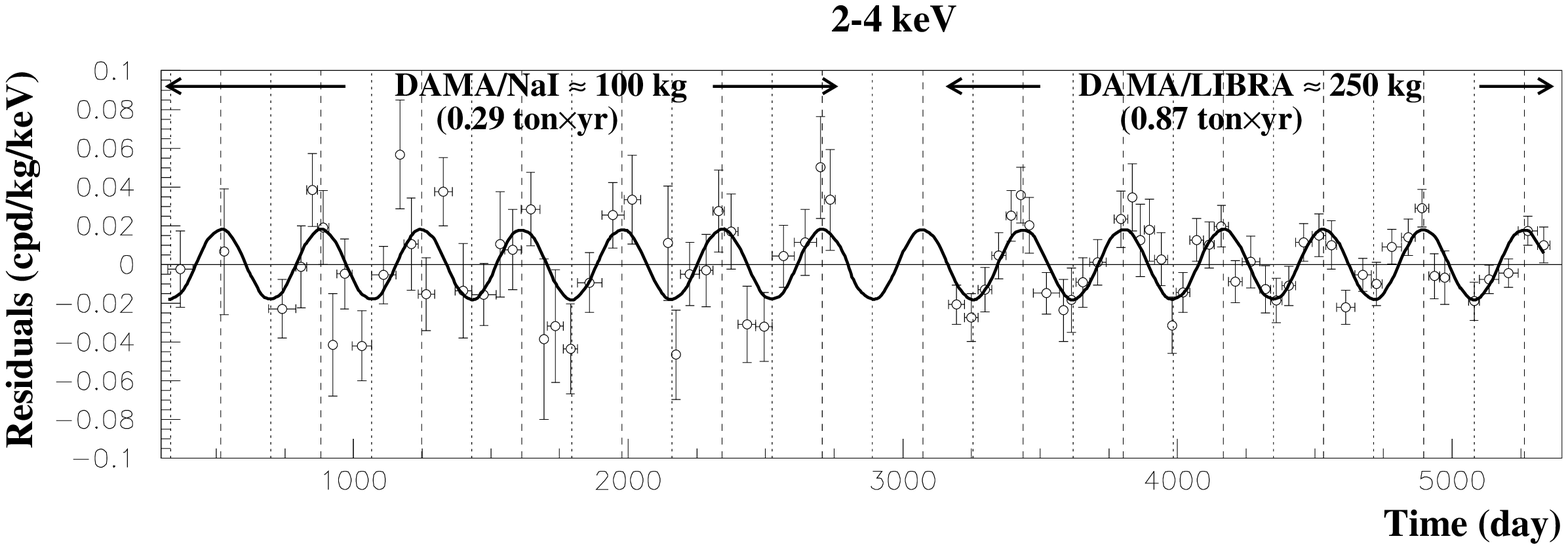}}
\resizebox{\columnwidth}{!}{\includegraphics{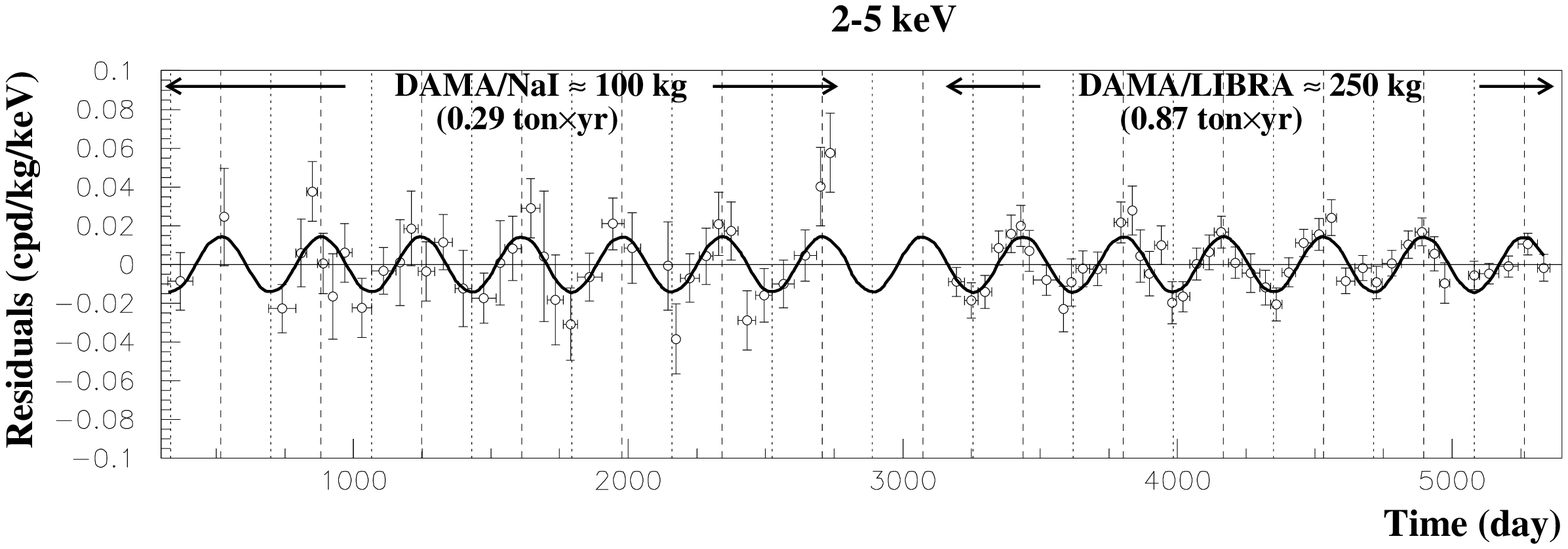}}
\resizebox{\columnwidth}{!}{\includegraphics{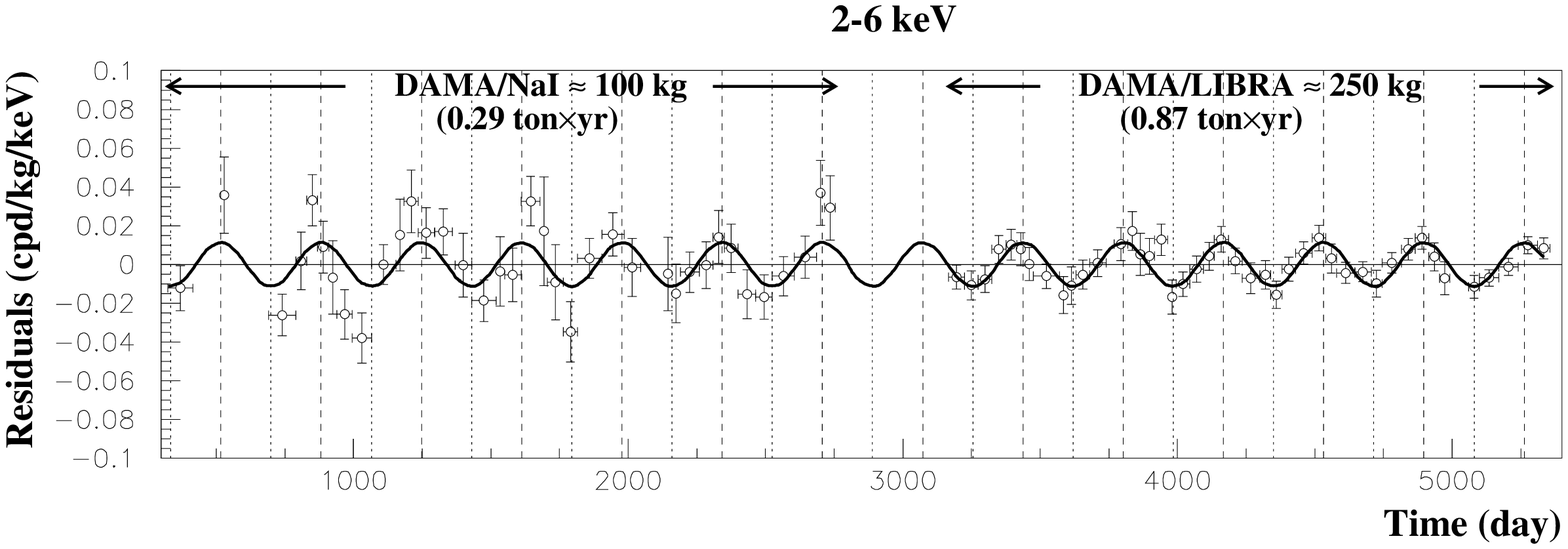}}
\caption{Experimental model-independent residual rate of the {\it single-hit} scintillation events,
measured by DAMA/NaI over seven and by DAMA/LIBRA over six annual cycles in the (2 -- 4), (2 -- 5), and
(2 -- 6) keV energy intervals
as a function of the time \protect\cite{RNC,ijmd,modlibra,modlibra2}. The zero of the time scale is January 1$^{st}$
of the first year of data taking.
The experimental points present the errors as vertical bars and the associated time bin width as horizontal bars.
The superimposed curves are $A \cos \omega(t-t_0)$
with period $T = \frac{2\pi}{\omega} =  1$ yr, phase $t_0 = 152.5$ day (June 2$^{nd}$) and
modulation amplitude, $A$, equal to the central value obtained by best fit over the whole data:
cumulative exposure is 1.17 ton $\times$ yr. The dashed vertical lines
correspond to the maximum expected for the DM signal (June 2$^{nd}$), while
the dotted vertical lines correspond to the minimum. See Refs.~\protect\refcite{modlibra,modlibra2} and text.}
\label{fg:res}
\end{figure}

Several analyses on the model-independent DM annual
modulation signature have been performed (see Refs.~\refcite{modlibra,modlibra2} and Refs. therein).
In particular, Fig. \ref{fg:res} shows the time behaviour of the experimental 
residual rates for {\it single-hit} 
events in the (2--4), (2--5), and (2--6) keV energy intervals, 
the only energy intervals where the modulation effect is present.
In fact, other possible energy intervals have also been studied, and no
modulation has been found (see later). This residual rate is 
calculated from the measured rate of the {\it single-hit} events (already corrected 
for the overall efficiency and for the acquisition dead time)
after subtracting the constant part: $\langle r_{ijk}-flat_{jk} \rangle_{jk}$.
Here $r_{ijk}$ is the rate in the considered $i$-th time interval for the $j$-th detector in the 
$k$-th energy bin, while $flat_{jk}$ is the rate of the $j$-th detector in the $k$-th energy bin 
averaged over the cycles. 
The average is made on all the detectors ($j$ index) and on all the energy bins ($k$ index) which 
constitute the considered energy interval. The weighted mean of the residuals must 
obviously be zero over one cycle. 
 
The hypothesis of absence of modulation in the data can be discarded, as reported in Table \ref{tb:mod_0}.

\begin{table}[ht]
\begin{center}
%\vspace{-0.6cm}
\tbl{Test of absence of modulation in the DAMA data. A null modulation amplitude 
is discarded by the data.}
{\resizebox{0.7\textwidth}{!}{
\vspace{-0.2cm}
\begin{tabular}{|c|c|c|}
\hline
 Energy interval & DAMA/NaI $+$ DAMA/LIBRA  \\
     (keV)       & (7+6 annual cycles) \\
\hline
 2-4 & $\chi^2$/d.o.f. = 147.4/80 $\rightarrow$ P = 6.8 $\times$ 10$^{-6}$ \\
 2-5 & $\chi^2$/d.o.f. = 135.2/80 $\rightarrow$ P = 1.1 $\times$ 10$^{-4}$ \\
 2-6 & $\chi^2$/d.o.f. = 139.5/80 $\rightarrow$ P = 4.3 $\times$ 10$^{-5}$ \\
\hline
\hline
\end{tabular}
\label{tb:mod_0}}}
\vspace{-0.2cm}
\end{center}
\end{table}

\begin{figure}[!h]
\resizebox{\columnwidth}{!}{%
\includegraphics {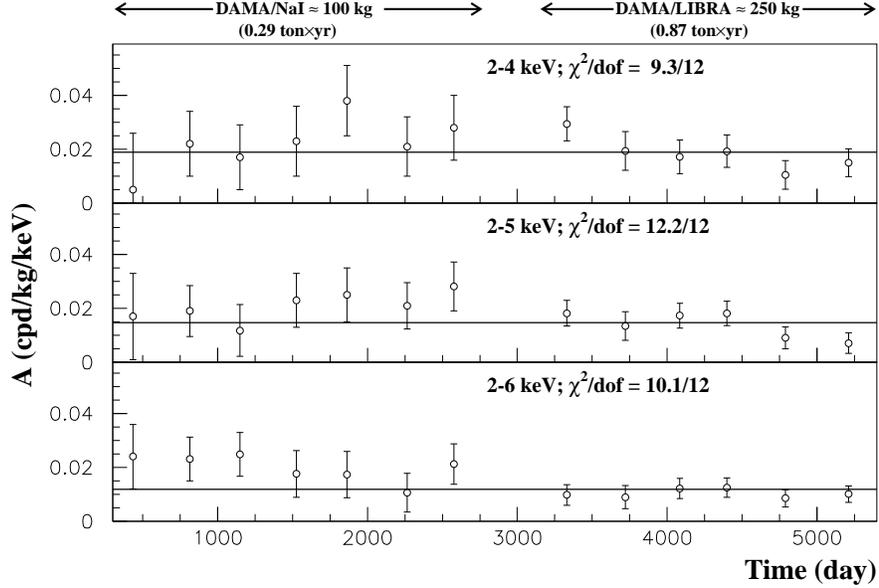}}
\caption{The data points are the modulation amplitudes of each
single annual cycle of DAMA/NaI and DAMA/LIBRA experiments for the (2--4), (2--5) and (2--6) keV 
energy intervals. The error bars are
the related 1$\sigma$ errors. The same time scale as in Fig. \ref{fg:res} is adopted. The solid horizontal
lines shows the central values obtained by best fit over the
whole data set (see Table \ref{tb:ampff}). The $\chi^2$ test and the run test accept
the hypothesis at 90\% C.L. that the modulation amplitudes are
normally fluctuating around the best fit values. See text.}
\label{fg:amp}
\normalsize
\end{figure}

The {\it single-hit} residual rate of DAMA/NaI and DAMA/LIBRA data of Fig. \ref{fg:res} can be 
fitted with the formula: $A \cos \omega(t-t_0)$ integrated over each time bin 
(i.e. $A \frac{\sin \omega(t_2-t_0) - \sin \omega(t_1-t_0)  }{\omega(t_2-t_1)}$, where $t_1$ and $t_2$
are the start and the stop time of each bin) and considering a
period $T = \frac{2\pi}{\omega} =  1$ yr and a phase $t_0 = 152.5$ day (June 2$^{nd}$), as 
expected by the DM annual modulation signature. The fit procedure takes into account also the 
experimental error of each data point; the results are shown in Table \ref{tb:ampff}.
\begin{table}[ht]
\begin{center}
\tbl{Modulation amplitude, A, obtained by fitting the {\it single-hit} residual rate  
of the DAMA/NaI and DAMA/LIBRA annual cycles 
(see Ref. \protect\refcite{modlibra,modlibra2} and Refs. therein)
for a total cumulative exposure of 1.17 ton $\times$ yr. It has been obtained by fitting
the data with the formula: 
$A \cos \omega(t-t_0)$ with $T = \frac{2\pi}{\omega} =  1$ yr and $t_0 = 152.5$ day (June 
2$^{nd}$), as
expected for a signal by the DM annual modulation signature. The corresponding $\chi^2$
value for each fit and the confidence level are also reported.}
{\resizebox{\textwidth}{!}{
\begin{tabular}{|c|c|c|}
\hline
 Energy interval & DAMA/NaI $+$ DAMA/LIBRA  \\
 (keV) & (cpd/kg/keV) \\
\hline
 2-4 & A=(0.0183$\pm$0.0022) \, \,  $\chi^2$/d.o.f. = 75.7/79 \, \, 8.3 $\sigma$ C.L. \\
 2-5 & A=(0.0144$\pm$0.0016) \, \, $\chi^2$/d.o.f. = 56.6/79  \, \, 9.0 $\sigma$ C.L. \\
2-6 & A=(0.0114$\pm$0.0013)  \, \, $\chi^2$/d.o.f. = 64.7/79  \, \, 8.8 $\sigma$ C.L. \\
\hline
\hline
\end{tabular}
\label{tb:ampff}}}
\end{center}
\end{table}
The compatibility among the 13 annual cycles can also be investigated. In particular,
the modulation amplitudes measured in each annual cycle of the whole 1.17 ton $\times$ yr exposure 
are reported in Fig. \ref{fg:amp}. Indeed these modulation amplitudes are normally 
distributed around their best fit value as pointed out by the $\chi^2$ test
($\chi^2 = 9.3$, 12.2 and 10.1 over 12 {\it d.o.f.} for the three energy 
intervals, respectively) and the {\it run test} (lower tail probabilities
of 57\%, 47\% and 35\% for the three energy intervals, respectively).

Table \ref{tb:ampfv} shows the results obtained for the cumulative 
1.17 ton $\times$ yr exposure when the period and phase parameters are 
kept free in the fitting procedure described above.
The period and the phase are well compatible with the expectations for a signal 
in the DM annual modulation signature. In particular, the phase -- whose better 
determination will be achieved in the following by using a maximum likelihood 
analysis -- is consistent with about June $2^{nd}$ within $2\sigma$; moreover, 
for completeness, we also note that a slight energy dependence of the phase 
could be expected in case of possible contributions of non-thermalized DM components 
to the galactic halo, such as e.g. the SagDEG stream \cite{sag06} 
and the caustics \cite{lin04}.
\begin{table}[!ht]
\begin{center}
\tbl{Modulation amplitude ($A$), period ($T = \frac{2\pi}{\omega}$)
and phase ($t_0$), obtained by fitting, with the formula: 
$A \cos \omega(t-t_0)$, the {\it single-hit} 
residual rate of the cumulative 1.17 ton $\times$ yr exposure.
The results are well compatible with expectations for a signal in the 
DM annual modulation signature. \cite{modlibra2}}
{\resizebox{\textwidth}{!}{
\begin{tabular}{|c|c|c|c|c|}
\hline
 Energy interval & $A$ (cpd/kg/keV)& $T = \frac{2\pi}{\omega}$ (yr) &  $t_0$ (days) & C. L. \\
\hline
 2-4 & (0.0194$\pm$0.0022) & (0.996$\pm$0.002) & 136$\pm$7 & 8.8$\sigma$\\
 2-5 & (0.0149$\pm$0.0016) & (0.997$\pm$0.002) & 142$\pm$7 & 9.3$\sigma$\\
 2-6 & (0.0116$\pm$0.0013) & (0.999$\pm$0.002) & 146$\pm$7 & 8.9$\sigma$\\
\hline
\hline
\end{tabular}
\label{tb:ampfv}
}}
\end{center}
\end{table}

The DAMA/NaI and DAMA/LIBRA {\it single-hit} residuals of Fig. \ref{fg:res} have also been 
investigated by Fourier analysis, following the Lomb-Scargle procedure \cite{Lomb,LombA}, 
accounting for the different time binning and the residuals' errors as mentioned
in DAMA literature \cite{RNC,modlibra,modlibra2}.
In particular, in the data analysis the procedure described below was applied.

\begin{figure}[!tbh]
\centering
\includegraphics[width=6.cm] {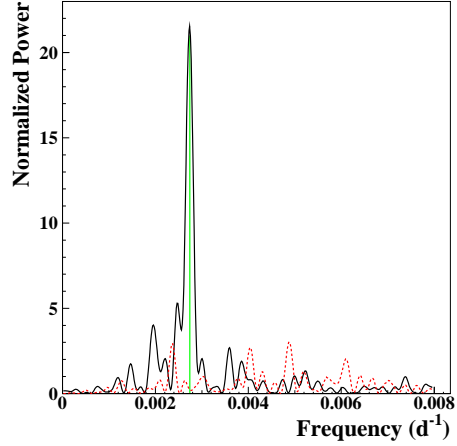}
\vspace{-0.2cm}
\caption{({\it Color online}). Power spectrum of the measured {\it single-hit} residuals in the (2--6) keV (black -- solid line) 
and (6--14) keV (red -- dotted line) energy intervals calculated according to Ref. \protect\refcite{Lomb,LombA}, including also the
treatment of the experimental errors and of the time binning (see text). The data refer to the cumulative 1.17 ton $\times$ yr
exposure (DAMA/NaI and DAMA/LIBRA).
As can be seen, the principal mode present in the (2--6) keV energy interval corresponds to a frequency
$2.735 \times 10^{-3}$ d$^{-1}$ (green -- vertical line). It corresponds to a period 
of $\simeq$ 1 year. A similar peak is not present in the (6--14) keV energy interval just above. \cite{modlibra2}}
\label{fg:pwr}
\normalsize
\end{figure}

In the Lomb-Scargle procedure, given a set of $N$ residuals values $r_{i}$
measured at times $t_{i}$ ($i=1,...,N$), 
one can firstly compute their mean and variance values as:
\begin{eqnarray}
\bar{r}= \frac{1}{N}\sum_{i=1}^{N} r_{i} \;\;\;\;  & 
& \sigma^{2}=\frac{1}{N-1} \sum_{i=1}^{N} \left( r_{i} - \bar{r} \right)^{2};
\end{eqnarray}
Then, for each angular frequency of interest $\omega = 2\pi f>0$, the time-offset $\tau$
is computed to satisfy the following equation:
\begin{equation}
 tan \left( 2\omega\tau \right) = \frac{\sum_{i} sin\left( 2\omega t_{i} \right) }{\sum_{i} cos\left( 2\omega t_{i} \right) }.
\end{equation}
Hence, the Lomb-Scargle normalized periodogram, spectral power as a function of $\omega$,
is defined as \cite{Lomb,LombA}:

\begin{equation}
P_N (\omega) = \frac{1}{2\sigma^2} \left\{ 
\frac{
\left[ \sum_i \left( r_i - \bar{r} \right)cos\omega (t_i - \tau)
\right]^2}
{\sum_i cos^2 \omega (t_i - \tau )}
+ \frac{
\left[ \sum_i \left( r_i - \bar{r} \right) sin\omega (t_i - \tau )\right]^2}
{\sum_i sin^2 \omega (t_i - \tau )}
\right\} 
\end{equation}
in order to take into account the different time binning ($2\Delta_{i}$) and 
the residuals' errors ($\Delta r_{i}$) this formula has to be rewritten, replacing:
\begin{eqnarray}
\sum_i (...) \rightarrow \frac{N}{ \sum_j \frac{1}{\Delta r_j^2}} \cdot \sum_i \frac{1}{\Delta r_i^2} (...) & \,\,\,\, \mbox{and} \,\,\,\, & \left\{
\begin{array}{c}
sin \omega t_i  \rightarrow \frac{1}{2\Delta_i} \int_{t_i - \Delta_i}^{t_i + \Delta_i} sin \omega t  dt \\
cos \omega t_i  \rightarrow \frac{1}{2\Delta_i} \int_{t_i - \Delta_i}^{t_i + \Delta_i} cos \omega t  dt 
\end{array}
\right..
\end{eqnarray}
The power spectrum of the residuals in the (2--6) keV energy interval is reported in Fig. \ref{fg:pwr};
for comparison also that in the nearby (6--14) keV energy interval is shown. In the plot a clear peak corresponding to a period of 1 year 
is present for the lower energy data; the same analysis in the other energy region shows instead only aliasing peaks.

\begin{figure}[!ht]
\begin{center}
\vspace{-0.4cm}
\includegraphics[width=6.0cm] {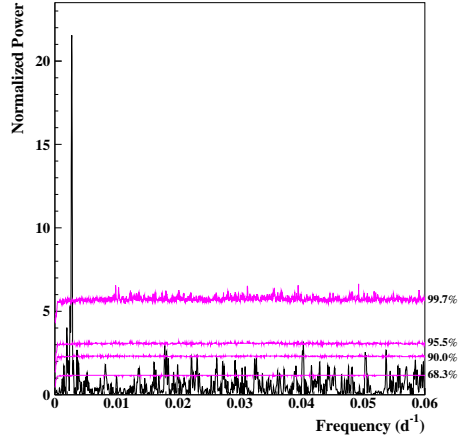}
\end{center}
\vspace{-0.5cm}
\caption{({\it Color online}). The Lomb-Scargle periodogram obtained by using DAMA/NaI and DAMA/LIBRA data with a cumulative exposure 
of 1.17 ton$\times$yr (black -- solid line) extended up to the frequency 0.06 d$^{-1}$. The violet curve represents 
the experimental sensitivity at various confidence levels: 68.3, 90, 95.5, 99.7 \%.}
\label{fg:LombL}
\vspace{-0.4cm}
\end{figure}

The Lomb-Scargle periodogram with the frequency extended up to 0.06 d$^{-1}$
is shown in Fig. \ref{fg:LombL} together with 
the C.L.'s obtained by Montecarlo procedure. 
Therefore, the only periodical signal at a confidence level much 
larger than 99.7\%  corresponds to a frequency  1 yr$^{-1}$; no other
periodical signal is experimentally observed (see also later).

\begin{figure}[!ht]
\begin{center}
\includegraphics[width=4.cm] {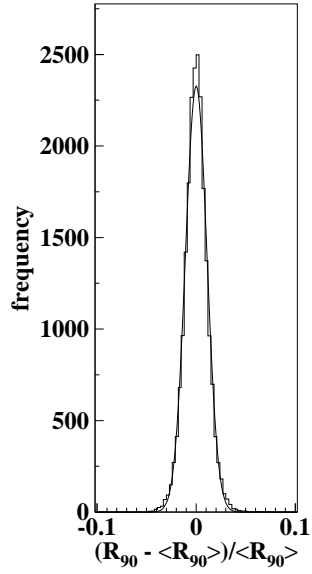}
\end{center}
\caption{Distribution of the percentage variations
of R$_{90}$ with respect to the mean values for all the detectors in the DAMA/LIBRA-1 to -6 
annual cycles (histogram); the superimposed curve is a gaussian fit. \cite{modlibra2}}
\label{fig_r90}
\end{figure}

\begin{figure}[!bth]
\centering
a) \includegraphics[width=0.45\textwidth] {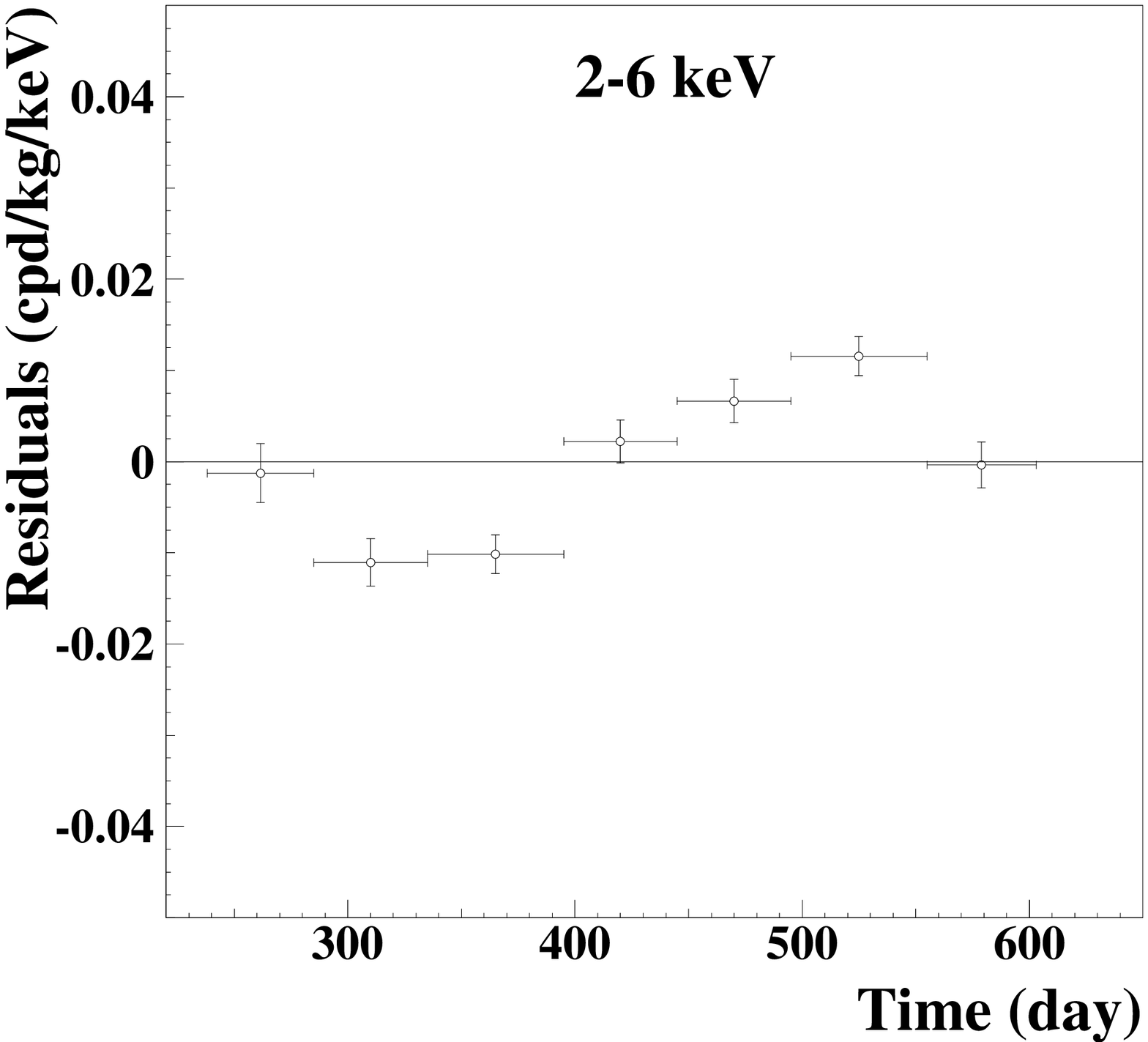}
b) \includegraphics[width=0.45\textwidth] {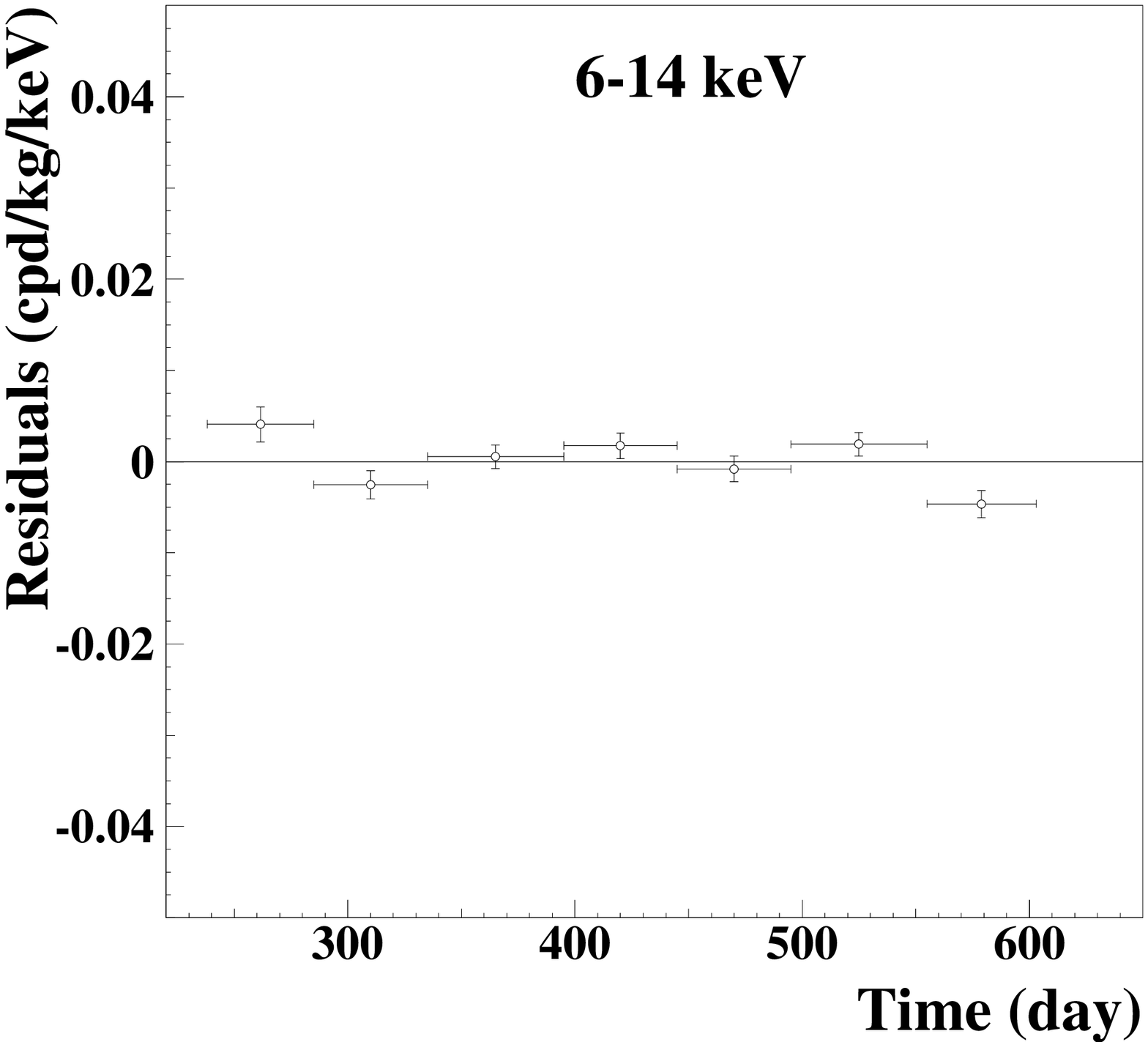}
\caption{Experimental residuals in the (2 -- 6) keV region and those in the
(6 -- 14) keV energy region just above
for the cumulative 1.17 ton $\times$ yr,  considered as collected in a
single annual cycle. The experimental points present the errors as vertical bars and the associated
time bin width as horizontal bars. The initial time of the figure is taken at August 7$^{th}$.
The clear modulation satisfying all the peculiarities of the DM annual modulation signature is present 
in the lowest energy interval,
while it is absent just above; in fact, in the latter case 
the best fitted modulation amplitude is:
(0.00007 $\pm$ 0.00077) cpd/kg/keV. \cite{modlibra2}}
\label{fg:res1}
\vspace{-0.5cm}
\end{figure}

The measured energy distribution has been investigated in
other energy regions not of interest for Dark Matter,
also verifying the absence of any significant background modulation\footnote{In fact, 
the background in the lowest energy region is
essentially due to ``Compton'' electrons, X-rays and/or Auger
electrons, muon induced events, etc., which are strictly correlated
with the events in the higher energy part of the spectrum.
Thus, if a modulation detected
in the lowest energy region would be due to
a modulation of the background (rather than to a signal),
an equal or larger 
modulation in the higher energy regions should be present.}.
Following the procedures described 
in Ref. \refcite{modlibra} and Refs. therein, the measured rate
integrated above 90 keV, R$_{90}$, as a function of the time has been analysed.
In particular, the distribution of the percentage variations
of R$_{90}$ with respect to the mean values for all the detectors shows a cumulative gaussian behaviour
with $\sigma$ $\simeq$ 1\%, well accounted by the statistical
spread expected from the used sampling time; see Fig. \ref{fig_r90}
for the DAMA/LIBRA annual cycles and e. g. Ref. \refcite{RNC} for the DAMA/NaI
cycles. 
Moreover, fitting the time behaviour of R$_{90}$
with phase and period as for DM particles, a modulation amplitude compatible with zero
is also found, for example:
$-(0.05 \pm 0.19)$ cpd/kg, 
$-(0.12 \pm 0.19)$ cpd/kg, 
$-(0.13 \pm 0.18)$ cpd/kg, 
$(0.15 \pm 0.17)$ cpd/kg, 
$(0.20 \pm 0.18)$ cpd/kg, 
$-(0.20 \pm 0.16)$ cpd/kg, 
in the six DAMA/LIBRA annual cycles, respectively.
Similar results have been obtained for the DAMA/NaI cycles (see Ref. \refcite{RNC} and Refs. therein).
This also excludes the presence of any background
modulation in the whole energy spectrum at a level much
lower than the effect found in the lowest energy region for the {\it single-hit} events.
In fact, otherwise -- considering the R$_{90}$ mean values --
a modulation amplitude of order of tens
cpd/kg, that is $\simeq$ 100 $\sigma$ far away from the measured value, would be present.
Similar result is obtained when comparing 
the {\it single-hit} residuals in the (2--6) keV with those 
in other energy intervals; see as an example Fig. \ref{fg:res1}. 
It is worth noting that the obtained results already account for whatever 
kind of background and, in addition, that no background process able to mimic
the DM annual modulation signature (that is able to simultaneously satisfy 
all the peculiarities of the signature and to account for the measured modulation amplitude)
is available (see later and also discussions e.g. in Ref. \refcite{modlibra,modlibra2,scineghe09}).

\begin{figure}[!ht]
\begin{center}
\includegraphics[width=0.9\textwidth] {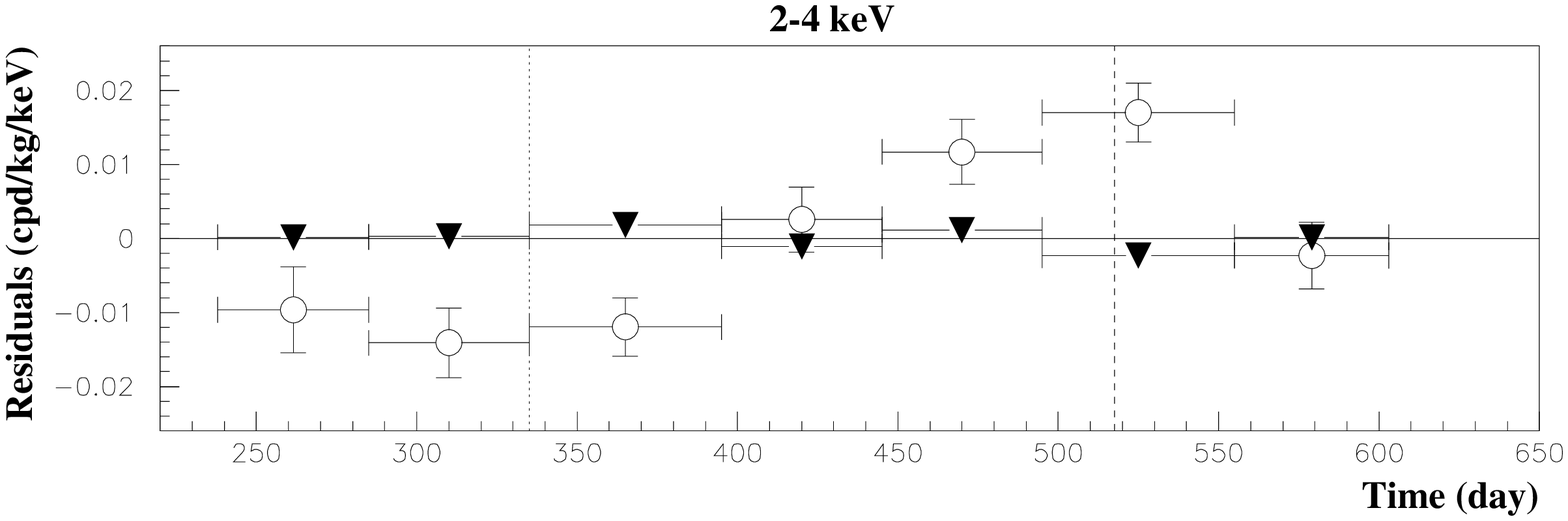}
\includegraphics[width=0.9\textwidth] {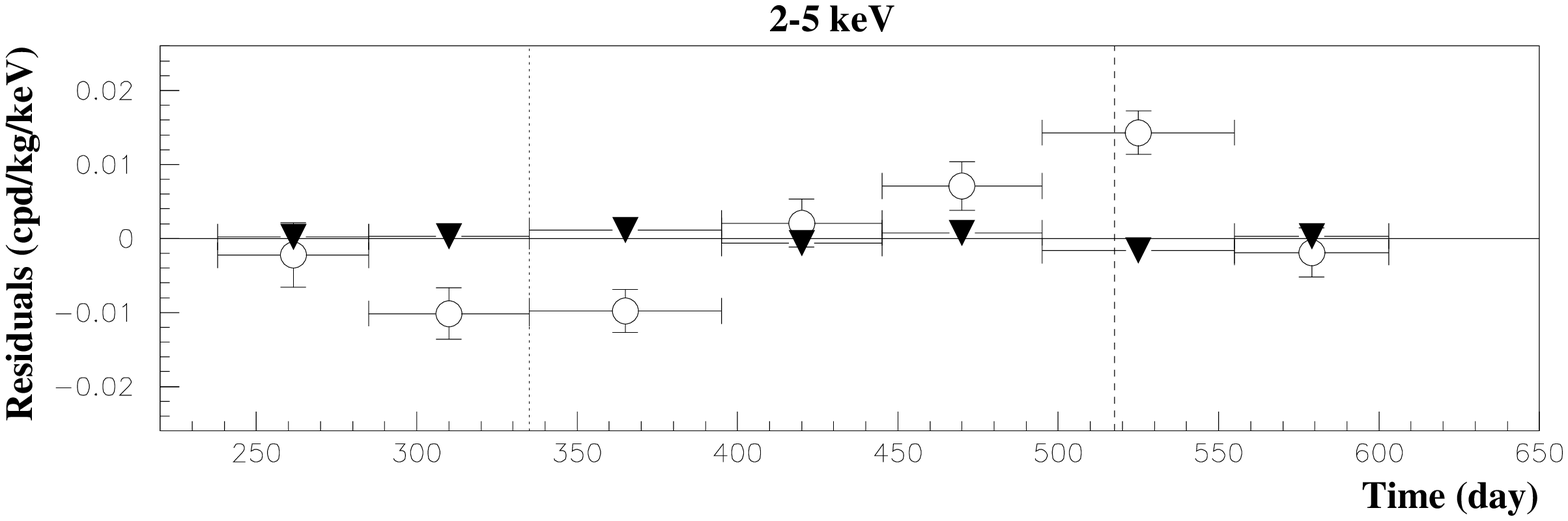}
\includegraphics[width=0.9\textwidth] {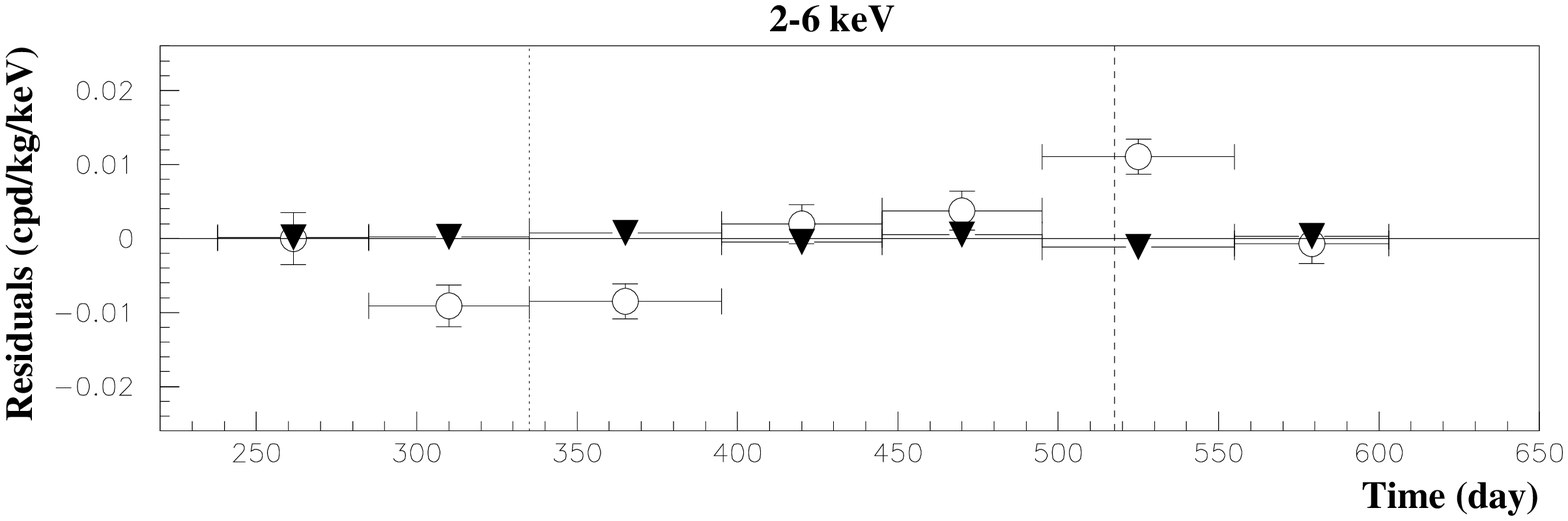}
\end{center}
\caption{Experimental residual rates over the six DAMA/LIBRA annual cycles for {\it single-hit} events 
(open 
circles) (class of events to which DM events belong) and for {\it multiple-hit} events (filled triangles)
(class of events to which DM events do not belong).
They have been obtained by considering for each class of events the data as collected in a 
single annual cycle 
and by using in both cases the same identical hardware and the same identical software procedures.
The initial time of the figure is taken on August 7$^{th}$.
The experimental points present the errors as vertical bars and the associated time bin width as horizontal 
bars. See text and Refs. \protect\refcite{modlibra,modlibra2}. 
Analogous results were obtained for the DAMA/NaI data 
\cite{ijmd}. See also Ref. \protect\refcite{modlibra2}}
\label{fig_mul}
\end{figure}

A further relevant investigation has been performed by applying to the {\it multiple-hit} events
the same hardware and software 
procedures used to acquire and to analyse the {\it single-hit} residual rate.
In fact, since the probability that a DM particle interacts in more than one detector 
is negligible, a DM signal can be present just in the {\it single-hit} residual rate.
Thus, the comparison of the results of the {\it single-hit} events with those of the  {\it 
multiple-hit} ones corresponds practically to compare between them the cases of DM particles beam-on 
and beam-off.
This procedure also allows an additional test of the background behaviour in the same energy interval 
where the positive effect is observed. 
In particular, in Fig. \ref{fig_mul} the residual rates of the {\it single-hit} events measured over 
the six DAMA/LIBRA annual
cycles are reported as collected in a single cycle, together with the residual rates 
of the {\it multiple-hit} events in the considered energy intervals.
As already observed, a clear modulation, satisfying all the peculiarities of the DM
annual modulation signature, is present in 
the {\it single-hit} events,
while the fitted modulation amplitudes for the {\it multiple-hit}
residual rate are well compatible with zero:
$-(0.0011\pm0.0007)$ cpd/kg/keV,
$-(0.0008\pm0.0005)$ cpd/kg/keV,
and $-(0.0006\pm0.0004)$ cpd/kg/keV
in the energy regions (2 -- 4), (2 -- 5) and (2 -- 6) keV, respectively.
Thus, again evidence of annual modulation with proper features as required by the DM annual 
modulation signature is present in the {\it single-hit} residuals (events class to which the
DM particle induced events belong), while it is absent in the {\it multiple-hit} residual 
rate (event class to which only background events belong).
Similar results were also obtained for the last two annual 
cycles of the DAMA/NaI experiment \cite{ijmd}, where an improvement of the electronics allowed such an analysis.
Since the same identical hardware and the same identical software procedures have been used to 
analyse the two classes of events, the obtained result offers an additional strong support for the 
presence of a DM particle component in the galactic halo.

The annual modulation amplitude, $S_{m}$
can also be analysed
by maximum likelihood method over the data considering $T=$1 yr and $t_0=$ 152.5 day,
as a function of the energy.

For such purpose the likelihood function of the {\it single-hit} experimental data
in the $k-$th energy bin is defined as: 
\begin{equation} 
{\it\bf L_k}  = {\bf \Pi}_{ij} e^{-\mu_{ijk}}
{\mu_{ijk}^{N_{ijk}} \over N_{ijk}!},
\label{eq:likelihood}
\end{equation} 
where $N_{ijk}$ is the number of events collected in the
$i$-th time interval (hereafter 1 day), by the $j$-th detector and in the
$k$-th energy bin. $N_{ijk}$ follows a Poisson's
distribution with expectation value
$\mu_{ijk} = \left[ b_{jk} + S_{ik} \right] M_j \Delta
t_i \Delta E \epsilon_{jk}$.
The b$_{jk}$ are the background contributions, $M_j$ is the mass of the $j-$th detector,
$\Delta t_i$ is the detector running time during the $i$-th time interval,
$\Delta E$ is the chosen energy bin,
$\epsilon_{jk}$ is the overall efficiency. Moreover, the signal can be written
as $S_{ik} = S_{0,k} + S_{m,k} \cdot \cos\omega(t_i-t_0)$, where $S_{0,k}$ is the constant part of 
the 
signal 
and $S_{m,k}$ is the modulation amplitude.
The usual procedure is to minimize the function $y_k=-2ln({\it\bf L_k}) - const$ for each energy bin;
the free parameters of the fit are the $(b_{jk} + S_{0,k})$ contributions and the $S_{m,k}$
parameter. Hereafter, the index $k$ is omitted when unnecessary. 

\begin{figure}[!ht]
\begin{center}
\includegraphics[width=\textwidth] {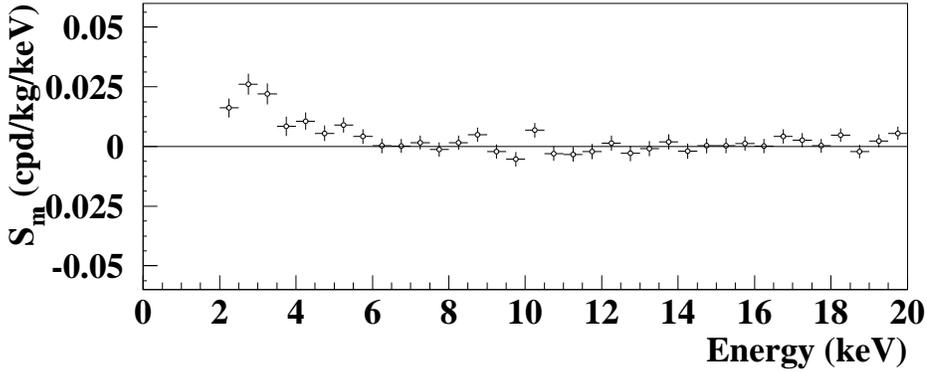}
\end{center}
\caption{Energy distribution of the $S_{m}$ variable for the
total cumulative exposure 1.17 ton$\times$yr. The energy bin is 0.5 keV.
A clear modulation is present in the lowest energy region,
while $S_{m}$ values compatible with zero are present just above. In fact, the $S_{m}$ values
in the (6--20) keV energy interval have random fluctuations around zero with
$\chi^2$ equal to 27.5 for 28 degrees of freedom. \cite{modlibra2}}
\label{sme}
\end{figure}

In Fig. \ref{sme} the obtained $S_{m}$  are shown 
for each considered energy bin (there $\Delta E = 0.5$ keV).
It can be inferred that positive signal is present in the (2--6) keV energy interval, while $S_{m}$
values compatible with zero are present just above. In fact, the $S_{m}$ values
in the (6--20) keV energy interval have random fluctuations around zero with
$\chi^2$ equal to 27.5 for 28 degrees of freedom.
All this confirms the previous analyses.

The method also allows the extraction of the the $S_{m}$ 
values for each detector, for each annual cycle and 
for each energy bin. Thus
we have also verified that the S$_{m}$ values
are statistically well distributed in all the six DAMA/LIBRA annual cycles and in all the
sixteen energy bins (here $\Delta$E = 0.25 keV in the 2--6 keV energy interval) for each detector.
For this purpose, the variable $x = \frac {S_m - \langle S_m \rangle}{\sigma}$  is considered;
here, $\sigma$ are the errors associated to $S_m$ and $\langle S_m \rangle$ 
are the mean values of the $S_m$ averaged over the detectors 
and the annual cycles for each considered energy bin.
Similar investigations have already been performed also previously for DAMA/NaI \cite{RNC,ijmd}.
\begin{figure}[!ht]
\begin{center}
\includegraphics[width=9.5cm] {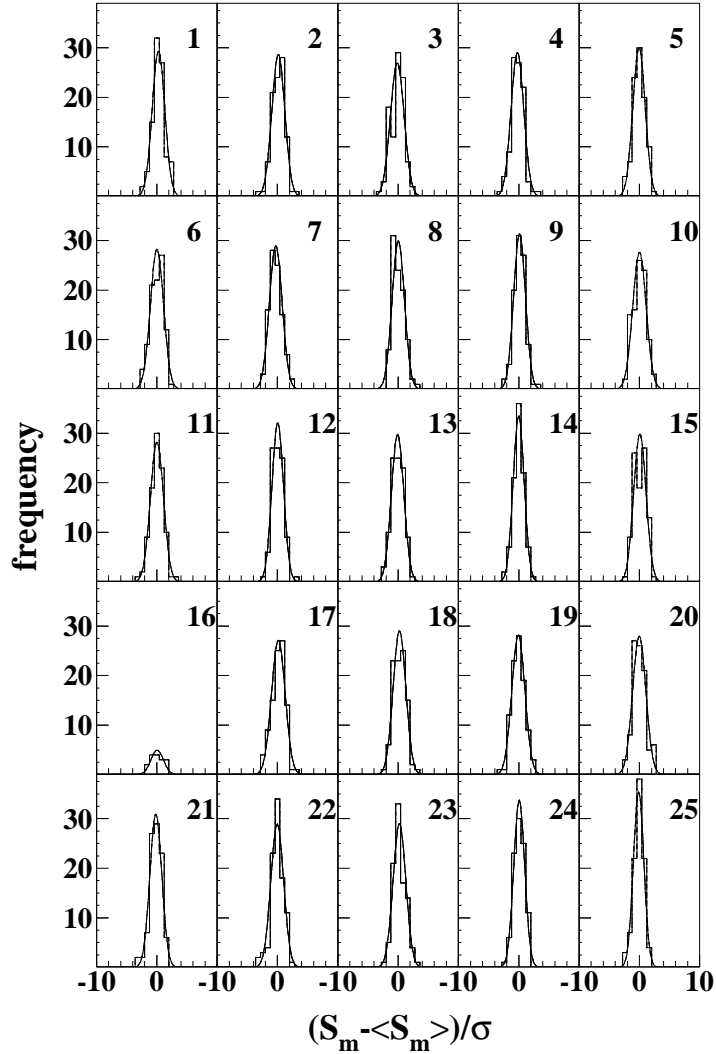}
\end{center}
\vspace{-0.4cm}
\caption{Distributions (histograms) of the variable $\frac {S_m - \langle S_m \rangle}{\sigma}$, where 
$\sigma$ are the errors
associated to the $S_m$ values and $\langle S_m \rangle$ 
are the mean values of the modulation amplitudes averaged over the detectors 
and the annual cycles for each considered energy bin (here $\Delta E = 0.25$ keV). 
The entries of each histogram are  
96 (16 in case of detector 16, see text), 16 energy bins in the (2--6) keV energy interval and 6 DAMA/LIBRA annual cycles;
the  r.m.s. values are reported in Fig. \ref{chi2}-{\it bottom}. 
The superimposed curves are gaussian fits. 
Each panel refers to a single DAMA/LIBRA detector.}
\label{gaus2}
\end{figure}
Fig. \ref{gaus2} shows the distributions of the variable $x$
for the DAMA/LIBRA data in the (2--6) keV energy interval
plotted for each detector separately.
The entries of each histogram are the 
96 (16 for the 16-th detector\footnote{As aforementioned, this detector 
has been restored in the trigger after the first upgrade in September 2008; 
thus, only the data of the last annual cycle are available for this detector.}) 
$x$ values, evaluated for the 16 energy bins in the considered
(2--6) keV energy interval and for the 6 DAMA/LIBRA annual cycles.
These distributions allow one to conclude that the observed annual modulation effect is well
distributed in all the detectors, annual cycles and energy bins. In fact, the standard deviations of 
the $x$ variable for the DAMA/LIBRA detectors
range from 0.87 to 1.14 (see also Fig. \ref{chi2}{\it --bottom}). 
Defining $\chi^2 = \Sigma x^2$ (where the sum is extended over 
all the 96(16) $x$ values), the $\chi^2/d.o.f.$ values  
range between 0.7 and 1.22 for twenty-four 
detectors, and the observed annual modulation effect is well
distributed in all these detectors at 95\% C.L. (see Fig. \ref{chi2}{\it --top}). A particular mention is deserved
to the remaining detector whose $\chi^2/d.o.f. = 1.28$  exceeds the value corresponding to that C.L.; this 
is also statistically 
consistent, considering that the expected number of detector exceeding this value over twenty-five
is 1.25. 

The mean value of the twenty-five $\chi^2/d.o.f.$ is 1.066, slightly larger than the expected value 1.
Although this can be still ascribed to statistical fluctuations (see before),
let us ascribe it to a possible systematics. In this case, one would
have an additional error $\leq 4 \times 10^{-4}$ cpd/kg/keV, if quadratically combined, or
$\leq 5 \times 10^{-5}$ cpd/kg/keV, if linearly combined, to the modulation amplitude
measured in the (2 -- 6) keV energy interval.
This possible additional error: $\leq 4\%$ or $\leq 0.5\%$, respectively, of the
DAMA/LIBRA modulation amplitude is an upper limit of possible systematic effects.
\begin{figure}[!ht]
\begin{center}
\vspace{-1.0cm}
\includegraphics[width=0.85\textwidth] {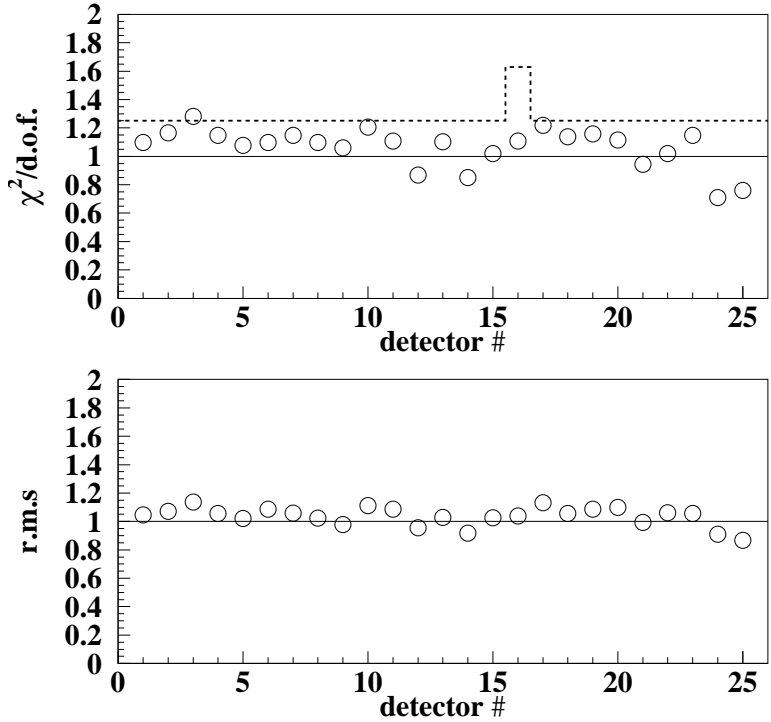}
\end{center}
\vspace{-0.4cm}
\caption{{\it Top:} $\chi^2/d.o.f.$ values of $S_m$ distributions around their mean value
for each DAMA/LIBRA detector in the (2--6) keV energy interval for the six annual cycles.
The dotted line corresponds to the upper tail probability of 5\%. 
The observed annual modulation effect is well
distributed in all the detectors at 95\% C.L. (see text). 
{\it Bottom:} standard deviations of the $x$ variable for the DAMA/LIBRA detectors;
see also Fig. \ref{gaus2}. See text.}
\label{chi2}
\end{figure}
In conclusion, the above arguments demonstrate 
that the modulation amplitudes are statistically well distributed in all the 
crystals, in all the data taking periods and in all the energy bins.

Among further additional tests, the analysis of the
modulation amplitudes as a function of the energy separately for
the nine inner detectors and the remaining external ones has been carried 
out for the DAMA/LIBRA data set\cite{modlibra,modlibra2}. 
The obtained values are fully in agreement; in fact,
the hypothesis that the two sets of modulation amplitudes as a function of the
energy belong to same distribution has been verified by $\chi^2$ test, obtaining:
$\chi^2/d.o.f.$ = 3.1/4 and 7.1/8 for the energy intervals (2--4) and (2--6) keV, 
respectively ($\Delta$E = 0.5 keV). This shows that the
effect is also well shared between inner and external detectors. 

Let us, finally, release the assumption of a phase $t_0=152.5$ day in the maximum likelihood procedure to 
evaluate the modulation amplitudes from the data of the 1.17 ton $\times$ yr. In this case
alternatively the signal has been written as:
\begin{equation}
S_{ik} = S_{0,k} + S_{m,k} \cos\omega(t_i-t_0) + Z_{m,k} \sin\omega(t_i-t_0) = 
S_{0,k} + Y_{m,k} \cos\omega(t_i-t^*).
\label{eqn} 
\end{equation}
For signals induced by DM particles one would expect: 
i) $Z_{m,k} \sim 0$ (because of the orthogonality between the cosine and the sine functions); 
ii) $S_{m,k} \simeq Y_{m,k}$; iii) $t^* \simeq t_0=152.5$ day. 
In fact, these conditions hold for most of the dark halo models; however, as mentioned above,
slight differences can be expected in case of possible contributions
from non-thermalized DM components, such as e.g. the SagDEG stream \cite{sag06} 
and the caustics \cite{lin04}.

\begin{figure*}[!ht]
\begin{center}
\includegraphics[width=0.48\textwidth] {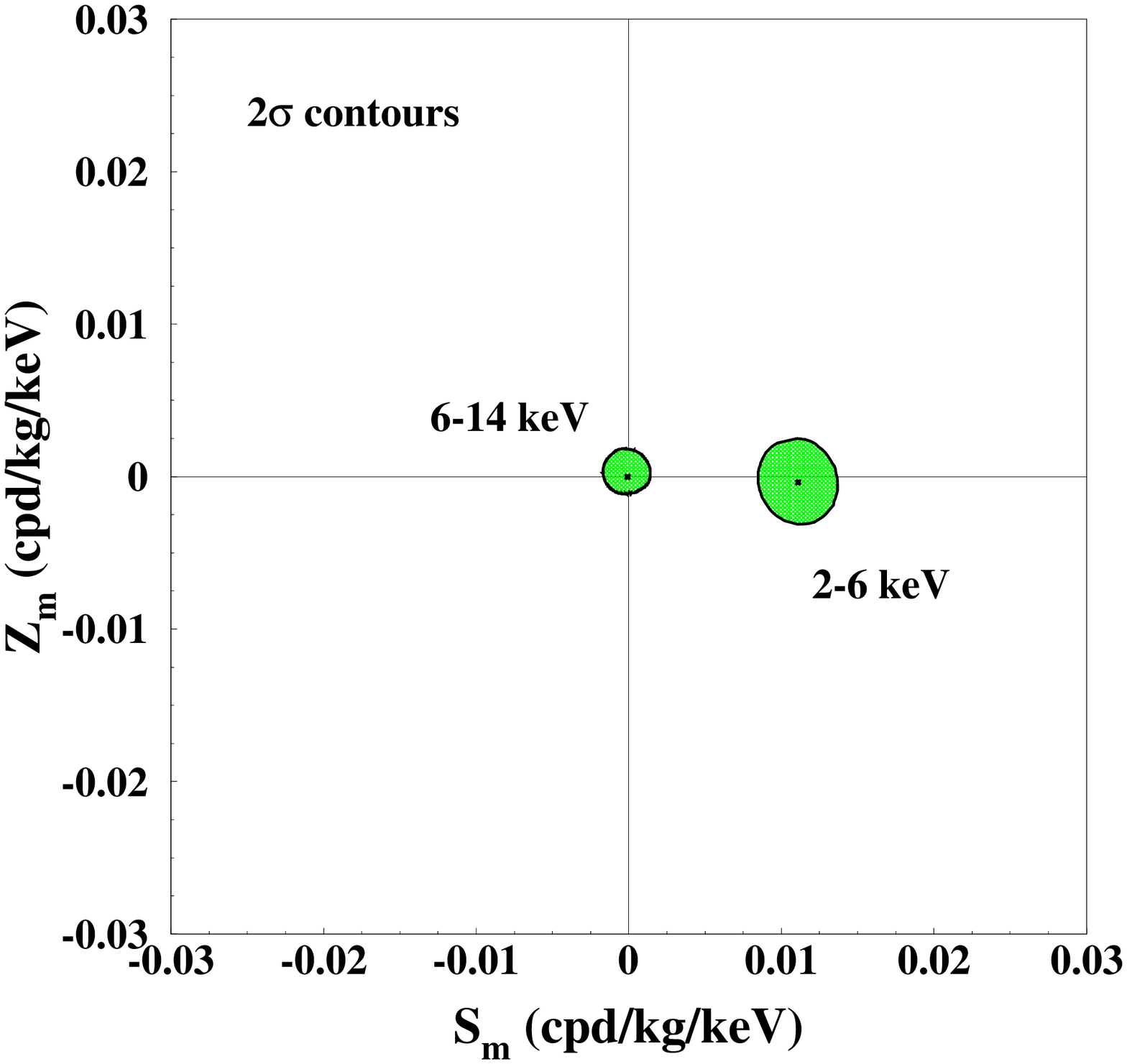}
\includegraphics[width=0.48\textwidth] {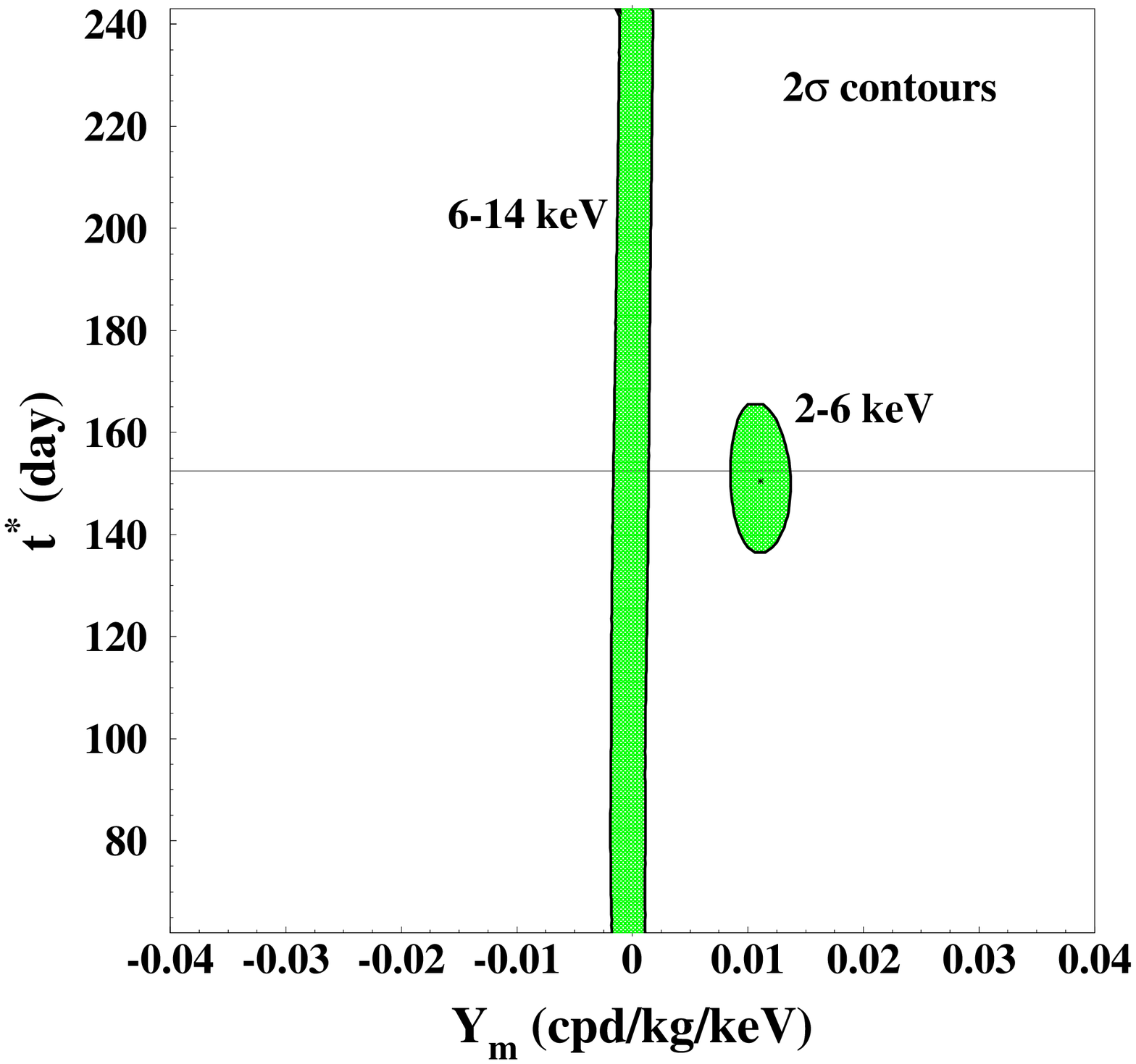}
\end{center}
\caption{$2\sigma$ contours in the plane $(S_m , Z_m)$ ({\it left})
and in the plane $(Y_m , t^*)$ ({\it right})
for the (2--6) keV and (6--14) keV energy intervals.
The contours have been  
obtained by the maximum likelihood method, considering 
the cumulative exposure of 1.17 ton $\times$ yr.
A modulation amplitude is present in the lower energy intervals 
and the phase agrees with that expected for DM induced signals. \cite{modlibra2}}
\label{fg:bid}
\end{figure*}

Fig. \ref{fg:bid}{\it --left} shows the 
$2\sigma$ contours in the plane $(S_m , Z_m)$ 
for the (2--6) keV and (6--14) keV energy intervals and 
Fig. \ref{fg:bid}{\it --right} shows, instead, those in the plane $(Y_m , t^*)$.
Table \ref{tb:bidbf} shows the best fit values for the (2--6) and (6--14) keV energy interval 
($1\sigma$ errors) for  S$_m$ versus  Z$_m$ and  $Y_m$ versus $t^*$.  

\begin{table}[ht]
\begin{center}
\tbl{Best fit values for the (2--6) and (6--14) keV energy interval
($1\sigma$ errors) for  S$_m$ versus  Z$_m$ and  $Y_m$ versus $t^*$,
considering the cumulative exposure of 1.17 ton $\times$ yr.
See also Fig. \ref{fg:bid}. \cite{modlibra2}}
{\resizebox{\textwidth}{!}{
\begin{tabular}{|c|c|c|c|c|}
\hline
E     & S$_m$        & Z$_m$        & $Y_m$        & $t^*$  \\
(keV) & (cpd/kg/keV) & (cpd/kg/keV) & (cpd/kg/keV) & (day) \\
\hline
  2--6 &  (0.0111 $\pm$ 0.0013) & -(0.0004 $\pm$ 0.0014) &  (0.0111 $\pm$ 0.0013) & (150.5 $\pm$ 7.0) \\
 6--14 & -(0.0001 $\pm$ 0.0008) &  (0.0002 $\pm$ 0.0005) & -(0.0001 $\pm$ 0.0008) &  undefined       \\
\hline
\hline
\end{tabular}}
\label{tb:bidbf}
}
\end{center}
\end{table}

\begin{figure*}[ht]
\begin{center}
\includegraphics[width=0.85\textwidth] {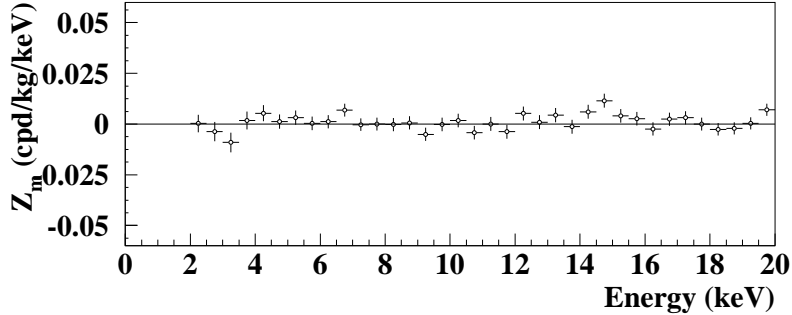}
\end{center}
\caption{Energy distribution of the $Z_{m}$ variable for the total exposure
(1.17 ton $\times$ yr, DAMA/NaI and DAMA/LIBRA), once forced 
to zero the contribution of the cosine function in Eq. (\ref{eqn}).
The energy bin is 0.5 keV.
The $Z_{m}$ values are expected to be zero in case of
presence of a DM particles' signal with $t^* \simeq t_0 = 152.5$ day.  By the fact, 
the $\chi^2$ test applied to the data supports the hypothesis that the $Z_{m}$ values are simply 
fluctuating around zero; see text. \cite{modlibra2}}
\label{fg:zm}
\end{figure*}

Finally, forcing to zero the contribution of the cosine function in Eq. (\ref{eqn}),
the $Z_{m}$ values as function of the energy have also been determined
by using the same procedure. The values of $Z_{m}$
as a function of the energy is reported in Fig. \ref{fg:zm}.
Obviously, such values are expected to be zero in case of
presence of a DM signal with $t^* \simeq t_0 = 152.5$ day.  By the fact, 
the $\chi^2$ test 
applied to the data supports the hypothesis that the $Z_{m}$ values are simply 
fluctuating around zero; in fact, for example 
in the (2--14) keV and (2--20) keV energy region the $\chi^2$/d.o.f.
are equal to 21.6/24 and 47.1/36 (probability of 60\% and 10\%), respectively.

\begin{figure*}[!ht]
\begin{center}
\includegraphics[width=0.78\textwidth] {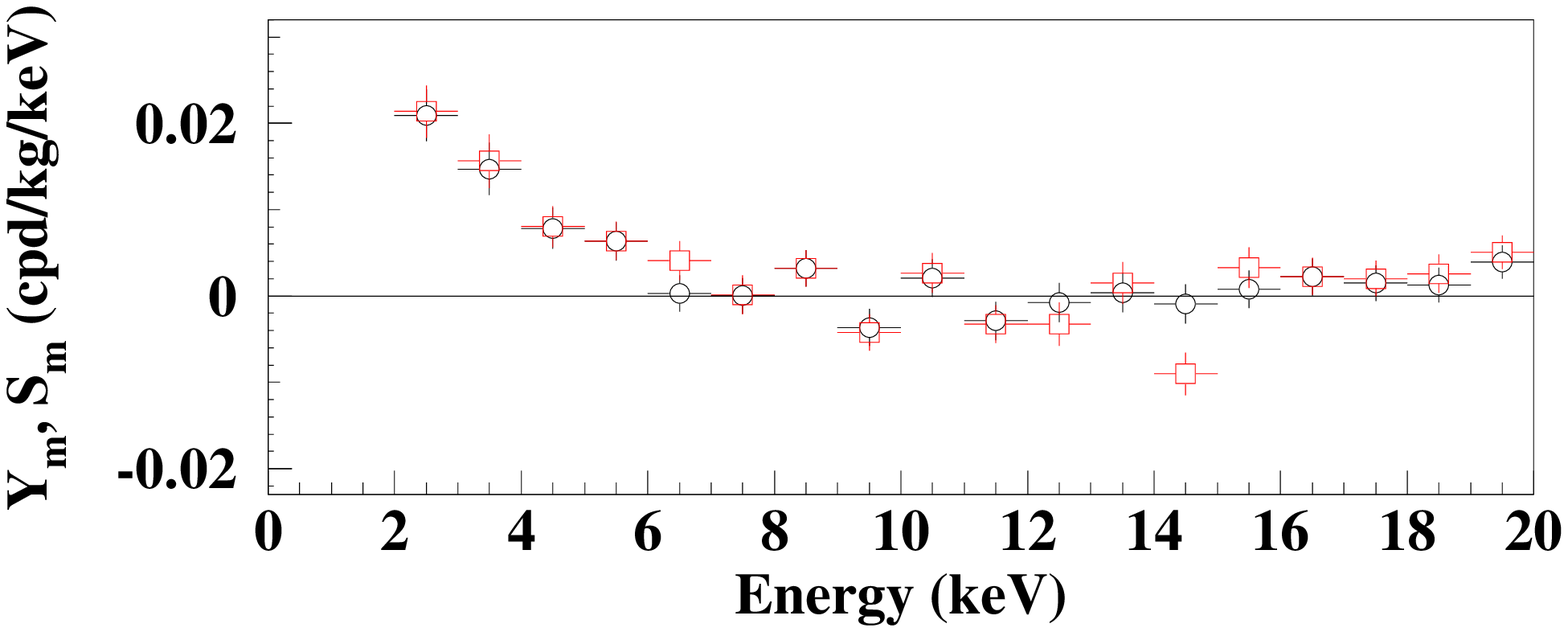}
\includegraphics[width=0.78\textwidth] {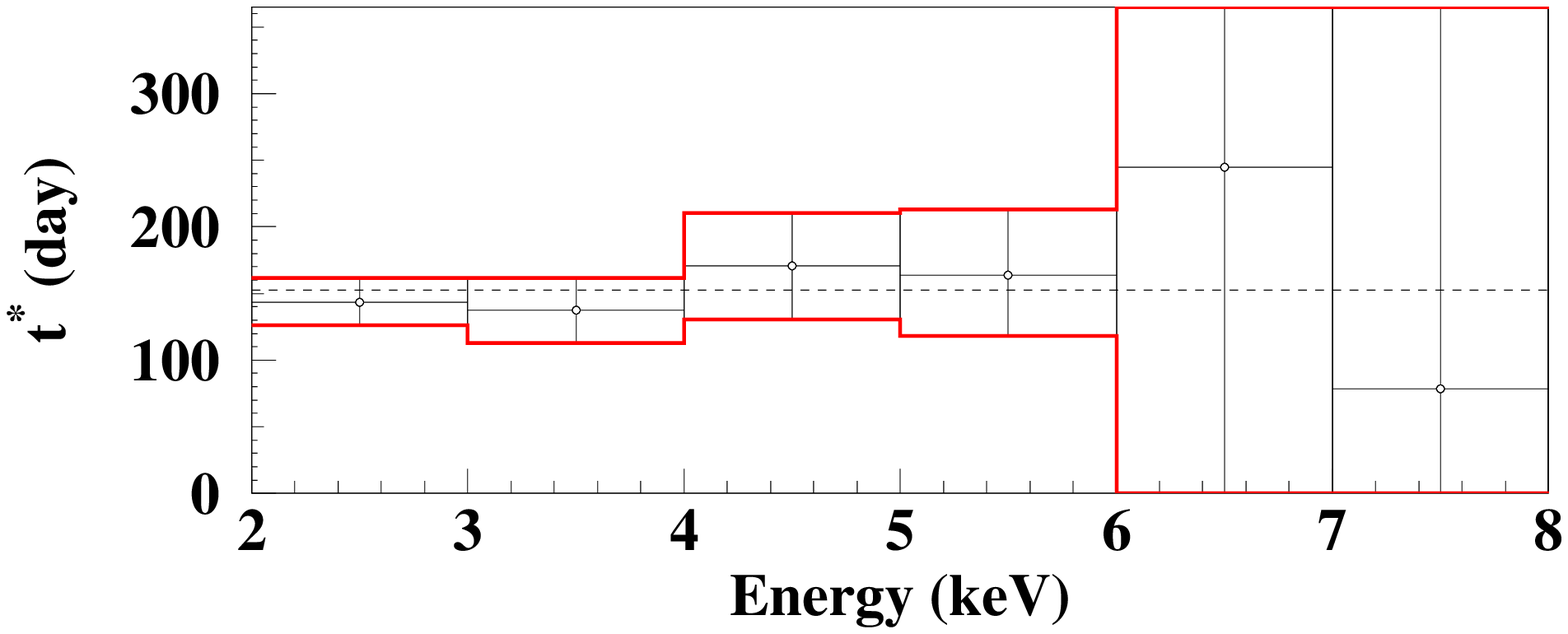}
\end{center}
\caption{{\em Top:} Energy distributions of the $Y_{m}$ variable (light data points; red color online)
and of the $S_{m}$ variable (solid data points) for the total exposure
(1.17 ton $\times$ yr, DAMA/NaI and DAMA/LIBRA). Here, unlike the data of
Fig. \ref{sme}, the energy bin is 1 keV.
{\em Bottom:} Energy distribution of the phase $t^*$ for the total exposure; here 
the errors are at $2\sigma$.
An annual modulation effect is present in the lower energy intervals 
up to 6 keV and the phase agrees with that expected for DM induced signals.
No modulation is present above 6 keV and the phase is undetermined. \cite{modlibra2}}
\label{fg:ymts}
\end{figure*}

The behaviours of the $Y_{m}$ and of the phase $t^*$ variables 
as function of energy are shown in Fig. \ref{fg:ymts} 
for the total exposure (1.17 ton $\times$ yr, DAMA/NaI and DAMA/LIBRA). 
The $Y_{m}$ are superimposed with the $S_{m}$ values 
with 1 keV energy bin (unlike Fig. \ref{sme} where the energy bin is 0.5 keV).
As in the previous analyses, an annual modulation effect is present in the lower energy intervals 
and the phase agrees with that expected for DM induced signals.

\vspace{0.2cm}

These results confirm those achieved by other kinds of analyses.

\vspace{0.3cm}

\section{No role for systematics and side processes in the DAMA annual modulation results}
\label{intro}

Sometimes naive statements were put forward as the fact that
in nature several phenomena may show some kind of periodicity.

\begin{sidewaystable}
\centering
\tbl{Modulation amplitudes ($1\sigma$ error) obtained by fitting the time behaviours of the main
running parameters including a possible annual modulation with phase and period
as for DM particles. These running parameters, acquired with the production data, 
are: i) the operating temperature of the detectors;
ii) the HP Nitrogen flux in the inner Cu box housing the detectors; 
iii) the pressure of the HP Nitrogen atmosphere of the inner Cu box
housing the detectors; iv) the environmental Radon in the inner part of the
barrack from which the detectors are however excluded (see text and Ref. \protect\refcite{perflibra} for details); 
v) the hardware rate above single photoelectron threshold. All the measured amplitudes are
compatible with zero.}
{
\resizebox{\textwidth}{!}{
\begin{tabular}{|c||c|c|c|c|c|c|} \hline
  & & & & & & \\
  & DAMA/LIBRA-1 & DAMA/LIBRA-2 & DAMA/LIBRA-3 & DAMA/LIBRA-4 & DAMA/LIBRA-5 & DAMA/LIBRA-6 \\
  & & & & & & \\
\hline
  & & & & & & \\
Temperature & $ -(0.0001 \pm 0.0061) ^{\circ}$C &
              $ (0.0026 \pm 0.0086) ^{\circ}$C  &
              $ (0.001 \pm 0.015) ^{\circ}$C    &
              $ (0.0004 \pm 0.0047) ^{\circ}$C  &
              $ (0.0001 \pm 0.0036) ^{\circ}$C  &
              $ (0.0007 \pm 0.0059) ^{\circ}$C  \\
  & & & & & & \\
Flux  & $ (0.13 \pm 0.22)$ l/h &
        $ (0.10 \pm 0.25)$ l/h &
        $-(0.07 \pm 0.18)$ l/h &
        $-(0.05 \pm 0.24)$ l/h &
        $-(0.01 \pm 0.21)$ l/h &
        $-(0.01 \pm 0.15)$ l/h \\
  & & & & & & \\
Pressure   & $ (15 \pm 30) 10^{-3}$ mbar &
              $ -(13 \pm 25) 10^{-3}$ mbar &
              $ (22 \pm 27) 10^{-3}$ mbar &
              $ (1.8 \pm 7.4) 10^{-3}$ mbar &
              $ -(0.8 \pm 1.2) 10^{-3}$ mbar &
              $ (0.7 \pm 1.3) 10^{-3}$ mbar \\
&  & & & & & \\
Radon       & $ -(0.029 \pm 0.029)$ Bq/m$^{3}$ &
              $ -(0.030 \pm 0.027)$ Bq/m$^{3}$ &
              $ (0.015 \pm 0.029)$ Bq/m$^{3}$ &
              $ -(0.052 \pm 0.039)$ Bq/m$^{3}$ &
              $ (0.021 \pm 0.037)$ Bq/m$^{3}$ &
              $ -(0.028 \pm 0.036)$ Bq/m$^{3}$ \\
&  & & & & & \\
Hardware rate & $-(0.20 \pm 0.18) 10^{-2}$ Hz &
              $ (0.09 \pm 0.17) 10^{-2}$ Hz &
              $ -(0.03 \pm 0.20) 10^{-2}$ Hz &
              $ (0.15 \pm 0.15) 10^{-2}$ Hz &
              $ (0.03 \pm 0.14) 10^{-2}$ Hz &
              $ (0.08 \pm 0.11) 10^{-2}$ Hz \\
&  & & & & & \\
\hline\hline
\end{tabular}}
\label{tb:par16}}
\end{sidewaystable}

\begin{table}[!ht]
\begin{center}
\tbl{Summary of the results obtained by investigating possible sources
of systematics or side processes \cite{perflibra,modlibra,modlibra2,scineghe09,taupnoz,vulca010,canj11,tipp11,muons,replica,replicaA}. 
None able to give
a modulation amplitude different from zero has been
found;
thus cautious upper limits (90\% C.L.)
on the possible contributions to the measured modulation amplitude
have been calculated and are shown here for the six annual cycles of DAMA/LIBRA as done before for
the seven annual cycles of DAMA/NaI \cite{RNC,ijmd}.}
{\begin{tabular}{|c|c|c|}
\hline \hline
 Source      & Main comment                       &  Cautious upper limit \\
             &    &       (90\%C.L.) \\
\hline\hline
             & Sealed Cu Box in         &  \\
  Radon      & HP Nitrogen atmosphere,  &  $<2.5 \times 10^{-6}$ cpd/kg/keV \\
             & 3-level of sealing       &  \\
\hline
Temperature  & Air conditioning         &  $<10^{-4}$ cpd/kg/keV \\
             & + huge heat capacity     &                        \\
\hline
Noise        & Efficient rejection      &  $<10^{-4}$ cpd/kg/keV \\
\hline
Energy scale & Routine                  &  $<1 - 2 \times 10^{-4}$ cpd/kg/keV \\
             & + intrinsic calibrations & \\
\hline
Efficiencies & Regularly measured       & $<10^{-4}$ cpd/kg/keV \\
\hline
             &  No modulation above 6 keV;       & \\
             &  no modulation in the (2 -- 6) keV  & \\
 Background  &  {\it multiple-hit} events;       & $<10^{-4}$ cpd/kg/keV \\
             &  this limit includes all possible & \\
             &  sources of background            & \\
\hline
Side reactions & From muon flux variation& $<3 \times 10^{-5}$ cpd/kg/keV \\
               & measured by MACRO  & \\
\hline
\multicolumn{3}{|c|} {In addition: no effect can mimic the signature} \\
\hline \hline
\end{tabular}
\label{tb:sist}}
\end{center}
\end{table}

It is worth noting that the point is whether they might
mimic the annual modulation signature in DAMA/NaI and in DAMA/LIBRA, i.e. whether they
might be not only quantitatively able to account for the observed
modulation amplitude but also able to contemporaneously
satisfy all the requirements of the DM annual modulation signature. The same is also for side reactions.
This has already been deeply investigated
\cite{perflibra,modlibra,modlibra2,RNC,ijmd,scineghe09,taupnoz,vulca010,canj11,tipp11,muons,replica,replicaA}; 
none has been found or suggested by anyone over more than a decade.

Firstly, in order to continuously monitor the running conditions, several pieces of information 
are acquired with the production data 
and quantitatively analysed.
In particular, all the time behaviours 
of the running parameters, acquired with the production data,
have been investigated. Table \ref{tb:par16} shows the modulation amplitudes obtained for each  
annual cycle when fitting the time behaviours of the values of the main parameters including a cosine 
modulation with the same 
phase and period as for DM particles.
As can be seen, all the measured amplitudes are
well compatible with zero.

No modulation has been found in any  
possible source of systematics or side reactions; thus, cautious upper limits (90\% C.L.)  
on possible contributions to the DAMA/LIBRA measured modulation amplitude
are summarized in Table \ref{tb:sist} (also see Ref. \refcite{modlibra,modlibra2}).
It is worth noting that they do not quantitatively account for the
measured modulation amplitudes, and also are not able to simultaneously satisfy all the many requirements of the signature. 
Similar analyses have also been done for
the seven annual cycles of DAMA/NaI \cite{RNC,ijmd}.

Although the arguments reported in \cite{perflibra,modlibra,modlibra2,RNC,ijmd,scineghe09,taupnoz,vulca010,canj11,tipp11,muons,replica,replicaA}
already rule out any contribution from known side processes or systematics, in the following 
some additional discussions are given also recalling already--published arguments.

\vspace{0.3cm}
\subsection{No role for muons and fast neutrons produced by muons interaction}   \label{muon}

The muons surviving the coverage of the Gran Sasso laboratory (3600 m w.e. depth) either can have direct interactions in 
the experimental set-up or can produce in the surroundings and/or inside the set-up secondary particles (e.g. fast neutrons, 
$\gamma$'s, electrons, spallation nuclei, hypothetical exotics, etc.) possibly depositing energy in the detectors. 
Let us demonstrate that muons cannot play any role in the achieved annual modulation results.

The surviving muon flux ($\Phi_\mu$) has been measured deep underground at LNGS by various experiments with very large exposures 
\cite{Mac97,LVD,borexino,borexinoA,borexino2,borexino2A}; 
its value is $\Phi_\mu \simeq 20 $ muons m$^{-2}$d$^{-1}$ \cite{Mac97}, that is about a factor 
$10^6$ lower than the value measured at sea level. The measured average single muon energy at the Gran Sasso laboratory is 
$\left[270 \pm 3 (stat) \pm 18(syst)\right]$ GeV \cite{Mac03}; this value agrees with the predicted values using different parameterizations 
\cite{hime}. A $\simeq$ 2\% yearly variation of the muon flux was firstly reported years ago by MACRO; when fitting the data of 
the period January 1991 -- December 1994 all together, a phase around middle of July was quoted \cite{Mac97}. It is worth 
noting that the variation of the muons flux is attributed to the variation of the temperature in the outer atmosphere, and its 
phase changes each year depending on the weather condition. 
Recently, other measurements have been reported by LVD, quoting a lower amplitude (about $1.5\%$) and a phase, when considering the 
data of the period January 2001 -- December 2008 all together, equal to (5 July $\pm$ 15 days) \cite{LVD}. Finally, the Borexino experiment 
has quoted a phase of (7 July $\pm$ 6 days), still considering the data taken in the period May 2007 -- May 2010 all together 
\cite{borexino,borexinoA}. 
More recently, the Borexino collaboration presented a modified phase evaluation (29 June $\pm$ 6 days)\footnote{It 
is worth noting that in Ref. \refcite{borexino2,borexino2A} 28 June (179.0 days) is instead quoted as measured phase; actually, 
in our convention -- coherent throughout the paper -- 179.0 days correspond to 00:00 of of 29 June (as, for example, $t=0.0$ days 
is 00:00 of 1st of January and $t=1.0$ days is 00:00 of 2nd of January).}, with a still lower modulation amplitude: about $1.3\%$ 
\cite{borexino2,borexino2A}, by adding the data collected in a further year; the appreciable difference in the fitted values further 
demonstrates the large variability of the muon flux feature year by year.

The measured muon variation at LNGS has no impact on the DAMA annual modulation results for many reasons; in the 
following we summarize the key items.

Let us firstly recall that a muon flux variation cannot play any role in the DAMA result
\cite{perflibra,modlibra,modlibra2,RNC,ijmd,scineghe09,taupnoz,vulca010,canj11,tipp11,muons},
not only  because it may give rise to quantitatively negligible effects, 
but also because it is unable to mimic the DM signature. In fact, it would fail some requirements of the signature; namely e.g.: 
(i) it would induce variation in the whole energy spectrum; (ii) it would induce variation also in the {\it multiple-hits} events 
(events in which more than one detector ``fires''), (iii) it would induce variation with a phase and amplitude 
distinctively different from the DAMA measured one.

\begin{figure}[!ht]
\centering
\includegraphics[width=0.6\textwidth] {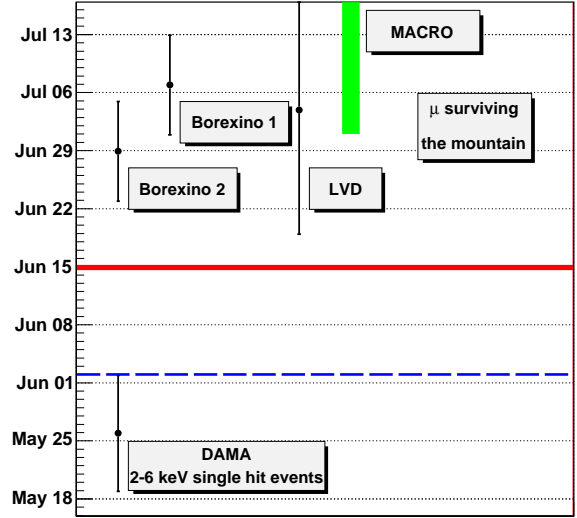}
\caption{The phase of the DAMA annual modulation signal \cite{modlibra2} 
and the muon phases quoted by Borexino in two analyses (May 2007 -- May 2010 \cite{borexino,borexinoA},
and May 2007 -- May 2011 \cite{borexino2,borexino2A}),
by LVD (January 2001 -- December 2008 \cite{LVD}), and by MACRO (January 1991 -- December 1994 \cite{Mac97}).
The muon phases quoted by those three experiments have 
to be regarded as mean values among the muon phases in all the considered years since the muon phase 
depends on the temperature of the outer atmosphere and, thus, it changes each year.
The phase of the DAMA observed effect has instead a stable value in the different years 
\cite{modlibra,modlibra2}. The 
horizontal dashed line corresponds to June 2$^{nd}$ (date around which the phase of the DM annual modulation is 
expected). The middle of June is also marked as an example; in fact, the maximum temperature of the $T_{eff}$
at the LNGS location (see text) cannot be as early as the 
middle of June (and for several years), date which is still 3 $\sigma$ far away from the phase of the DAMA 
observed effect. \cite{muons}}
\label{fg:phase}
\end{figure}

The phase of muons surviving the LNGS coverage and the phase of the (2--6) keV {\it single-hit} events measured by DAMA
are distinctively different (see Fig. \ref{fg:phase}). In particular, the values quoted by MACRO, LVD and Borexino experiments 
for the muon phase have to be regarded as mean values of the muon phases among the analyzed years and the associated errors 
are not simply due to statistical fluctuation of the data, but rather to the variations of the muon phase in the different years.
The phase of the DAMA observed effect has instead a stable value in the different years \cite{modlibra,modlibra2} and is 5.7 
(5.9, 4.7) $\sigma$ from the LVD (Borexino, first and recently modified evaluations, respectively) ``mean'' phase of muons 
(7.1 $\sigma$ from the MACRO one).
This latter simple approach does not consider that the experimental errors in the muon flux are not completely gaussian; 
however, it gives the right order of magnitude of the confidence level for the incompatibility between the DAMA 
phase and the phase of muons and of the muon-induced effects. Analyses carried out by different authors confirm 
these outcomes; for example, a disagreement in the correlation analysis between the LVD data on muon flux and the
DAMA (2--6) keV residuals with a confidence level greater than 99.9\% has been reported in Ref. \refcite{chang11}, where 
it has been shown that they differ in their power spectrum, phase, and amplitude.

In particular, the expected phase for DM is significantly different than the expected phase 
of muon flux at Gran Sasso: in fact, while the first one is always about 152.5 day of the year, the second one is 
related to the variations of the atmospheric temperature above the site location, $T_{eff}$.
The behaviour of $T_{eff}$ at the LNGS location as function of time has been determined e.g. in Ref. 
\refcite{borexino2,borexino2A}; as first order approximation $T_{eff}$ was fitted with a cosinusoidal behaviour and the phase
turned out to be (24 June $\pm$ 0.4 days) \cite{borexino2,borexino2A} (this is later than e.g. the middle of June, date     
which is still $3 \sigma$ far away from the DAMA measured phase, see Fig. \ref{fg:phase}).
In addition, fitting the temperature values at L'Aquila in the years 1990-2011 \cite{tempaq} with a 
cosinusoidal function, a period of $(365.1 \pm 0.1)$ days and a phase of (25 July $\pm$ 0.6 days) are obtained
\cite{muons}. 

Thus, in conclusion, in addition to the previously mentioned arguments, also
the phase of the DAMA annual modulation signal \cite{modlibra2} is significantly different 
than the phases of the surface temperature and of the $T_{eff}$, on which the muon flux measured deep underground 
is dependent, and than the phases of the muon flux measured by MACRO, LVD and Borexino experiments. 

The above arguments also holds for 
every kind of cosmogenic product (even hypothetical exotics) due to muons \cite{muons}. 
In particular, when the decays or the de-excitations of any hypothetical cosmogenic product have mean-life time $\tau$,
the expected phase, $t_{side}$, would be (much) larger than the muon phase (of each considered year) $t_{\mu}$
and even more different from the one measured by the DAMA experiments and expected from the DM annual modulation signature 
($\simeq$ June 2$^{nd}$) \cite{muons}.
In fact, the number of the cosmogenic products, $N(t)$, satisfies the following equation \cite{muons}:
\begin{center}
$dN = -N(t)\frac{dt}{\tau} + \left[ a + b \; cos \omega (t-t_\mu) \right] dt$,
\end{center}
where the $N(t)$ variation is given by the sum of two contributions: the former due to the decay of the species and the latter due to their production, 
showing the typical pattern of muon flux with $b/a \simeq 0.015$, and period $T=2\pi/\omega=1$ year; $a$ is the mean production 
rate. Solving this differential equation, one has \cite{muons}:
\begin{center}
$N(t) = A e^{-t/\tau} + a \tau + \frac{b}{\sqrt{(1/\tau)^2 + \omega^2}} \; cos \omega (t-t_{side}), $
\end{center}
where $A$ is an integration constant, and $t_{side} = t_\mu + \frac{arctg(\omega\tau)}{\omega}$. In condition of secular 
equilibrium (obtained for time scale greater than $\tau$), the first term vanishes and the third term shows an annual modulation pattern 
with phase $t_{side}$. The relative modulation amplitude of the effect is: $\frac{b/a}{\sqrt{1 + (\omega\tau)^2}}$.
Two extreme cases can be considered:
if $\tau \ll T/2\pi$, one gets $t_{side} \simeq t_{\mu}+\tau$; else if $\tau \gg T/2\pi$, one gets $t_{side} \simeq t_{\mu} + T/4$ 
($\simeq t_{\mu} + 90$ days) and the relative modulation amplitude of the effect is $\ll 1.5\%$.

In conclusion, the phase of muons and of whatever (even hypothetical) muon-induced effect is inconsistent
with the phase of the DAMA annual modulation effect.

In addition to the previous arguments, the direct interaction of muons crossing the DAMA set-ups cannot give rise to any 
appreciable variation of the measured rate. In fact, the exposed NaI(Tl) surface of DAMA/LIBRA is about 0.13 m$^2$ (and smaller 
in the former DAMA/NaI); thus the total muon flux in the $\simeq$ 250 kg DAMA/LIBRA set-up is about 2.5 muons/day;
moreover, the impinging muons give mainly {\it multiple-hits} events and over the whole energy spectrum.

\begin{figure}[!ht]
\centering
\includegraphics[height=7.5cm] {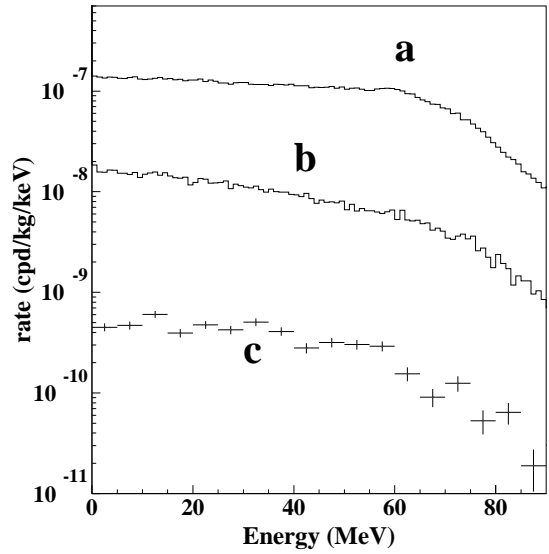}
\caption{{\it Single-hit} event rate as a function of the energy, expected 
for different sets of DAMA/LIBRA detectors from the 
direct interaction of muons crossing the DAMA set-up, taking into account the muon 
intensity distribution, the Gran Sasso rock overburden map,
and the geometry of the set-up. 
Case a): average contribution of the 5 upper and 5 lower detectors
in the $5\times5$ matrix. Case b):
average contribution of the remaining 15 detectors. Case c) average contribution of
the 9 inner detectors \cite{fond40}. See also Ref. \protect\refcite{muons}
}
\label{fig_mu}
\end{figure}

The order of magnitude of contribution due to the direct interaction of muons crossing the DAMA set-up 
has been estimated by a Montecarlo calculation \cite{fond40,papep} where 
three topologies of the detectors (depending on their locations in the $5\times5$ DAMA/LIBRA detectors' matrix) 
have been considered (see Fig.~\ref{fig_mu}).
One can easily infer that the contributions from muons interacting in the detectors directly -- for {\it single-hit} events in 
the (2--6) keV energy region -- to the DAMA total counting rate (that is around 1 cpd/kg/keV) and to the observed annual modulation 
amplitude (that is around $10^{-2}$ cpd/kg/keV) are negligible (many orders of magnitude lower).
In addition, as mentioned above, this contribution would also fail some requirements of the DM annual modulation signature.

The surviving muons and the muon-induced cascades or showers can be sources of neutrons in the underground laboratory.
Such neutrons produced by cosmic rays are substantially harder (extending up to several hundreds MeV energies \cite{hime}) 
than those from environmental radioactivity; their typical flux is about $10^{-9}$ neutrons/cm$^2$/s \cite{kud08}, 
that is three orders of magnitude smaller than the neutron flux produced by radioactivity.
In particular, the fast neutron rate produced by muons interaction is given by: $R_n = \Phi_\mu \cdot Y \cdot M_{eff}$
where $M_{eff}$ is the effective mass where muon interactions can give rise to events detected in the DAMA set-up and $Y$ is the 
integral neutron yield, which is normally quoted in neutrons per muon per g/cm$^2$ of the crossed target material. 
The integral neutron yield critically depends on the chemical composition and on the density of the medium through which the 
muons interact (see e.g. Ref. \refcite{muons} and Refs. therein for more details). 
The modulation amplitude of the {\it single-hit} events in the lowest energy region induced in DAMA/LIBRA by a muon 
flux modulation can be estimated according to:
      \begin{center}
          $S_m^{(\mu)} = R_n \cdot g \cdot \epsilon \cdot f_{\Delta E} \cdot
          f_{single} \cdot 1.5\% / (M_{set-up} \cdot \Delta E)$,
      \end{center}
where $g$ is a geometrical factor, $\epsilon$ is the detection efficiency for neutrons, 
$f_{\Delta E}$ is the acceptance of the considered energy window 
(E $\ge$ 2 keV), $f_{single}$ is the {\it single-hit} efficiency and 1.5\% is the 
muon modulation amplitude \cite{muons}.
Since $M_{set-up} \simeq$ 250 kg, $\Delta E \simeq$  4 keV, assuming the very cautious values:
$g \simeq \epsilon \simeq f_{\Delta E} \simeq f_{single} \simeq 0.5$ and taking for $M_{eff}$ the total 
mass of the heavy shield, 15 ton, one obtains:
      \begin{center}
      $S_m^{(\mu)} < (0.3 - 2.4) \times 10^{-5}$ cpd/kg/keV 
      \end{center}
that is, $S_m^{(\mu)} \ll 0.5\%$ of the observed {\it single-hit} events modulation amplitude \cite{muons}.
In conclusion, any appreciable contribution from fast neutrons produced by the muon interactions
can be quantitatively excluded. In addition, it also would fail some of the requirements of the DM annual modulation signature. 

In addition to the previous arguments, it has been demonstrated in Ref. \refcite{muons} that any hypothetical effect  
due to muons crossing the NaI(Tl) detectors and/or the surroundings of the set-up cannot give any appreciable contribution to 
the observed (2--6) keV {\it single-hit} event rate, already just owing to statistical considerations.
In fact, because of the poissonian fluctuation on the number of muons, the standard deviation, $\sigma(A)$, of any 
hypothetically induced (2--6) keV {\it single-hit} modulation amplitude would be much larger than measured by DAMA, thus, 
giving rise to no statistically-significant effect (see Ref. \refcite{muons}).
In fact, $\sigma(A)$ is expected to be (much) larger than $\approx 0.005$ cpd/kg/keV for any possible (known and exotic) 
contribution of muons, as demonstrated in Ref. \refcite{muons}.
In particular, for muons interacting directly in the NaI(Tl) DAMA/LIBRA detectors it is expected to be $\sigma(A)=0.017$ cpd/kg/keV.
Therefore, these fluctuations are much larger than the value experimentally observed by DAMA for $\sigma(A)$, 
0.0013 cpd/kg/keV \cite{modlibra2}.
We can conclude that all (standard and exotic) mechanisms, 
because of the low number of the involved muons, provide too high fluctuations of the data, not observed in DAMA.
Even just this last argument alone is enough to discard any kind of hypothesis about muons.

\subsection{No role for environmental neutrons}

Environmental neutrons cannot give any significant contribution to the annual modulation measured by the DAMA 
experiments \cite{modlibra,modlibra2,RNC,ijmd,Sist}.
In fact, the thermal neutron flux surviving the multicomponent DAMA/LIBRA shield has 
been determined by studying the possible presence of $^{24}$Na from neutron activation of $^{23}$Na in NaI(Tl).
In particular, $^{24}$Na presence has been investigated by looking for triple coincidences induced
by a $\beta$ in one detector and by the two $\gamma$'s in two adjacent
ones. An upper limit on the thermal neutron flux surviving
the multicomponent DAMA/LIBRA shield has been
derived as: $< 1.2 \times 10^{-7}$ n cm$^{-2}$ s$^{-1}$ (90\% C.L.) \cite{modlibra}, that is at least one order of magnitude 
lower than the value of the environmental neutrons measured at LNGS. 
The corresponding capture rate is: $< 0.022$ captures/day/kg. 
Even assuming cautiously a 10\% modulation (of whatever origin\footnote{For 
the sake of correctness, it is worth noting that
a variation of the neutron flux in the underground Gran Sasso laboratory
has never been suitably proved. 
In particular, besides few speculations, there is just an unpublished 2003 short internal report of 
the ICARUS collaboration, TM03-01, that seemingly reports a 5\% environmental neutron variation 
in hall C by exploiting the pulse shape discrimination (PSD) in commercial BC501A liquid scintillator.
However, the stability of the data 
taking and of the applied PSD procedures over the whole data taking period and also the nature of 
the discriminated events
are not fully demonstrated. 
Anyhow, even assuming the existence of   
a similar neutron variation, it cannot quantitatively contribute to the DAMA observed modulation 
amplitude \cite{modlibra,RNC,ijmd} as well as satisfy all the peculiarities of the DM annual modulation
signature.}) of the thermal 
neutrons flux, and with the same phase and period as for the DM case, the corresponding modulation amplitude in the 
lowest energy region would be \cite{modlibra,RNC}: $<0.01 \%$ of the DAMA observed modulation amplitude. 
Similar outcomes have also been achieved for the case of fast neutrons; the fast neutrons have been measured in 
the DAMA/LIBRA set-up by means of the inelastic reaction $^{23}$Na$(n,n')^{23}$Na$^*$ (2076 keV)
which produces two $\gamma$'s in coincidence (1636 keV and 440 keV). An upper
limit -- limited by the sensitivity of the method -- has been found:
$< 2.2 \times 10^{-7}$ n cm$^{-2}$ s$^{-1}$ (90\% C.L.) \cite{modlibra},
well compatible with the value measured at LNGS;
a reduction at least an order of magnitude is expected due to the neutron shield of the set-up.
Even when cautiously assuming a 10\% modulation (of whatever origin) of the fast neutrons flux, and with the same phase 
and period as for the DM case, the corresponding modulation amplitude 
is $<0.5 \%$ of the DAMA observed modulation amplitude \cite{modlibra,RNC}. 

Moreover, in no case the neutrons can mimic the DM annual modulation signature since some of the peculiar 
requirements of the signature would fail.

\subsection{No role for $^{128}$I decay}

It has been claimed in Ref. \refcite{ralston} that 
environmental neutrons (mainly thermal and/or epithermal), occasionally 
produced by high energy muon interactions, once captured by Iodine might contribute to the modulation observed by DAMA 
through the decay of activated $^{128}$I (that produces -- among others -- low energy X-rays/Auger electrons). But such an 
hypothesis is already excluded by several arguments given above and has already been 
rejected in details in Ref. \refcite{vulca010,canj11,muons}. Let us give the main arguments in the following.

\begin{figure}[!hb]
\centering
\vspace{-1.0cm}
\includegraphics[width=0.6\textwidth] {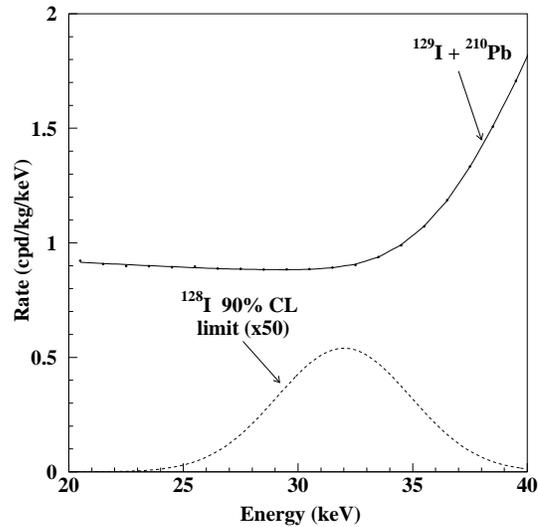}
\vspace{-0.2cm}
\caption{Energy distribution of the events measured by DAMA/LIBRA
in the region of interest for the $K$-shell EC decay of $^{128}$I;
the exposure here is 0.53 ton $\times$ yr. The solid line represents the result
of the fit described in Ref. \protect\refcite{papep}, including the contributions
of $^{129}$I and $^{210}$Pb to the background.
The gaussian (dashed) line is 50 times the limit of the $^{128}$I contribution,
0.074 cpd/kg, excluded at 90\% C.L. \protect\cite{papep}. See also Ref. \protect\refcite{muons}}
\label{fg:128ia}
\end{figure}

The $^{128}$I decays via EC (6.9\%) and $\beta^-$ (93.1\%) channels\cite{iod128}.
When it decays via EC, it produces low energy X-rays and Auger electrons, totally contained inside the NaI(Tl) detectors;
thus, the detectors would measure the total energy release of all the X-rays and Auger electrons, that is 
the atomic binding energy either of the $K$-shell (32 keV) or of the $L$-shells (4.3 to 5 keV) of the 
$^{128}$Te.
In particular considering the branching ratios of the EC processes in the $^{128}$I decay, the $K$-shell contribution 
(around 30 keV) must be about 8 times larger than that of $L$-shell; while no modulation has been observed 
by DAMA above 6 keV (see \cite{modlibra,modlibra2} and Refs. therein) and, in particular, around 30 keV.
Moreover the $^{128}$I also decays by $\beta^-$ with much larger branching ratio (93.1\%) than EC (6.9\%) and with 
a $\beta^-$ end-point energy at 2 MeV. Again, no modulation has instead been observed in DAMA experiments at 
energies above 6 keV \cite{modlibra,modlibra2}.
Finally the $L$-shell contribution would be a gaussian centered around 4.5 keV; this shape is excluded by the 
behaviour of the measured modulation amplitude, $S_m$, as a function of energy (see  Fig. \ref{sme}).

\begin{figure}[!ht]
\begin{center}
\includegraphics[width=8.cm] {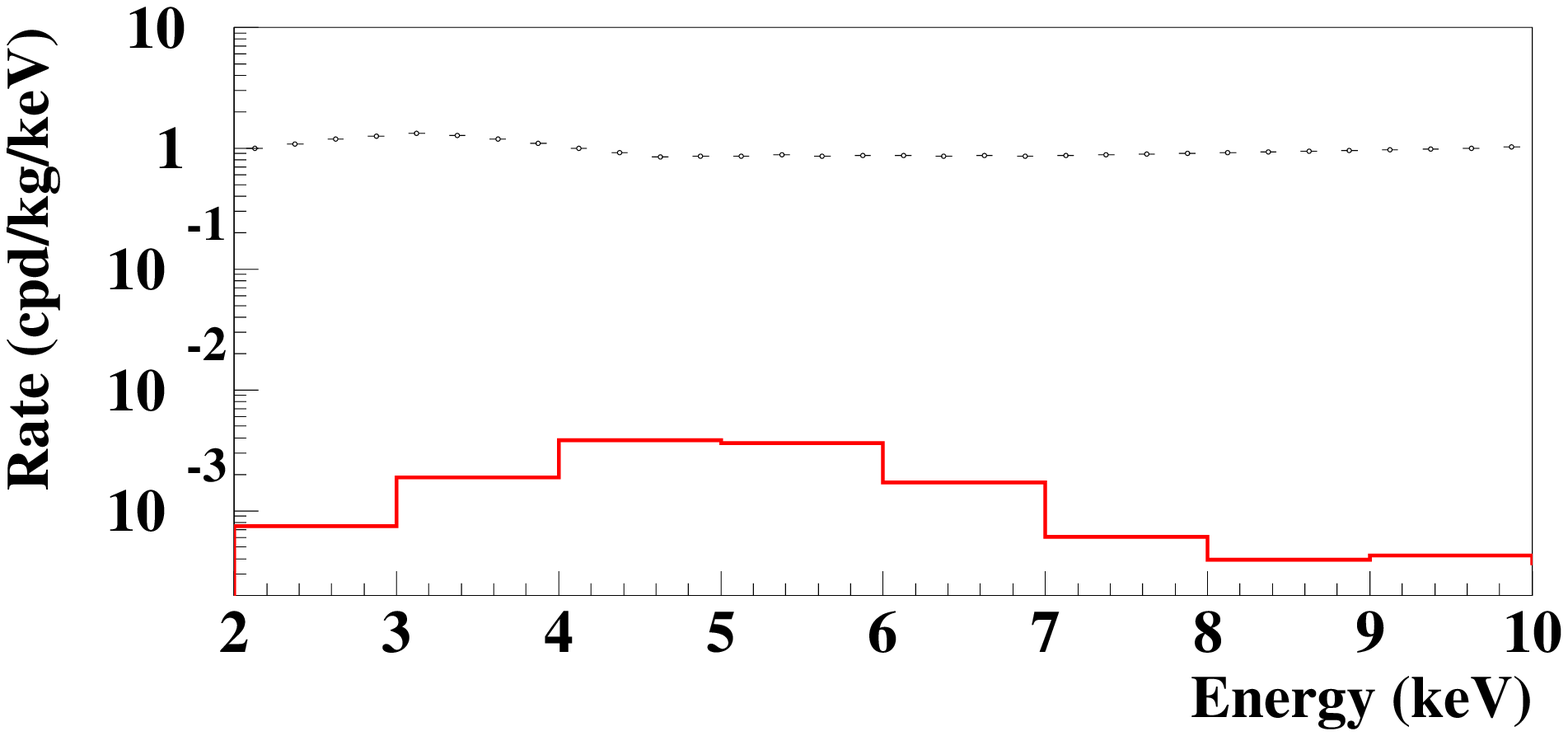}
\includegraphics[width=8.cm] {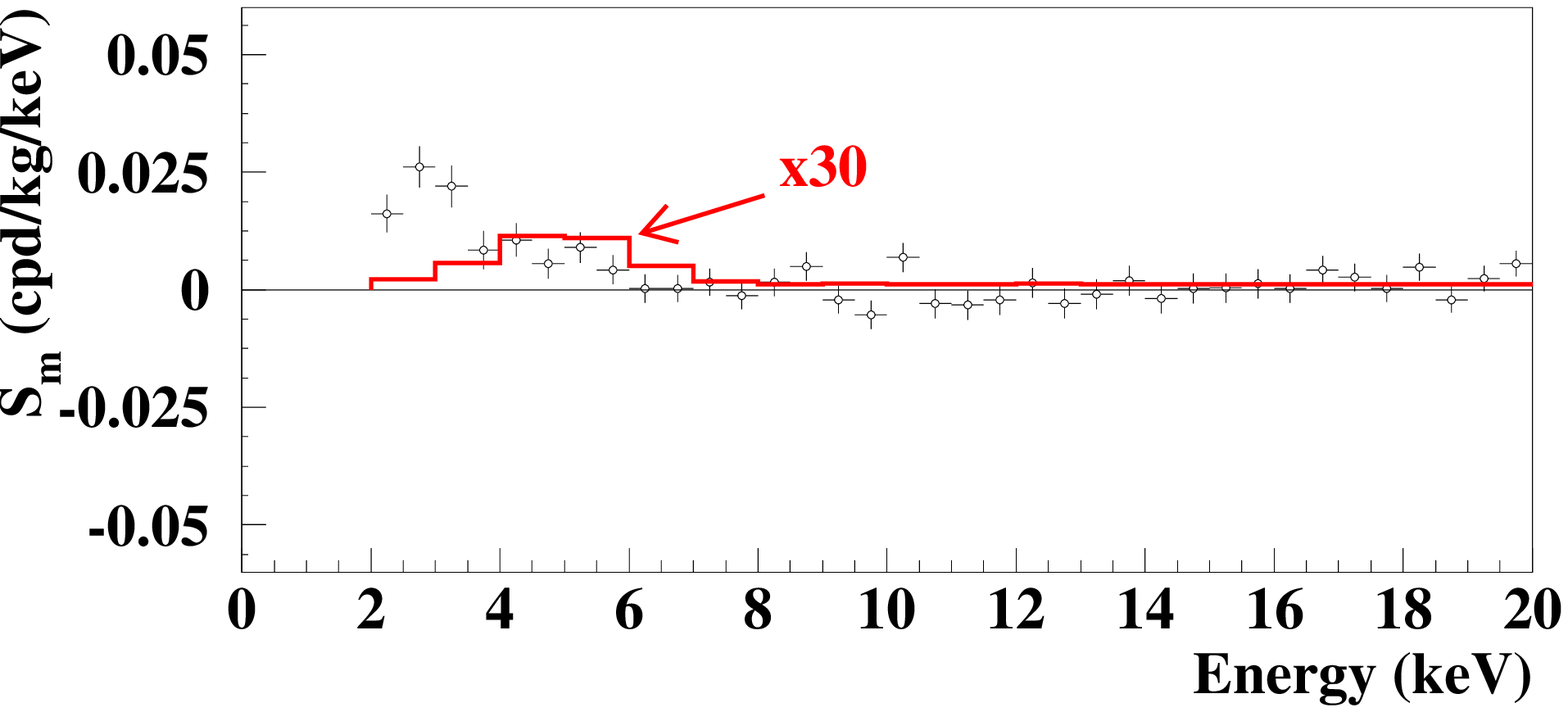}
\end{center}
\vspace{-0.5cm}
\caption{{\it Top -} Data points: cumulative low-energy distribution of the {\it single-hit} scintillation events
measured by DAMA/LIBRA \cite{perflibra} above the 2 keV software energy threshold of the experiment. Histogram (color 
online): maximum expected counting rate from $^{128}$I decays corresponding to the measured upper limit on $^{128}$I 
activity in the NaI(Tl) detectors: $<$15 $\mu$Bq/kg (90\% C.L.); see the data in Ref. \protect\refcite{papep} and the text.
{\it Bottom -} Data points: the DAMA measured modulation amplitude as a function of the energy. Histogram (color 
online): maximum expected modulation amplitude multiplied by a factor 30 as a function of the energy from $^{128}$I 
decays corresponding to the measured upper limit on $^{128}$I activity given above when assuming an hypothetical 10\% 
neutron flux modulation, and with the same phase and period as a DM signal.}
\label{128I}
\end{figure}

The data collected by DAMA/LIBRA allow also the determination of the possible presence of $^{128}$I in the 
detectors. In fact, neutrons would generate $^{128}$I homogeneously distributed in the NaI(Tl) detectors; therefore 
studying the characteristic radiation of the $^{128}$I decay and comparing it with the experimental data, one can obtain
the possible $^{128}$I concentration. The most sensitive way to perform such a measurement is to study the possible 
presence of the 32 keV peak ($K$-shell contribution) in the region around 30 keV \cite{papep,muons}.
As it can be observed in Fig. \ref{fg:128ia}, there is no evidence of such a peak in the DAMA/LIBRA data; hence 
an upper limit on the area of a peak around 32 keV can be derived to be: 0.074 cpd/kg (90\% C.L.) \cite{papep}. 
Considering the branching ratio for $K$-shell EC, the efficiency to detect events in the energy interval around 
30 keV for one $^{128}$I decay has been evaluated by the Montecarlo code to be 5.8\%. Hence, one can obtain a limit on 
possible activity of $^{128}$I: $a_{128} < 15 \; \mu$Bq/kg (90\% C.L.) \cite{muons}.
This upper limit allows us to derive the maximum counting rate which may be expected from $^{128}$I in the keV region;
it is reported in Fig. \ref{128I}--{\it Top} together with the cumulative low-energy distribution of the {\it single-hit} 
scintillation events measured by DAMA/LIBRA \cite{perflibra}. It can be noted that any hypothetical contribution from 
$^{128}$I would be negligible; even arbitrarily assuming the hypothetical case of a 10\% environmental neutron flux 
modulation, and with the same phase and period as the DM signal, the contribution to the DAMA measured (2--6) keV 
{\it single-hit} modulation amplitude would be $<3\times10^{-4}$ cpd/kg/keV at low energy, as reported in Fig. 
\ref{128I}--{\it Bottom}, that is $<2\%$ of the DAMA observed modulation amplitudes. 
In conclusion, any single argument given in this section excludes a role played by $^{128}$I decay.

\subsection{No role for potassium in the DAMA annual modulation result}

Let us here address in details some arguments about the presence of potassium in the detectors and 
the exclusion of any hypothetical role of it.
As first we remark that the only potassium isotope contributing to the background is the radioactive $^{40}$K (natural abundance 
$1.17 \times 10^{-4}$ and half life $1.248 \times 10^{9}$ yr), which is at ppt ($10^{-12}$g/g) level in the detectors\cite{perflibra}.

\begin{figure}[!ht]
\centering
\includegraphics[width=0.85\textwidth] {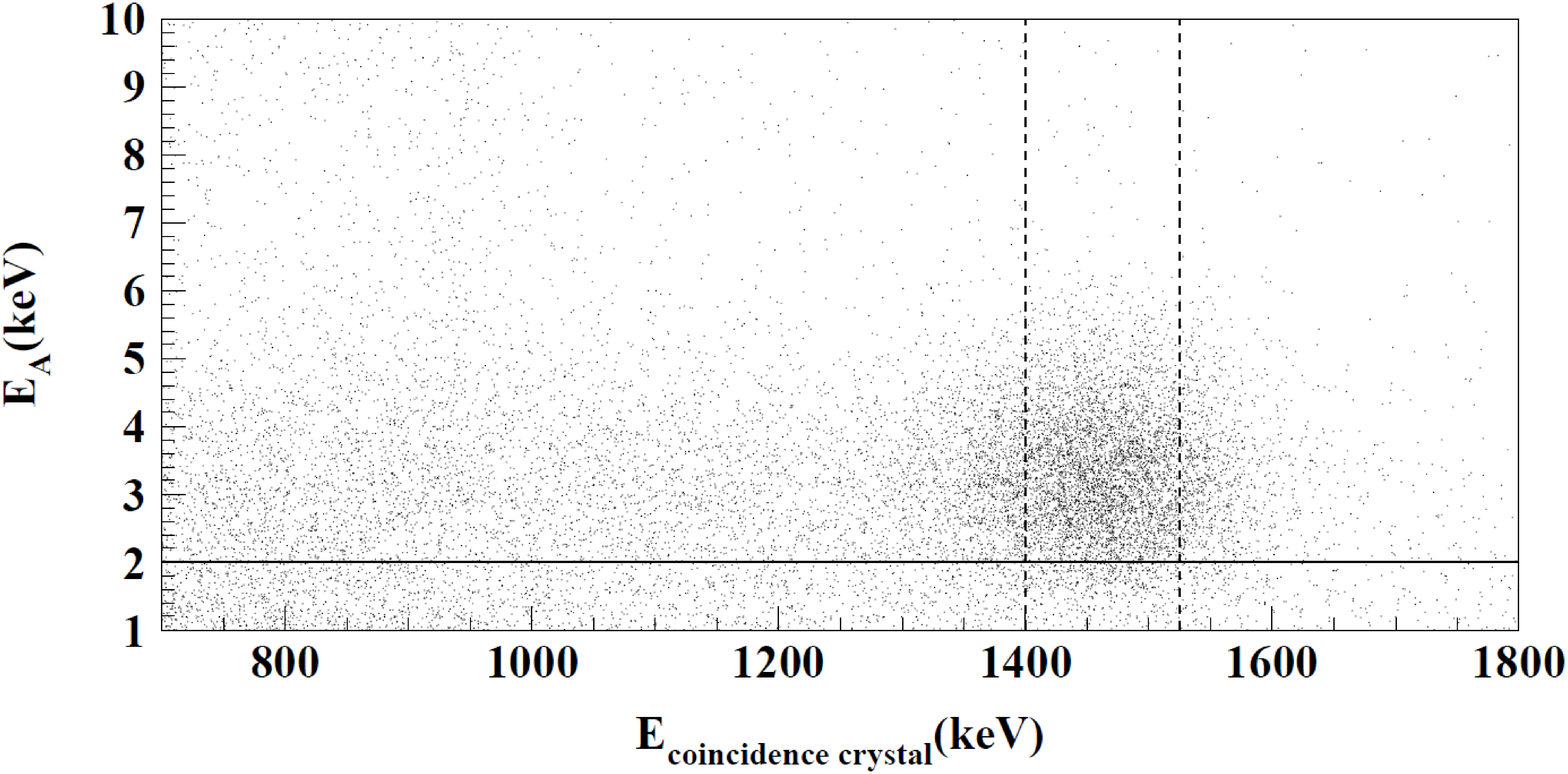}
\caption{Scatter plot of double coincidence events between a DAMA/LIBRA detector, $A$, (low energy region) and an adjacent one
(higher energy region). The threshold of each PMT is at single photoelectron level. For comparison, the software energy threshold
used in the data analyses of the {\it single-hit} events for Dark Matter particle investigation: 2 keV, is shown as continuous line. The 
double coincidence events due to $^{40}$K decay (3.2 keV -- 1461 keV) are well identified. For details see Ref. \protect\refcite{perflibra}.}
\vspace{-0.4cm}
\label{40k}
\end{figure}

The $^{40}$K content of each crystal has been quantitatively determined through the investigation of double coincidences 
(see below and Fig. \ref{40k}). The measured value of $^{nat}$K content averaged on all the crystals is 13 ppb as reported e.g. in 
Ref. \refcite{taupnoz}.
It is worth noting that it appears difficult to do better in NaI(Tl), considering e.g. the chemical affinity of Na and K.

Any hypothetical role played by $^{40}$K in the annual modulation results is discarded by the data, 
by the published analyses and also by simple considerations. 
Let us summarize that:
\begin{itemize}

\item $^{40}$K decay cannot give any modulation at all, unless evoking new exotic physics (see later);

\item although the peak around 3 keV in the cumulative energy spectrum (see Fig. 1 of Ref. \refcite{modlibra} and the discussion in Ref. 
      \refcite{taupnoz}) can partially be ascribed to $^{40}$K decay, there is no evidence for any 3 keV peak in the S$_m$ distribution 
      (see Fig. \ref{sme}). 
      At the present level of sensitivity the S$_m$ behaviour is compatible within the uncertainties both with a monotonic behaviour and 
      with a {\it kind} of structure, as expected for many DM candidates, also including those inducing just nuclear recoils;
      see e.g. Sect. \ref{compatibility};

\item no modulation has been observed in other energy regions where $^{40}$K decays also contribute \cite{modlibra};

\item no modulation has been observed in {\it multiple-hits} events (events where more than one detector fires) in the same energy region 
      where DAMA observes the peculiar modulation of the {\it single-hit} events (events where just one detector fires). In fact, 
      $^{40}$K can also give rise to double events in two adjacent detectors when: i) $^{40}$K decays in a detector, $A$, by EC of K 
      shell to the 1461 keV level of $^{40}$Ar; ii) the 1461 keV $\gamma$ escapes from the $A$ detector and hits an adjacent one causing 
      a double coincidence. The 3.2 keV X-rays/Auger electrons from K shell of $^{40}$Ar are fully contained in the $A$ detector with 
      efficiency $\simeq 1$, giving rise in $A$ to a 3.2 keV peak. These double coincidence events, shown in Fig. \ref{40k}, and also 
      possible multi-site events due to Compton scatterings, are {\it multiple-hits} events, and are not modulated, as fore-mentioned;

\begin{figure}[!ht]
\centering
\includegraphics[width=0.45\textwidth] {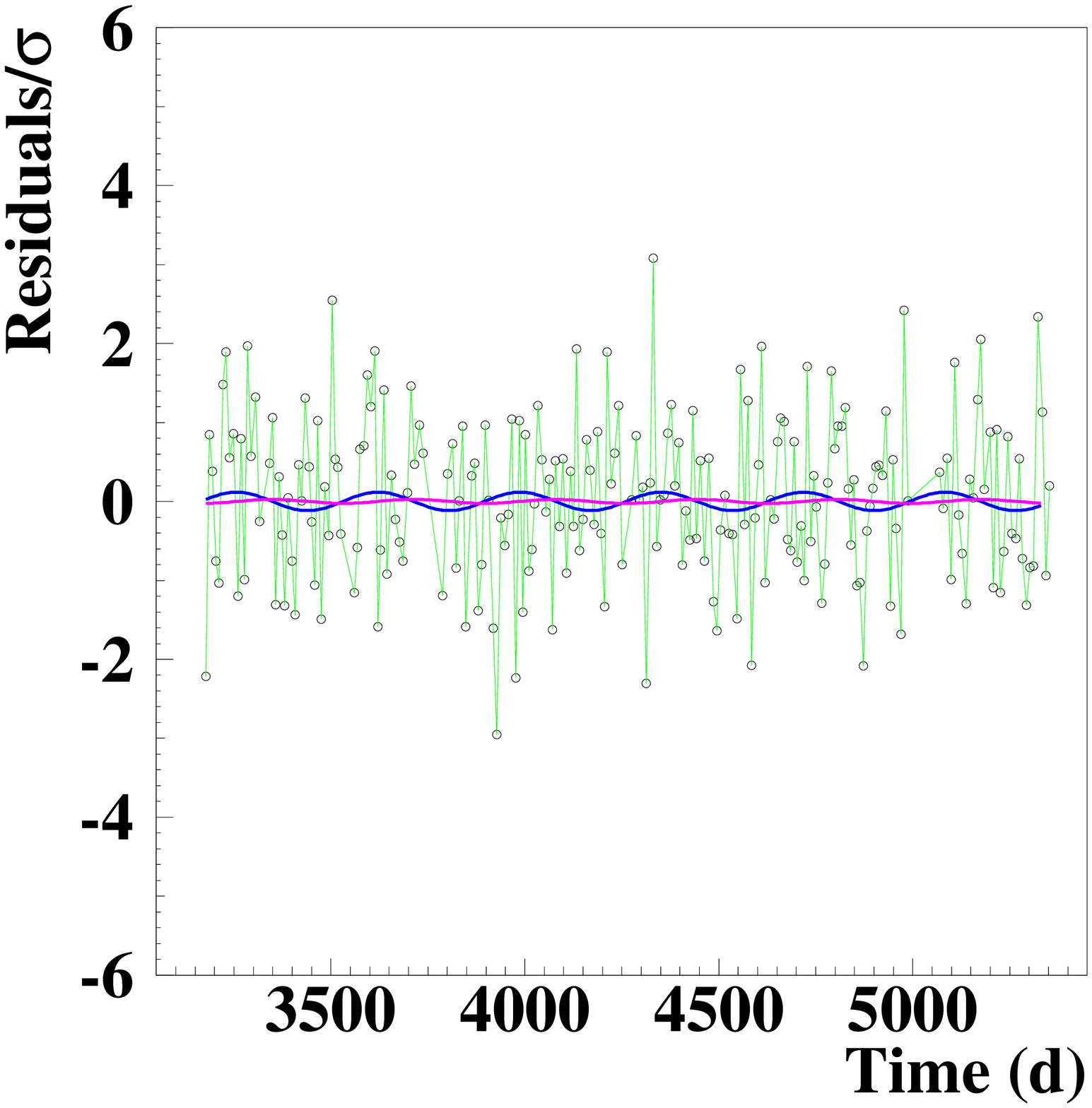}
\includegraphics[width=0.45\textwidth] {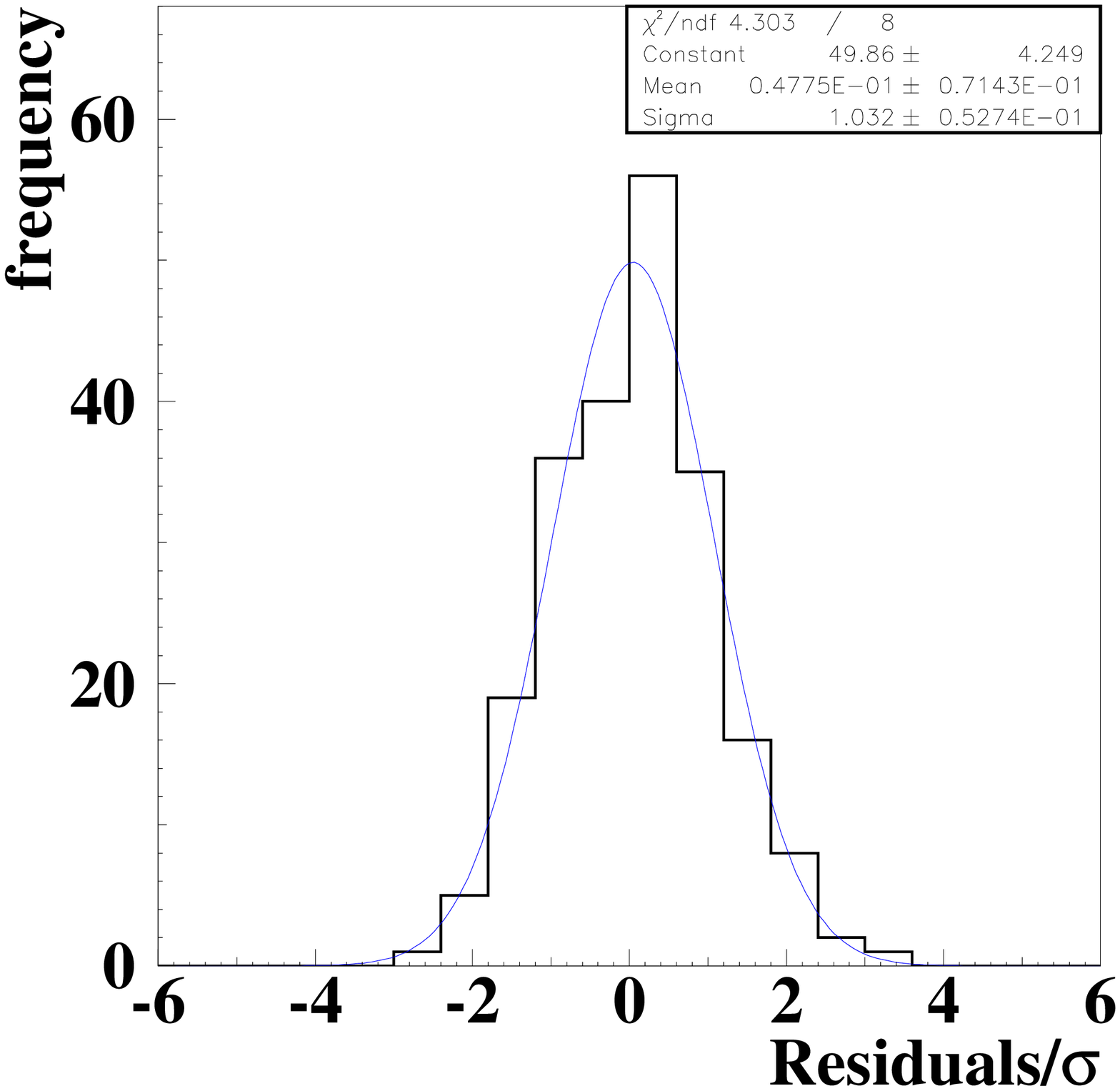}
\caption{
{\it Left:} residuals -- normalized to the error -- of the double coincidence events (3.2 keV -- 1461 keV) due to $^{40}$K decay 
as function of the time throughout the six yearly cycles of DAMA/LIBRA \protect\cite{modlibra,modlibra2} (the data points are connected by
straight segments). The two superimposed curves represent 
the best-fit modulation behaviours $A cos \omega (t-t_0)$ and $A sin \omega (t-t_0)$ with 1 year period and phase 
$t_0 = $ June 2$^{nd}$; the best fit values of the modulation amplitudes are: $A=-(0.117\pm0.094)$ and $A=-(0.025\pm0.098)$, respectively. 
Both values are well compatible with absence of any annual modulation in the double coincidence events due to $^{40}$K decay. Moreover, 
the data show a random fluctuation around zero ($\chi^2/dof=1.04$) and their distribution ({\it right panel}) is well represented 
by a gaussian with r.m.s. $\simeq 1$.}
\label{fg:40k}
\end{figure}

\item no modulation is present in the double coincidence events (3.2 keV -- 1461 keV) due to $^{40}$K decay.
      Their residuals\footnote{They are calculated averaging the counting rate of the double coincidence events 
      (3.2 keV -- 1461 keV) in every pair of adjacent detectors, once subtracted
      its mean value.} -- normalized to the error -- as function of the time are shown in Fig. \ref{fg:40k}{\it --left} throughout 
      the six yearly cycles of DAMA/LIBRA \protect\cite{modlibra,modlibra2}. The two superimposed curves represent the best-fit modulation 
      behaviours $A cos \omega (t-t_0)$ and $A sin \omega (t-t_0)$ with 1 year period and phase $t_0 = $ June 2$^{nd}$;
      the best fit values of the modulation amplitudes are: $A=-(0.117\pm0.094)$ and $A=-(0.025\pm0.098)$, respectively.
      These latter values are well compatible with absence of any annual modulation.
      Moreover, the data show a random fluctuation around zero ($\chi^2/dof=1.04$) and their distribution (see Fig. \ref{fg:40k}{\it --right}) is well 
      represented by a gaussian with r.m.s. $\simeq 1$.
      Hence, in conclusion, there is absence of any annual modulation in the double coincidence events due to $^{40}$K decay;

\item no modulation has been observed in the data points of Fig. \ref{fg:40k}{\it --left} also for the case
      of the phase of cosine to the perihelion ($\simeq$ Jan 3rd). This case is considered owing to the effect in Ref. \refcite{fish08},
      where a 0.3\% yearly modulation of nuclear decays of two nuclides has been reported \footnote{Let us remark that in Ref. \refcite{coop}
      this effect has already been confuted by searching for modifications to the exponential radioactive decay law with the Cassini 
      spacecraft. Moreover, direct measurements also rule it out \cite{bellotti12,bellotti12A}.} with a phase roughly equal to perihelion.
      In particular, the modulation amplitude obtained by best fit is compatible with zero: $A=(0.10\pm0.12)$. 
      Let us note that the effect in Ref. \refcite{fish08} is well below this sensitivity, and that
      this effect in every case is not able to account for 
      the DAMA signal because of the well-different phase, of the marginal modulation amplitude and of several other arguments given in 
      this list. 

\item the analysis of $^{40}$K double coincidences also rules out at level more than 10$\sigma$ C.L. (see Fig. \ref{fg:40k_2}) 
\begin{figure}[!ht]
\centering
\includegraphics[width=0.45\textwidth] {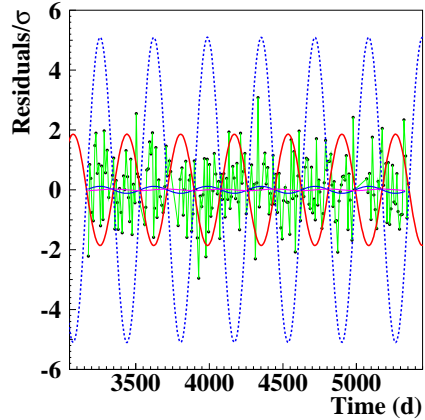}
\caption{({\it Color online}). 
The residuals -- normalized to the error and shown in Fig. \ref{fg:40k} -- of the double coincidence events 
(3.2 keV -- 1461 keV) due to $^{40}$K decay 
as function of the time throughout the six yearly cycles of DAMA/LIBRA \protect\cite{modlibra,modlibra2}. The two superimposed curves represent 
the modulations expected when assuming the DAMA effect given by  the two hypothetical cases of: 
i) $^{40}$K ``exotic'' modulation decay with the same phase and period of the DM particles (red line); 
ii) spill-out from double to single events and viceversa along the time (blue line). While the former effect is expected to be 
in phase with the observed {\it single-hit} modulation, the latter one is in counterphase.
This analysis of $^{40}$K double coincidences rules out the two hypotheses 
at level more than 10 $\sigma$ C.L.: by the fact,
the $\chi^2$ are equal to 674 and 3063, respectively, over 219 degrees of freedom.
However, these exotic hypothetical arguments were already excluded by several other arguments given in the text.}
\label{fg:40k_2}
\end{figure}
      the modulation amplitudes expected 
      when assuming that the DAMA effect might be due to the two hypothetical cases of: i) $^{40}$K ``exotic'' modulation decay (also see 
      above) with the same phase and period of the DM particles; 
      ii) spill-out from double to single events and viceversa along the time; on the other hand these exotic hypothetical 
      arguments were already excluded also by several other arguments given in this list.

\item the behaviour of the overall efficiency during the whole data taking period is highly stable. A quantitative investigation of the 
      role of the efficiency as possible systematics leads to a cautious upper limit less than 1\% of the modulation amplitude, as 
      reported e.g. in Ref. \refcite{modlibra,modlibra2};

\item the annual modulation signal -- observed by DAMA -- is present both in the outer and in the inner detectors \cite{modlibra,modlibra2}. 
      Hence, there is no dependence on the veto capability, that is different -- by geometrical reasons -- among the outer and the inner 
      detectors. In particular, if the $^{40}$K decay hypothetically played some role in the annual modulation of the low-energy 
      {\it single-hit} events, the effect would be larger in the outer detectors where the 1461 keV $\gamma$'s accompanying the 3.2 keV 
      X-rays/Auger electrons have lower probability to be detected by closer detectors; 

\item the annual modulation signal -- observed by DAMA -- is equally distributed over all the detectors \cite{modlibra,modlibra2}; see also the 
      previous considerations.

\end{itemize}

Thus, for all the above reasons (and just one of them is enough), no role can be played by $^{40}$K in the DAMA annual modulation results.

\subsection{Further arguments and miscellanea}

In the following we will discuss further arguments erroneously put forward in literature.

\subsubsection{No necessity for a large modulation fraction ($S_m/S_0$) in the inter\-pre\-ta\-tion of the DAMA result}
The necessity for a large modulation fraction ($S_m/S_0$) in corollary model dependent 
interpretation of the DAMA result has been claimed in Refs. \refcite{Pra12,Pra12A,Pra12B}.

In particular, in Ref. \refcite{Pra12,Pra12A,Pra12B} is claimed that the presence of $^{nat}$K also at $\sim10$ ppb level poses a challenge to any 
interpretation of the DAMA results in terms of a Dark Matter model, requiring a 20\% modulation fraction or more; these papers contain 
several erroneous claims already confuted by DAMA coll.\cite{replica,replicaA}.
On the contrary of what is claimed in Ref. \refcite{Pra12,Pra12A,Pra12B}, the obtained DAMA model independent evidence is compatible with a 
wide set of scenarios regarding the nature of the Dark Matter candidate and related astrophysical, nuclear and particle physics (see Sect. 
\ref{compatibility} and literature).

It is worth noting that the procedure used by DAMA in the model-dependent analyses accounts for all the experimental 
information carried out by the data.
In particular a maximum likelihood, ML, procedure is used; the ML function can be defined as the product of the
ML functions for single energy bin given in Eq. \ref{eq:likelihood}:
\begin{equation} 
{\it\bf L}  = 
{\bf \Pi}_k {\it \bf L_{\it k}} =
{\bf \Pi}_{ijk} e^{-\mu_{ijk}}
{\mu_{ijk}^{N_{ijk}} \over N_{ijk}!}.
\end{equation} 
We remind that $N_{ijk}$ is the number of events collected in the
$i$-th time interval (hereafter 1 day), by the $j$-th detector and in the
$k$-th energy bin,
and $\mu_{ijk} = \left[ b_{jk} + S_{0,k} + S_{m,k} \cdot cos \omega(t_i - t_0) \right] M_j \Delta
t_i \Delta E \epsilon_{jk}$ is its expectation. 
Since $S_{0,k}$ and $S_{m,k}$ depend on the theoretical scenario (DM candidate, astrophysical parameters, ...)
{\bf L} will be a function of all these parameters, generally indicated as $DM$, and of the whole $b_{jk}$ and $N_{ijk}$ sets, 
shortly indicated as $b$ and $N$, respectively:
${\bf L} = {\bf L}(N|b,DM)$. 
The maximum likelihood estimates of the parameters $b$ and $DM$ are obtained by maximizing {\bf L} for a 
given set of observations, $N$, or by minimizing the function $y= -2\mbox{ln}{\bf L} -const$, with $const$
chosen so as to have $y(no DM) = 0$. In the minimization procedure $b_{jk} \geq$ a cautious estimate of each 
detector background. Thus, this maximum likelihood procedure
requires the agreement:
i) of the expectations for the modulated part of the signal
with the measured modulated behavior for each
detector and for each energy bin;
ii) of the expectations for the unmodulated component of
the signal with respect to the measured differential
energy distribution for each detector.

A visual indication can be found in Ref. \refcite{taupnoz}, where the cumulative energy spectrum over the 25 detectors has been
reported up to 10 keV, just as an example, showing that there was room for a sizeable constant part of the signal: namely $S_0 \leq 0.25$
cpd/kg/keV in the (2--4) keV energy region, considering the measured $^{40}$K residual contamination in the crystals and the 
remaining background estimated from the data behaviours in the energy region up to $\sim 20 $ keV, 
requiring consistency with the behaviour of each detector.
Since the measured modulation amplitude ($S_m$) is around $10^{-2}$ cpd/kg/keV, there is no reason to claim the necessity to have 
``a 20\% modulation fraction or more''.

Moreover, the necessity for a large modulation fraction ($S_m/S_0$) claimed in Ref. \refcite{Pra12,Pra12A,Pra12B} is based on the erroneous 
assumption of a 0.85 cpd/kg/keV flat background in the (2--7) keV energy region. This approach is completely arbitrary  \footnote{This
procedure is not the same as the one performed by DAMA, on the contrary of what claimed there.} because it is
not based on a knowledge of the background contributions, but it is the result of a fitting procedure -- among others -- also based
on incorrect hypotheses, as discussed in Ref. \refcite{replica,replicaA}.
In the approach followed by the DAMA experiments, the Dark Matter annual modulation signature, the experimental observable is not S$_0$, 
but the modulation amplitude, S$_m$, as a function of energy.
Thus, the Dark Matter annual modulation approach does not require any identification of S$_0$ from the total {\it single-hit} counting 
rate, in order to establish the presence of DM particles in the galactic halo. 
Thus, such signature allows one to overcome the large uncertainties associated to the exploitation of many 
data selections/subtractions/statistical-discrimination procedures, to the modeling of surviving background in keV region and 
to the {\it a priori} assumption on the nature and interaction type of the DM particle(s), which affect other approaches.
In particular, as already mentioned e.g. in Ref. \refcite{perflibra}, a precise modeling of background in the keV region counting rate is always 
unlikely because e.g. of 
(i) the limitation of Montecarlo simulations at very low energies; 
(ii) the fact that often just upper limits for residual contaminants are available (and thus the real amount is unknown); 
(iii) the unknown location of each residual contaminant in each component of the set-up; 
(iv) the possible presence of non-standard contaminants, generally unaccounted;
(v) etc..
In conclusion, this explains why the conclusions derived from some fitting procedures used in Ref. \refcite{Pra12,Pra12A,Pra12B} are untenable.

Finally, just for completeness, let us note that in any case scenarios and Dark Matter candidates exist which can provide relatively 
large modulation fraction as well (see f.i. Ref. \refcite{Freese12}).

\subsubsection{No role for hypothetical phosphorescence induced by muons}     \label{phos}

The possible contribution to the (2--6) keV {\it single-hit} events from delayed phosphorescent pulses induced by the muon 
interaction in the NaI(Tl) crystals, as proposed in Ref. \refcite{nygren}, can be excluded according to the following arguments
(see also Refs. \refcite{muons,tipp11}).

Since the total $\mu$ flux in DAMA/LIBRA is about 2.5 $\mu$/day (see Sect. \ref{muon}), the total $\mu$ modulation amplitude 
in DAMA/LIBRA is about: $0.015 \times 2.5$ $\mu$/day $\simeq 0.0375$ $\mu$/day (1.5\% muon modulation has been 
adopted, see Sect. \ref{muon}). The {\it single-hit} modulation amplitude measured in DAMA/LIBRA in the (2--6) keV energy 
interval is instead \cite{modlibra,modlibra2}: $S_{m}(2-6 \; \textrm{keV}) \times \Delta E \times M_{set-up} \sim 
10 \;\textrm{cpd}$ \cite{muons}. Thus, the number of muons is too low to allow a similar effect to contribute 
to the DAMA observed amplitude; in fact, to give rise to the DAMA measured modulation amplitude each $\mu$ should give 
rise to about 270 {\it single-hit} correlated events in the (2--6) keV energy range in a relatively short period 
\cite{muons}. 

Such an hypothesis would imply dramatic consequences for every NaI(Tl) detector at sea level (where the $\mu$ flux is 
$10^{6}$ times larger than deep underground at LNGS), precluding its use in nuclear and particle physics.
Moreover, phosphorescence pulses (as afterglows) are single and spare photoelectrons with very short time decay 
($\sim$10 ns); they appear as ``isolated'' uncorrelated spikes. On the other side, scintillation events are the sum 
of correlated photoelectrons following the typical time distribution with mean time equal to the scintillation 
decay time ($\sim$240 ns). Pulses with short time structure are already identified and rejected in the noise 
rejection procedure described in detail, e.g. in Ref. \refcite{perflibra} (the information on the pulse profile is recorded).
Thus, in addition, phosphorescence pulses are not present in the DAMA annual modulation data.

Furthermore, because of the poissonian fluctuation on the number of muons, the standard deviation of the (2--6) keV {\it single-hit} 
modulation amplitude due to an effect of hypothetical phosphorescence induced by muons
would be 13 times larger than that measured by DAMA (see Ref. \refcite{muons})
and therefore no statistically-significant effect, produced by any correlated events, could be singled out.
Even just this argument is enough to discard the hypothesis of Ref. 
\refcite{nygren} (similar considerations are also reported in Ref. \refcite{blum}).

Thus, the argument regarding a possible contribution from delayed phosphorescent pulses induced by muons can be safely rejected.

\subsubsection{No role for long term modulation}

In Ref. \refcite{blum} it is argued that high-energy muons measured by LVD might show a long term modulation with 
a period of about 6 years, suggesting that a similar long term modulation might also be present in the DAMA data.
Many arguments already addressed in the present paper discard such a suggestion. However, for completeness such a 
long term modulation has also been looked for in the DAMA data\cite{muons}.

\begin{figure}[!ht]
\begin{center}
\includegraphics[width=6.5cm] {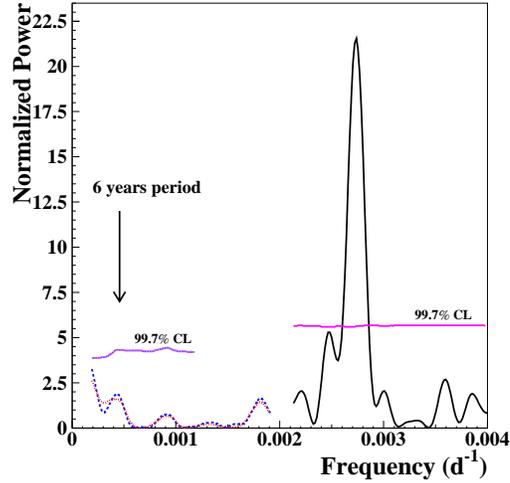}
\end{center}
\caption{({\it Color online}).
Power spectrum of the annual baseline counting rates for (2--4) keV (dashed -- blue -- curve) and 
(2--6) keV (dotted -- red -- curve) {\it single-hit} events. Also shown for comparison is the power spectrum 
obtained by considering the 13 annual cycles of DAMA/NaI and DAMA/LIBRA for the {\it single-hit} 
residuals in (2--6) keV (solid -- black -- line) \cite{modlibra2}. The calculation has been performed according to 
Ref. \protect\refcite{Lomb,LombA}, including also the treatment of the experimental errors and of the time binning. As can 
be seen, a principal mode is present at a frequency of $2.735 \times 10^{-3}$ d$^{-1}$, that corresponds to a 
period of $\simeq$ 1 year. The 99.7\% confidence lines for excluding the white noise hypothesis are also shown (see text). 
No statistically-significant peak is present at lower frequencies and, in particular,
at frequency corresponding to a 6-year period. This implies that no evidence for a long term modulation in the 
counting rate is present. \cite{muons}}
\label{fg:lomb}
\end{figure}

For each annual cycle of DAMA/NaI and DAMA/LIBRA, the annual baseline counting rates, that is the averages on all 
the detectors ($j$ index) of $flat_{j}$, have been calculated for the (2--4) keV and (2--6) keV energy intervals, respectively.
Their power spectra (dashed -- blue online -- and dotted -- red online -- curves, respectively) 
in the frequency range 0.0002--0.0018 d$^{-1}$ (corresponding to a period range 13.7--1.5 year) are reported in Fig. 
\ref{fg:lomb}.
The power spectrum (solid black line) above 0.0022 d$^{-1}$ of Fig. \ref{fg:pwr}, 
obtained when considering the (2--6) keV {\it single-hit} residuals,
is reported for comparison. To evaluate the statistical significance of these power spectra 
we have performed a Montecarlo simulation imposing constant null expectations for residuals; from the simulated power spectra the 
probability that an apparent periodic modulation may appear as a result of pure white noise has been evaluated.
The 99.7\% confidence lines for excluding the white noise hypothesis are shown in Fig. \ref{fg:lomb}. 
A principal mode is present in the power spectrum of the experimental data for a frequency equal to $2.735 \times 10^{-3}$ d$^{-1}$ 
(black solid curve), corresponding to a period of $\simeq$ 1 year, while no statistically-significant peak 
is present at lower frequencies and, in particular, at frequency corresponding to a 6-year period. 

\begin{figure}[!ht]
\begin{center}
\includegraphics[width=6.0cm] {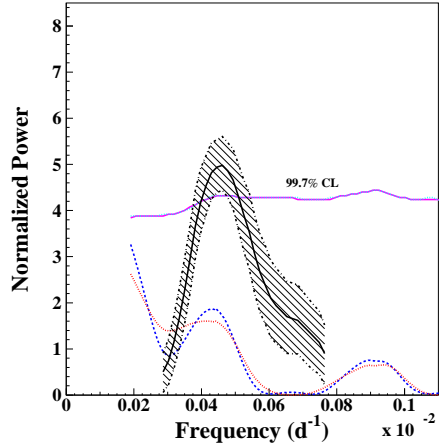}
\end{center}
\caption{({\it Color online}). A detail of Fig. \ref{fg:lomb}, with superimposed power spectrum
(solid black curve),
expected in the hypothesis of a contribution from muons with 6-year period. 
The shaded region is the $1 \sigma$ (68\% C.L.) band.
The peak at 6 year period would be in this hypothesis well evident and above the threshold 
of detectability at $3\sigma$ C.L.. On the contrary, the power spectrum of the 
experimental data (dashed -- blue -- and dotted -- red -- curves)
is completely outside the $1 \sigma$ band.
For simplicity, the calculations are shown just for the cumulative (2--6) keV energy interval. 
This further shows that no evidence for a long term modulation in the 
counting rate is present, as it should already be expected on the basis of 
the many other arguments discussed in this paper. 
See text. \cite{muons}}
\label{fg:lomb_cl}
\end{figure}

A further investigation of any hypothetical 6-year period has been performed in Ref. \refcite{muons}, taking into account that the LVD 
muon data have a 1-year period modulation amplitude equal to 1.5\% \cite{LVD}, and -- according to the claim of Ref. \refcite{blum} --
a 6-year period modulation amplitude equal to 1\% (actually $\gsim 1 \%$, as reported in Ref. \refcite{blum}). Thus, in the case 
that muons might contribute to the DAMA effect (which is excluded as already discussed in Sect. \ref{muon}), a 6-year period modulation would be present in the DAMA data with amplitude 
$\simeq 0.008$ cpd/kg/keV, that is a $1\%/1.5\%$ fraction of the 1-year period modulation amplitude measured by DAMA.
A simulation of $10^6$ experiments has been performed and the power spectrum of each simulation has been derived. 
The average of all the simulated power spectra is reported in Fig. \ref{fg:lomb_cl} as solid black curve; the shaded 
region is the $1 \sigma$ (68\% C.L.) band. The hypothetical peak at 6-year period would be under these assumptions 
well evident and above the threshold of detectability at $3\sigma$ C.L.; on the contrary, the power spectrum of the 
experimental data is completely outside the $1 \sigma$ band.

This further shows that no evidence for a long term modulation in the counting rate is 
present, as -- on the other hand -- it should already be expected on the basis of the many other arguments 
(and just one suffices) discussed in this paper, further demonstrating that there is no role for muons.

\subsubsection{No role for other modulations with frequencies larger than the annual frequency}

In Ref. \refcite{sturrock} it has been argued, with wrong arguments, the presence in the DAMA data of other modes
with higher frequencies ($\simeq 11.4$ y$^{-1}$), corresponding to about 30 day period.
The authors use these arguments to infer that the DAMA results might at some extent have 
a similar nature as the claimed measurements of $^{36}$Cl and of $^{32}$Si data in Ref. \refcite{alb86},
and be correlated with the solar rotation.

Let us comment that the Lomb-Scargle procedure, used in Ref. \refcite{sturrock}, does not take into account either 
the experimental errors or the time binning (i.e. the start and stop times of each measurement), as instead always done 
in DAMA analysis (see Sect. \ref{dama_res}).
Moreover, the frequencies studied in Ref. \refcite{sturrock} largely exceed the Nyquist one
(which is for DAMA data of the order of 3 y$^{-1}$).

Thus, the underlying problem is that it is not possible to extract from the DAMA residual rate
(having 40-60 days time bins) an oscillation 
with 30 day period (that is 11-13 y$^{-1}$) as that claimed in Ref. \refcite{sturrock}.
In fact, it is quite easy to understand that e.g. a 30 days oscillation is washed out by the integration over the 
time bin used by DAMA.
As mentioned, the integration over the time bin is not considered in the procedure adopted by those authors, and this 
invalidates their outcome.
The Lomb-Scargle periodogram obtained with the correct procedure and treatment of experimental errors and time bin
is shown in Fig. \ref{fg:LombL}, where no peak is present at the frequency of $\simeq 11.4$ y$^{-1}$.
Therefore, their conclusion about hypothetical influence of solar rotation with 10--15 y$^{-1}$ frequency is meaningless.

\section{Corollary model dependent analyses and the compatibility of the DAMA results with many scenarios}
\label{compatibility}

The DM model independent results pointed out by DAMA experiments give evidence with high C.L.
for the presence of DM particles in the galactic halo.
In order to perform corollary investigation on the nature of the DM particles, model dependent 
analyses are necessary \footnote{It is worth noting that it does not exist in direct and indirect DM detection experiments 
approaches which can offer such information independently on the assumed models.}; thus,
many theoretical and experimental parameters and models appear 
and many hypotheses must also be exploited. A wide set of scenarios is possible 
considering the lack of knowledge about the real nature of the candidate, its
distribution in the Galaxy, its coupling to target materials, etc. This reflects the fact that
there is no ``Standard Model'' for DM particles and whatever corollary interpretation
and comparison in the field is model-dependent.

\begin{figure}[!t]
\begin{center}
\includegraphics[width=6.cm] {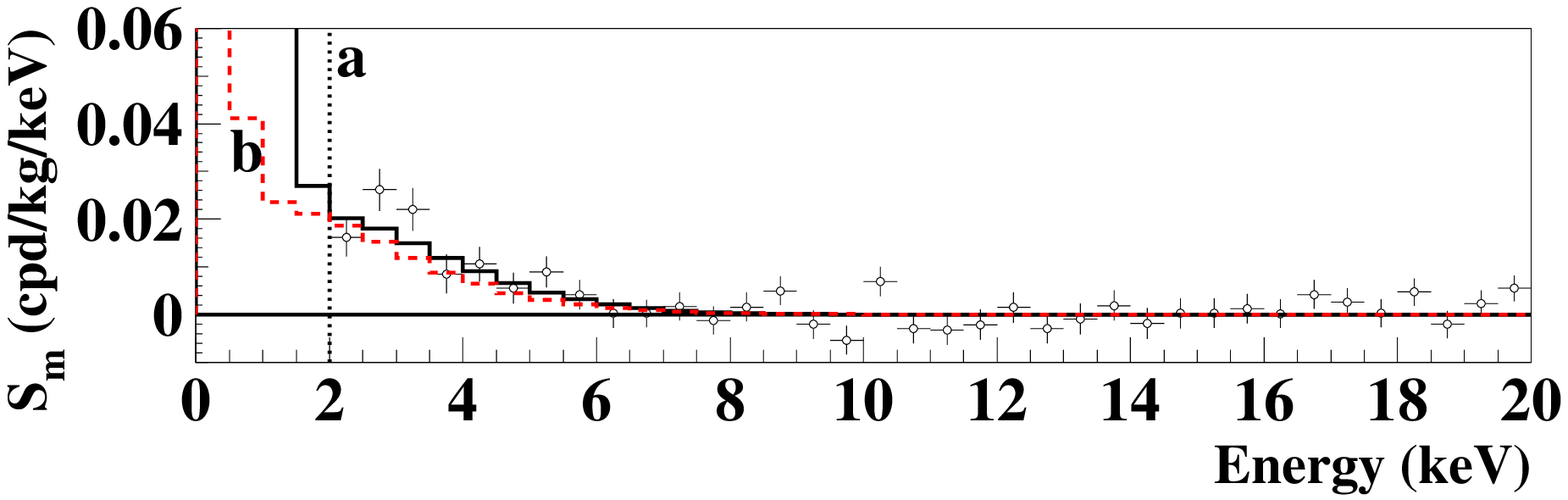}
\includegraphics[width=6.cm] {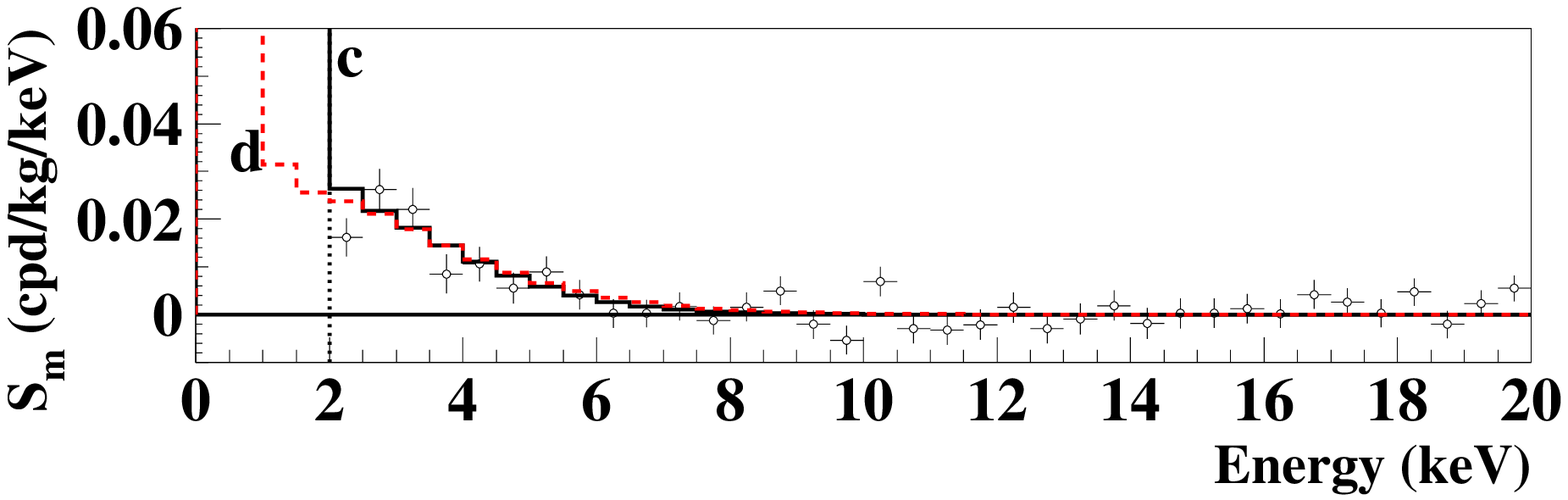}
\includegraphics[width=6.cm] {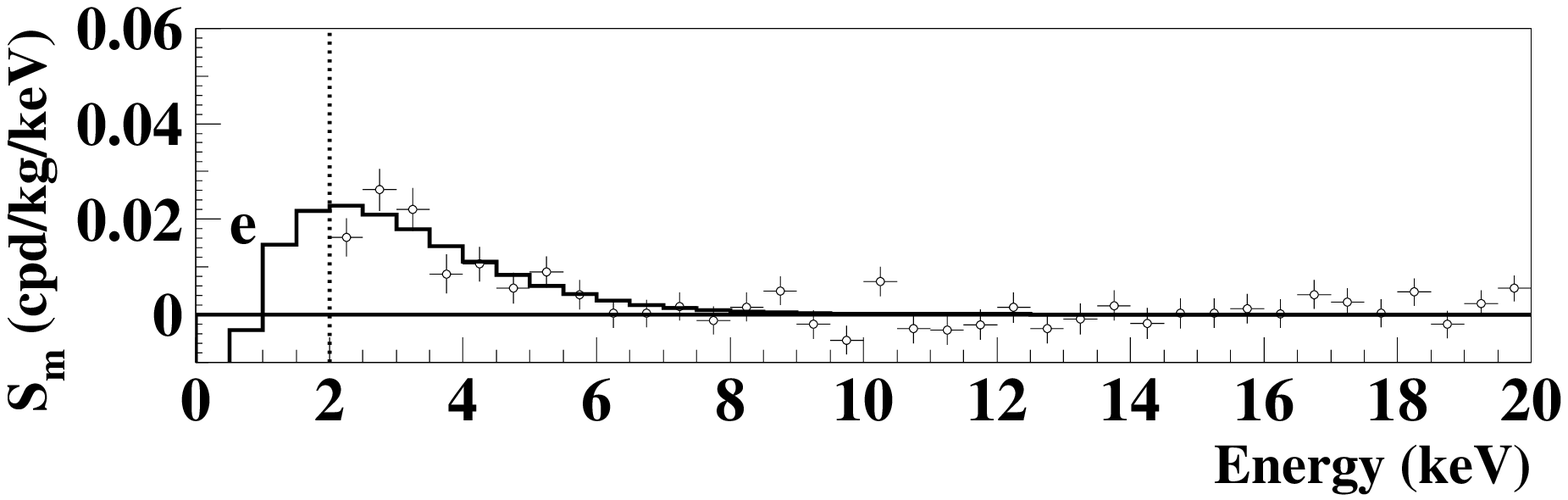}
\includegraphics[width=6.cm] {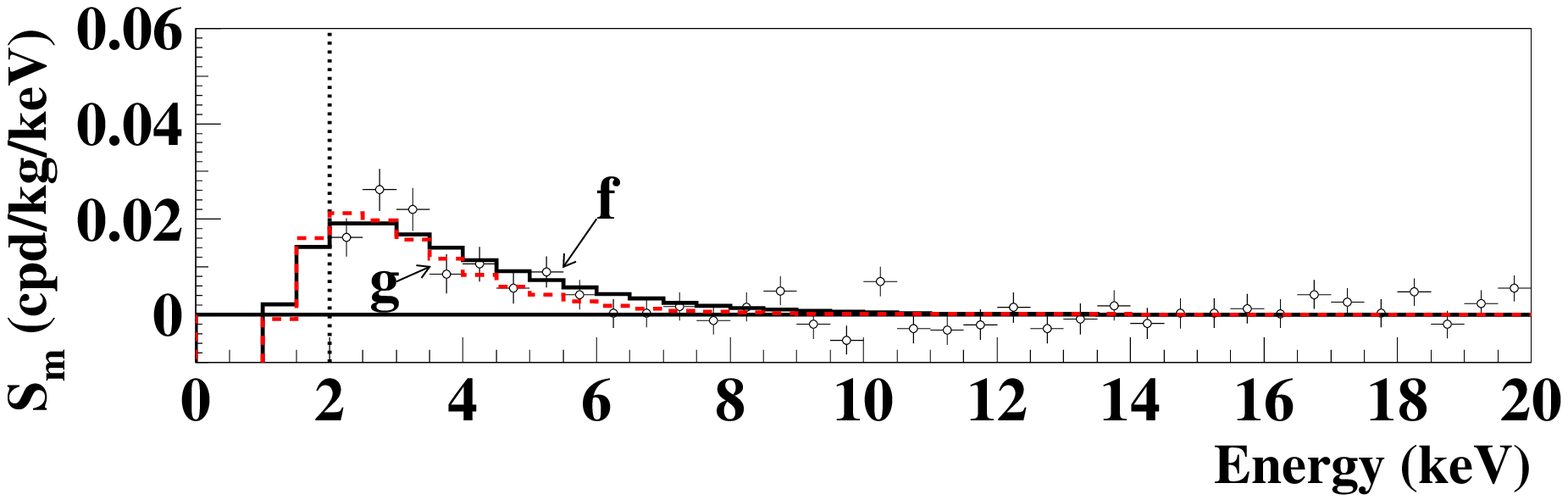}
\includegraphics[width=6.cm] {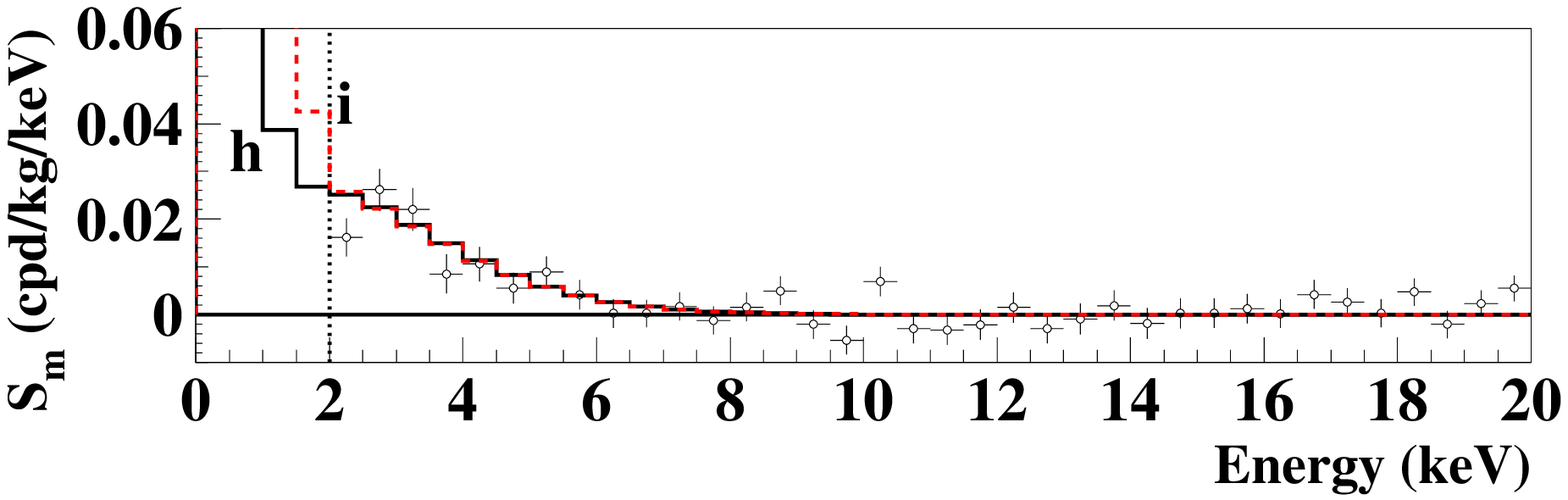}
\includegraphics[width=6.cm] {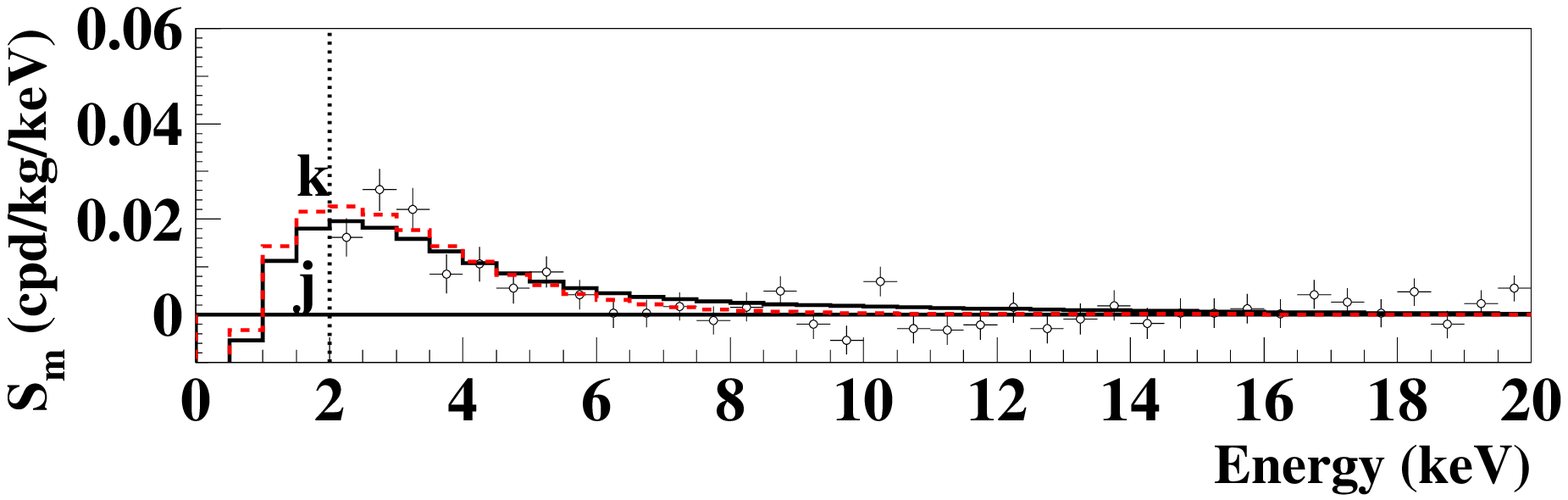}
\includegraphics[width=6.cm] {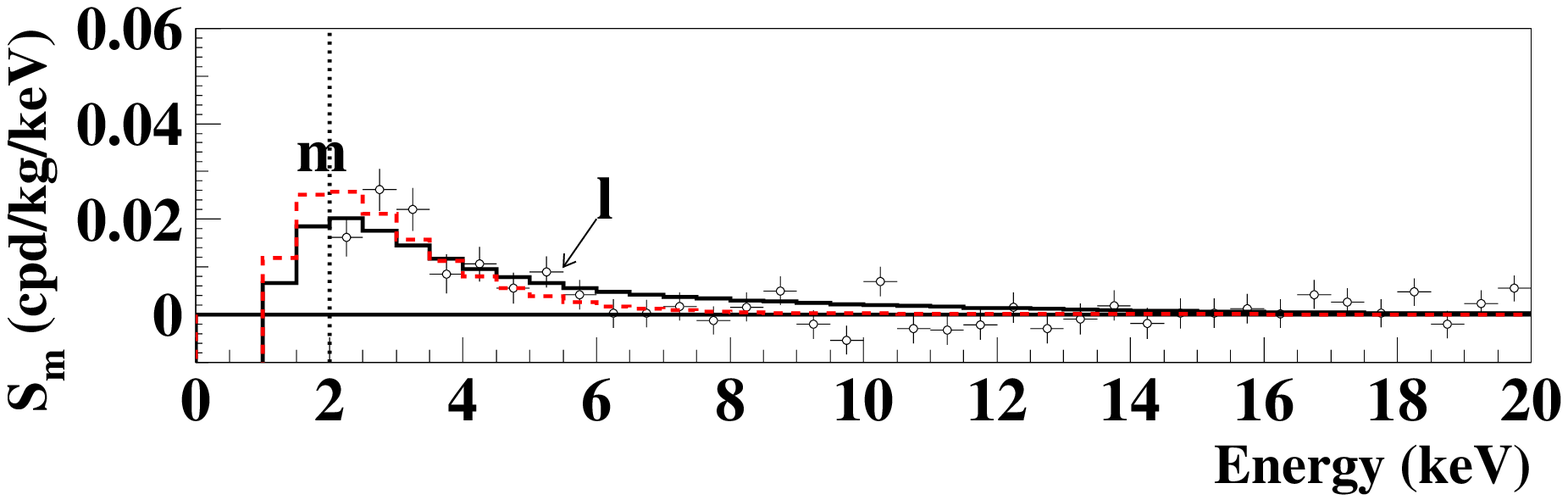}
\includegraphics[width=6.cm] {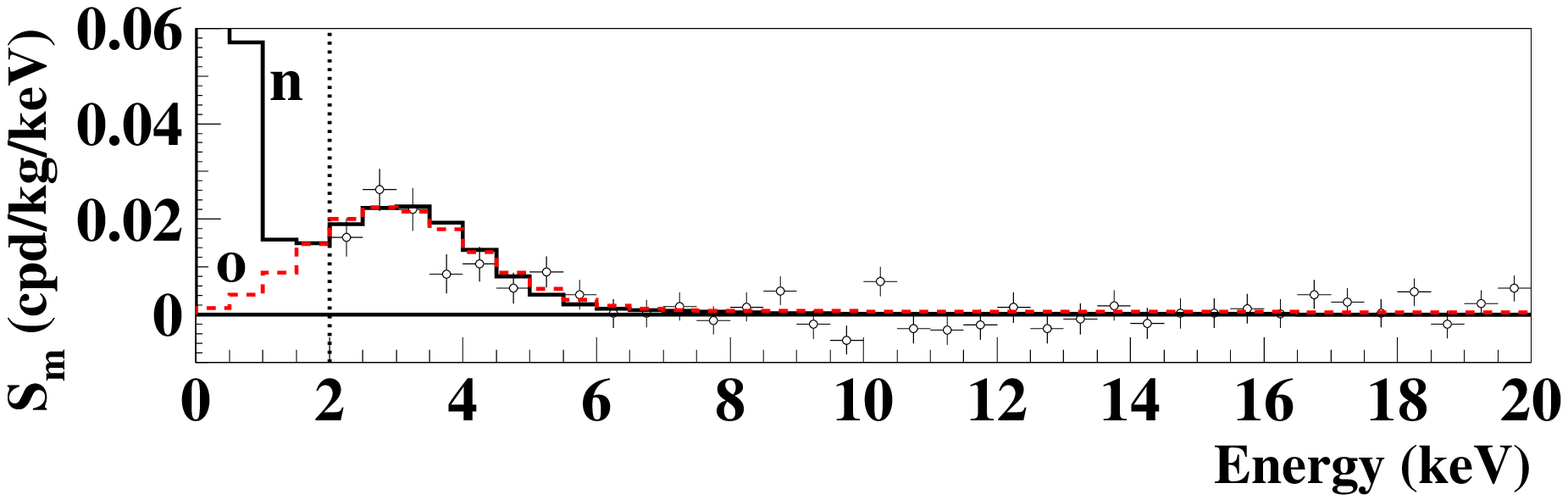}
\includegraphics[width=6.cm] {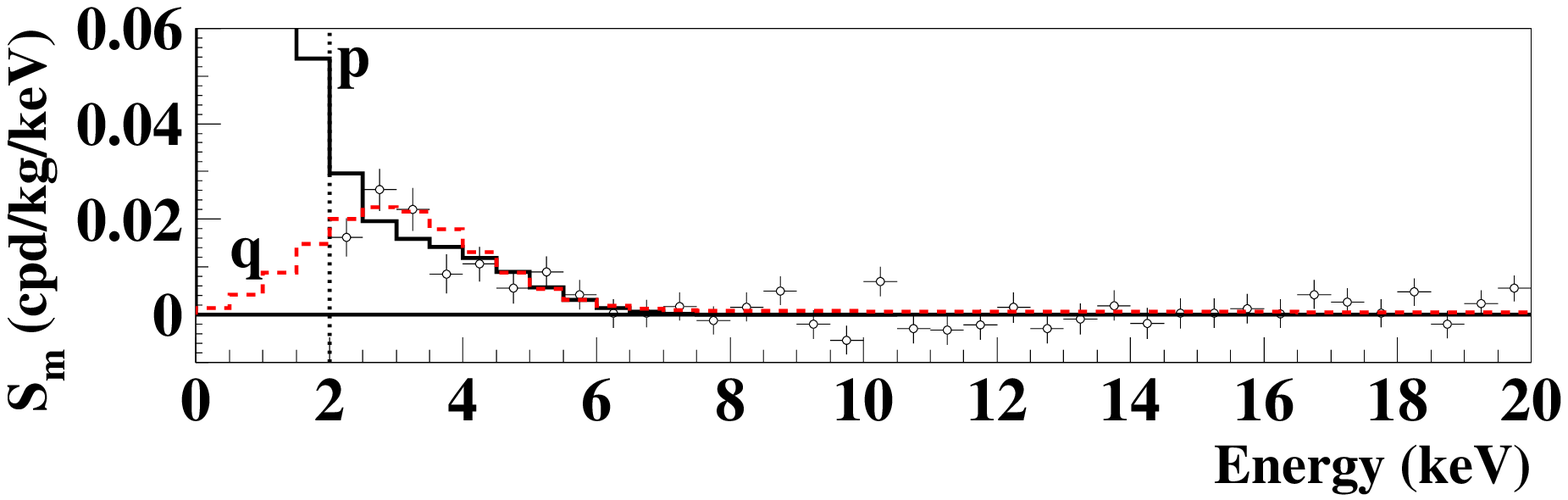}
\includegraphics[width=6.cm] {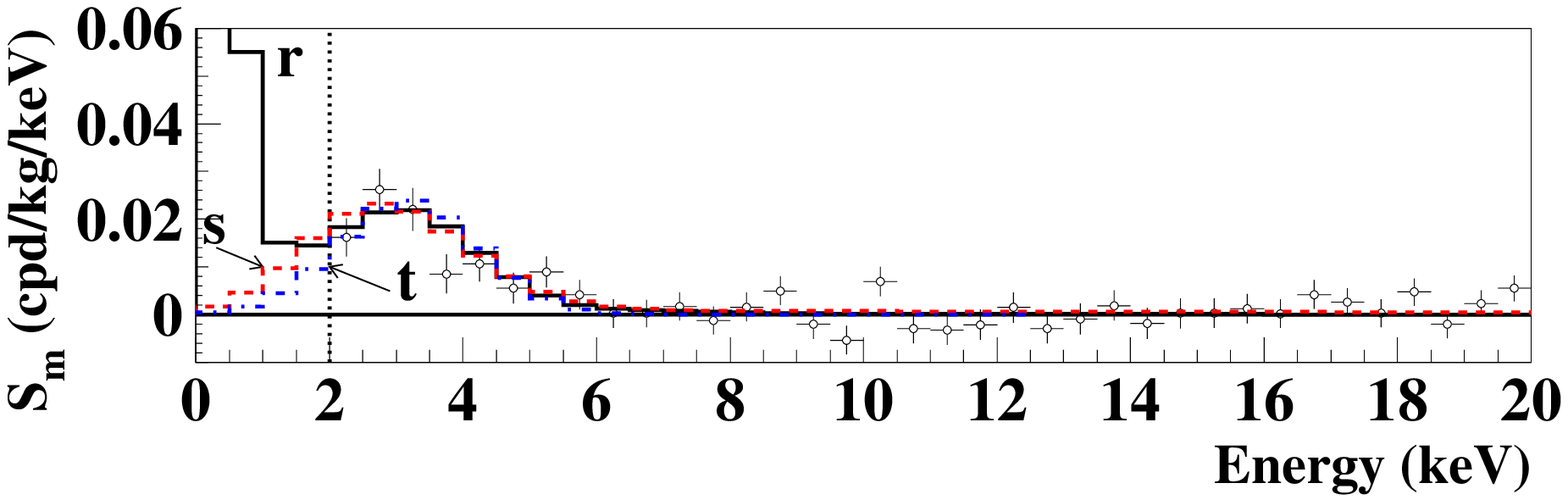}
\end{center}
\caption{Examples of expected behaviors for some of the many possible scenarios 
(see Table \ref{tb:legenda}), superimposed to the measured $S_{m}$ values of Fig. 
\ref{sme}. The shown behaviours have not been obtained by the maximum likelihood method (see
in our quoted literature) and are shown just for illustrative purpose; they all give practically 
about the same C.L.. 
The full treatment of
the data by maximum likelihood method to update the volumes/regions allowed at given C.L. 
by the cumulative DAMA/NaI and DAMA/LIBRA data
for the considered scenarios 
will be presented elsewhere following the full analysis method of Refs. 
\protect\refcite{RNC,ijmd,ijma,sag06,em_int,chan,wimpele,ldm}.}
\label{fg:tem}
\end{figure}   

The obtained model independent evidence of DAMA is compatible with a wide 
set of existing scenarios regarding the nature of the DM candidate
and related astrophysical, nuclear and particle Physics. Some interpretations
and examples can be found  e.g. in
Refs.~\refcite{wimpele,em_int,ldm,ijma,RNC,sag06,allDM,allDMA,allDMB,allDMC,allDMD,allDME,allDMF,allDMG,allDMH,allDMI,allDMJ,allDMK,allDML,allDMM,ijmd,chan} 
and in Appendix A of Ref.~\refcite{modlibra}. 
Fig. \ref{fg:tem} shows some particular cases where 
the experimental $S_m$ values of Fig. \ref{sme} are superimposed with 
their expected behaviors, estimated for the DM candidates in  the
scenarios and for the parameters' values reported in Table \ref{tb:legenda}. 
\begin{table}[!ht]
\begin{center}
\tbl{Scenarios and parameters values used in the expected energy distributions shown in 
Fig. \ref{fg:tem}; they have been chosen
among the many possible ones \cite{RNC,ijmd,ijma,sag06,em_int,chan,wimpele,ldm}.
In the fourth column the considered Set -- as in Ref. \protect\refcite{RNC} --
of nuclear form factors and/or of nuclear quenching factors is reported.
Here:
i)    $\sigma_{SI}$ is the spin independent point-like cross section;
ii)   $\sigma_{SD}$ is the spin dependent point-like cross section;
iii)  $\theta$ is an angle defined in the [0,$\pi$) interval, whose tangent is
      the ratio between the DM-neutron and the DM-proton effective SD coupling strengths,
      respectively \cite{RNC};
iv)   $m_H$ is the mass of the LDM particle;
v)    $\Delta$ is the mass splitting \cite{ldm};
vi)   $g_{aee}$ is the bosonic axion-like particle coupling to electrons.
For the cross sections of the LDM particle see Ref. \protect\refcite{ldm}
and for the channeling effect see Ref. \protect\refcite{chan}. See also Ref. \protect\refcite{modlibra}.
\vspace{0.3cm}
}
{\resizebox{\textwidth}{!}{
\begin{tabular}{|c|c|c|c|c|c|c|c|c|}
\hline
\multicolumn{9}{|c|}{DM particle elastic scattering on nuclei, spin-independent (SI) and spin-dependent (SD) couplings,} \\
\multicolumn{9}{|c|}{local velocity = 170 km/s (220 km/s for the cases $a$ and $b$) and nuclear cross section scaling laws as in Ref. \protect\refcite{RNC}} \\
\hline
 Curve & Halo model & Local density  & Set as & DM particle & $\xi\sigma_{SI}$&$\xi\sigma_{SD}$& $\theta$ & Channeling \\
 label & (see Ref. \refcite{RNC,Hep}) & (GeV/cm$^{3}$) & in Ref. \protect\refcite{RNC} & mass & (pb) & (pb)     & (rad)    & \cite{chan} \\
\hline
\hline
 $a$ & A5 (NFW)          & 0.33 & A &  10 GeV & $ 1.6 \times 10^{-4}$ & 0 & -- & no \\
 $b$ & A5 (NFW)          & 0.33 & A &  10 GeV & $ 7.1 \times 10^{-6}$ & 0 & -- & yes \\
 $c$ & A5 (NFW)          & 0.2  & A &  15 GeV & $ 3.1 \times 10^{-4}$ & 0 & -- & no \\
 $d$ & A5 (NFW)          & 0.2  & A &  15 GeV & $ 1.3 \times 10^{-5}$ & 0 & -- & yes \\
 $e$ & A5 (NFW)          & 0.2  & B &  60 GeV & $ 5.5 \times 10^{-6}$ & 0 & -- & no \\
 $f$ & B3 (Evans         & 0.17 & B & 100 GeV & $ 6.5 \times 10^{-6}$ & 0 & -- & no \\
     & power law)        &      &   &         &                       &   &    &    \\
 $g$ & B3 (Evans         & 0.17 & A & 120 GeV & $ 1.3 \times 10^{-5}$ & 0 & -- & no \\
     & power law)        &      &   &         &                       &   &    &    \\
\hline
 $h$ & A5 (NFW)          & 0.2  & A &  15 GeV & $ 10^{-7}$            & 2.6 & 2.435 & no \\
 $i$ & A5 (NFW)          & 0.2  & A &  15 GeV & $ 1.4 \times 10^{-4}$ & 1.4 & 2.435 & no \\
 $j$ & A5 (NFW)          & 0.2  & B &  60 GeV & $ 10^{-7}$            & 1.4 & 2.435 & no \\
 $k$ & A5 (NFW)          & 0.2  & B &  60 GeV & $ 8.7 \times 10^{-6}$ & $8.7 \times 10 ^{-2}$ & 2.435 & no \\
 $l$ & B3 (Evans         & 0.17 & A & 100 GeV & $ 10^{-7}$            & 1.7 & 2.435 & no \\
     & power law)        &      &   &         &                       &     &       &    \\
 $m$ & B3 (Evans         & 0.17 & A & 100 GeV & $ 1.1 \times 10^{-5}$ & 0.11& 2.435 & no \\
     & power law) &      &      &             &                       &     &       &    \\
\hline
\hline
\end{tabular}}}
{\resizebox{\textwidth}{!}{
\begin{tabular}{|c|c|c|c|c|c|c|c|}
\hline
\multicolumn{8}{|c|}{Light Dark Matter (LDM) inelastic scattering and bosonic axion-like interaction as in Ref. \protect\refcite{ijma,ldm},} \\
\multicolumn{8}{|c|}{A5 (NFW) halo model as in Ref. \protect\refcite{RNC,Hep}, local density = 0.17 GeV/cm$^{3}$, local
velocity = 170 km/s } \\
\hline
 Curve & DM particle & Interaction  & Set as        & $m_H$ & $\Delta$ & Cross & Channeling \\
 label &             &              & in Ref. \protect\refcite{RNC} &       &          & section (pb) & \cite{chan} \\
\hline
\hline
 $n$ & LDM & coherent   & A &  30 MeV & 18 MeV & $\xi\sigma_m^{coh} = 1.8 \times 10^{-6}$ & yes \\
     &     & on nuclei  &   &         &        &                               &     \\
 $o$ & LDM & coherent   & A & 100 MeV & 55 MeV & $\xi\sigma_m^{coh} = 2.8 \times 10^{-6}$ & yes \\
     &     & on nuclei  &   &         &        &                               &     \\
 $p$ & LDM & incoherent & A &  30 MeV &  3 MeV & $\xi\sigma_m^{inc} = 2.2 \times 10^{-2}$ & yes \\
     &     & on nuclei  &   &         &        &                               &     \\
 $q$ & LDM & incoherent & A & 100 MeV & 55 MeV & $\xi\sigma_m^{inc} = 4.6 \times 10^{-2}$ & yes \\
     &     & on nuclei  &   &         &        &                               &     \\
 $r$ & LDM & coherent   & A &  28 MeV & 28 MeV & $\xi\sigma_m^{coh} = 1.6 \times 10^{-6}$ & yes \\
     &     & on nuclei  &   &         &        &                               &     \\
 $s$ & LDM & incoherent & A &  88 MeV & 88 MeV & $\xi\sigma_m^{inc} = 4.1 \times 10^{-2}$ & yes \\
     &     & on nuclei  &   &         &        &                               &     \\
 $t$ & LDM & on electrons & -- & 60 keV & 60 keV & $\xi\sigma_m^{e} = 0.3 \times 10^{-6}$ & --  \\
\hline
 $t$ & pseudoscalar & see Ref. \refcite{ijma} & -- & \multicolumn{2}{|c|}{Mass = 3.2 keV} &
$g_{aee} = 3.9 \times 10^{-11}$ & -- \\
     & axion-like   &                      &    & \multicolumn{2}{|c|}{}               &   & \\
\hline
\hline
\end{tabular}}
\label{tb:legenda}
}
\end{center}
\end{table}
It is worth noting that the increase of the exposure and the lowering 
of the used 2 keV energy threshold will improve the discrimination capability among different
astrophysical, nuclear and particle Physics scenarios; DAMA/LIBRA--phase2 (see later) will exploit that.

\begin{figure}[!ht]
\centering
\vspace{-0.6cm}
\resizebox{0.5\columnwidth}{!}{%
\includegraphics{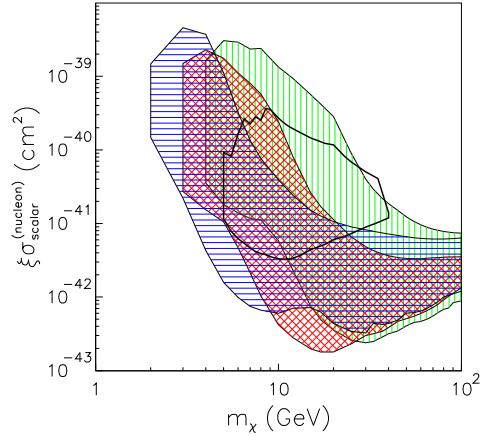} }
\vspace{-0.3cm}
\caption{Regions in the nucleon cross section vs DM particle mass plane 
allowed by DAMA in three different instances for the Na and I quenching 
factors: i) without including the channeling effect [(green) 
vertically-hatched region], ii) by including the channeling effect 
[(blue) horizontally-hatched region)], and iii) without the channeling 
effect using the energy-dependent Na and I quenching factors \cite{tret,bot11} [(red) 
cross-hatched region]. The velocity distributions and the same uncertainties 
as in Refs. \protect\refcite{RNC,ijmd} are considered here. 
These regions represent the domain where the likelihood-function values differ
more than $7.5\sigma$ from the null hypothesis (absence of modulation).
The allowed region obtained for 
the CoGeNT experiment, including the same astrophysical models as in 
Refs. \protect\refcite{RNC,ijmd} and assuming for simplicity a fixed value for the Ge 
quenching factor and a Helm form factor with fixed parameters, is 
also reported and denoted by a (black) thick solid line. 
This region is meant to include configurations whose
likelihood-function values differ more than $1.64\sigma$ from the null
hypothesis (absence of modulation). This corresponds roughly to
90\% C.L. far from zero signal. See text and
for details see Ref. \protect\refcite{bot11}.
}
\label{fig_que}
\end{figure}

Another example is the analysis in terms of DM particles inducing nuclear recoils, reported in 
Ref. \refcite{bot11},
where also some uncertainties have been discussed, taken into account
and properly treated. 
In this analysis also some other uncertainties 
due to the description of the halo models are accounted for (see Ref. \refcite{bot11} and Refs. therein).
Fig. \ref{fig_que} reports the regions in the nucleon cross section vs DM particle mass plane 
allowed by DAMA in three different instances for the Na and I quenching 
factors: i) without including the channeling effect, ii) by including the channeling effect, 
and iii) without the channeling effect using the energy-dependent Na and I quenching factors \cite{tret,bot11}.
The velocity distributions and the same uncertainties 
as in Refs. \refcite{RNC,ijmd} are considered as well. 
These regions represent the domain where the likelihood-function values differ
more than $7.5\sigma$ from the null hypothesis (absence of modulation).
The allowed region obtained for 
the CoGeNT experiment (see later), including the same astrophysical models as in 
Refs. \refcite{RNC,ijmd} and assuming for simplicity a fixed value for the Ge 
quenching factor and a Helm form factor with fixed parameters, is 
also reported.
This latter region is meant to include configurations whose
likelihood-function values differ more than $1.64\sigma$ from the null
hypothesis (absence of modulation); this corresponds roughly to
90\% C.L. far from zero signal.
It is worthwhile to comment that the inclusion of other uncertainties, not considered there, can 
further enlarge these allowed regions.

A large literature on corollary analyses of the DAMA results in terms of various astrophysical, nuclear and particle physics
scenarios is available (see e.g. Refs.
\cite{neutr,neutrA,neutrB,neutrC,neutrD,neutrE,neutrF,neutrG,neutrH,neutrI,neutrJ,sneu,sneuA,sneuB,sneuC,inel,inel_other,inel_otherA,inel_otherB,wimpele,em_int,ldm,ijma,mirror,mirrorA,mirrorB,resonant,4gen,Al,AlA,AlB,AlC,composite,compositeA,andr10,complDM,exoth,seclDM,asymDM,isoDM,isoDMA,singletDM,singletDMA,specGU,specGUA,candidates,candidatesA,candidatesB,candidatesC,candidatesD,candidatesE,Wei01,Wei01A}).

\section{Some arguments on comparisons}
\label{moddep}

No other experiment, whose result can be directly compared
in a model-independent way with that of DAMA, is available so far in the field
of Dark Matter direct detection. 
On the other hand, many theoretical and experimental parameters,
models and assumptions appear in the model dependent investigation of the experimental results of direct 
search experiments. 
In particular, one has to consider that:
i)   each model requires its parameters;
ii)  each parameter has an allowed range of values and not just a single value;
iii) uncertainties in the models and in the parameters can play a relevant role
in the derived model dependent results and in comparisons.
Below just few of these quantities, models and parameters
-- mostly for the case of DM interactions on nuclei -- are recalled as examples:

\vspace{0.3cm}

\begin{itemize} 

\item  cross section, mass, and other quantities describing the phenomenology and the
       space parameter of the considered DM candidate particle.

\item  Spin-Independent (SI) and/or Spin-Dependent (SD) interaction:
       elastic scatterings  with electromagnetic contribution 
       arising from excitation of bound electrons, inelastic scatterings on target nuclei with either 
       SI and/or SD coupling in various scenarios, 
       interaction of light DM either on electrons or on nuclei with production 
       of a lighter particle, preferred interaction with electrons, conversion of DM 
       particles into e.m. radiation, etc..

\item  Effective couplings to nucleon: possible isospin dependence of the
       couplings can be considered;  
       effective DM particle-nucleon coupling strengths for either SI
       and/or SD interaction can be defined, and this is important 
       in the comparison of results
       obtained with experiments using different target nuclei depending on the unpaired
       nucleon (compare e.g. odd spin isotopes of Xe, Te, Ge, Si, W
       with $^{23}$Na and $^{127}$I).

\item  Form Factor: it depends on the target nucleus; there is not a universal formulation
       for it, many profiles are available in literature. In these profiles some parameters 
       -- whose value is not fixed -- appear. In case of SD interaction  there is no 
       decoupling between nuclear and DM particles degrees of freedom and  
       it depends on adopted nuclear potential. The form factor profiles can 
       differ by order of magnitude; the value strongly affects the 
       expected signal and the model dependent interpretation of the results.

\item  Spin Factor: in the SD interaction it is a crucial quantity. It depends on the
       nuclear potential\cite{RNC}. In fact, not only large differences
       in the measured rate can be expected for
       target nuclei sensitive to the SD component of the interaction (such as e.g.
       $^{23}$Na and $^{127}$I) with respect to those largely insensitive to such a
       coupling (such as e.g. $^{nat}$Ge and $^{nat}$Si), but also when using different
       target nuclei, although all -- in principle --
       sensitive to such a coupling. Large differences exist in different models and 
       even within the same model, as the case e.g. of $^{23}$Na and $^{127}$I, having a proton as unpaired nucleon,
       and $^{29}$Si, $^{73}$Ge, $^{129}$Xe, $^{131}$Xe, having a neutron as unpaired nucleon, 
       where the sensitivities are almost complementary, depending 
       on the $tg \theta = a_n/a_p$ value (ratio between the neutron and the
       proton effective SD coupling strengths) (see e.g. \cite{RNC}).

\item  Scaling Law: the experimental observable in direct detection when searching for scatterings of DM particles on 
       target nuclei is the nucleus cross section
       of the interaction; a scaling law for the cross sections is required 
       in order to compare the results obtained by using different target nuclei. For example, 
       it has been proposed that two-nucleon currents from pion exchange in the nucleus
       can give different contribution for nuclei with different atomic number \cite{scalaw}; as
       a consequence the cross section for some nuclei can be enhanced with respect to others.
       Also similar arguments have a great relevance in the model dependent comparisons.

\item  Halo model and DM velocity distribution: they are an open problem
       of the field (see section \ref{p:halo}); the existing uncertainties in these models
       affect the expected counting rate and they must  
       be taken into account. 
      
\item etc.

\end {itemize}

\vspace{0.3cm}

Moreover an important role, when a DM candidate inducing just nuclear recoils is considered,
is played by the quenching factor (namely the ratio between the detected
energy in keV electron equivalent [keVee] and the
kinetic energy of the recoiling nucleus in keV).
As is widely known, the quenching factor is a specific
property of the employed detector(s) and not an universal 
quantity for a given material. For example, in
liquid noble-gas detectors, it depends -- among other aspects --
on the level of trace contaminants which can vary in time
and from one liquefaction process to another, on the
cryogenic microscopic conditions, etc.; in bolometers it
depends for instance on specific crystal properties, trace contaminants,
cryogenic conditions, etc. of each specific detector,
while generally it is assumed very favorably exactly equal to unity. In crystal 
scintillators, the quenching factor depends, for example,
on the dopant concentration, on the growing method/procedures, 
on residual trace contaminants, etc., and is
expected to have an energy dependence \cite{tret}. Channeling can play a role too.
Thus, all these aspects are relevant sources of uncertainties
when interpreting and/or comparing whatever result in terms of DM
candidates inducing just nuclear recoils.
The precise determination of the quenching factor
is quite difficult for all kinds of used detectors.
In fact, generally the direct measurements of quenching
factors are performed with reference detectors, that in
some cases can have features quite
different from the detectors used in the running conditions; 
in some other cases,
these quenching factors are even not measured at all.
Moreover, the real nature of these measurements and the
used neutron beam/sources may not point out all the possible
contributions or instead may cause uncertainties
because e.g. of the presence of spurious effects due
to interactions with dead materials as e.g. housing or
cryogenic assembling, different electronic assembling (light response, energy threshold, energy resolution, etc.); 
therefore, they are intrinsically
more uncertain than generally quoted.

In particular, the values of quenching factors for Na and I, used by DAMA in the corollary model-dependent analyses
concerning candidates inducing just nuclear recoils had, as a first reference, the values measured in Ref. \refcite{plb96}.
This measurement was performed with a small NaI(Tl) crystal from the same growth of the used detectors,
irradiated by a $^{252}$Cf source, by applying the
same method previously employed in Ref. \refcite{fu93} where 
quenching factors equal to $(0.4 \pm 0.2)$ for Na and $(0.05 \pm 0.02)$ for I (integrated over the 5--100 keV
and the 40--300 keV nuclear recoil energy range, respectively) were obtained.
Using the same parameterization as in Ref. \refcite{fu93}, DAMA measured in Ref. \refcite{plb96} quenching factors 
equal to 0.30 for Na and 0.09 for I, integrated over the 6.5--97 keV and the 22--330 keV
nuclear recoil energy ranges (corresponding approximately to the 2--30 keV electron equivalent range for both cases), respectively.
The associated errors
derived from the data were quoted as one unity in the least
significant digit; then, considering also both the large
variation available in the literature\cite{RNC,fu93,plb96,otherque,otherqueA,otherqueB,otherqueC,otherqueD,otherqueE}
and the use of a test detector, a 20\%
associated error has been included, when deriving corollary model dependent results.

In Ref. \refcite{tret} has been pointed out that the
quenching factors for nuclear recoils in scintillators can be
described by a semi-empirical formula having only one
free parameter: the Birks constant, $k_B$, which depends on
the specific set-up. Applying this procedure to the DAMA
detectors operating underground and fixing the $k_B$ parameter
to the value able to reproduce the light response to alpha
particles in these detectors, the expected Na and I quenching
factors are established as a function of the energy with
values ranging from 0.65 to 0.55 and from 0.35 to 0.17 in
the 2--100 keV electron equivalent energy interval for Na
and I nuclear recoils, respectively.
These considerations point out
the energy dependence of quenching factors for various
recoiling ions in the same detector and they have called our attention
to the fact that the large uncertainties in the method
of Ref. \refcite{fu93} could be due, in a significant part,
to uncertainties in the parameterization itself, which we
also adopted. Moreover, an increase of
the quenching factors towards lower energies can
be expected\cite{tret}, as observed in some crystal detectors as e.g. CsI; 
therefore, the use of an integrated value of the quenching factor
in the interpretation and in the comparison of results 
is another source of uncertainty.

An additional argument on uncertainties of quenching
factors in crystals (as also is NaI(Tl)) is the
presence and the amount of the well known channeling
effect of low energy ions along the crystallographic axes
and planes of the crystals. Such an effect can have a
significant impact in the corollary model dependent analyses,
in addition to the uncertainties discussed above, 
since a fraction of the nuclear recoil events would have a
much larger quenching factor than that derived from the neutron
calibrations. Since the channeling effect cannot be reliably pointed out
in usual neutron measurements,
as discussed in details in Ref. \refcite{chan}, generally just theoretical modeling
can be applied. In particular, the modeling
of the channeling effect described by DAMA in
Ref. \refcite{chan} is compatible with the nuclear recoil spectrum measured
at neutron beam by some other groups \cite{otherque,otherqueA,otherqueB,otherqueC,otherqueD,otherqueE}.
We also mention alternative channeling models, as that of
Ref. \refcite{mat08}, where larger probabilities of the planar channeling
are expected. For completeness, we mention the analytical
calculation claiming that the channeling effect holds for
nuclear recoils coming from outside a crystal and not from nuclear recoils
produced inside it, due to the blocking effect \cite{boz10};
nevertheless, although some amount of blocking effect
may be present, the precise description of the crystal
lattice with dopant and trace contaminants is quite difficult
and analytical calculations require some simplifications
which affect the result.

All these arguments hold also for the measurement 
recently reported in Ref. \refcite{collar_que,collar_queA},
where a decrease of the quenching factor 
for Na and I recoils at lower energy has been obtained in a specific detector;
in these measurements absence of significant channeling effect in NaI(Tl) 
for low energy nuclear recoils is also claimed.
It is worth noting that, for the reasons reported above, also in these results 
uncertainties are present: i) the quenching factor
values can be different in different crystals (see e.g. Ref. \refcite{tret} and references therein); 
ii) no $\alpha$ light yield has been
measured for the used NaI(Tl) crystal for comparison; iii) efficiency for low energy
nuclear recoils in the set-up; iv) etc.. Thus, it is not methodologically correct to apply
these results to whatever NaI(Tl) detector.

Obviously, as mentioned above, similar or even larger uncertainties exist on the quenching factor values 
adopted in the DM direct detection field for whatever kind of detector (see e.g. Ref. \refcite{RNC}).

In conclusion, there are a lot of relevant sources of uncertainties
when interpreting and/or comparing whatever result in terms of DM
candidates inducing just nuclear recoils.
More generally, each interpretation of experimental results is affected by the uncertainties 
due to the models and the considered scenarios: by the fact,
no unique nor ``standard'' nor ``reference'' scenario (as sometimes is claimed in literature) is available. 
In this view the comparisons of the results obtained by different
experiments using different target materials and/or different experimental approaches are very complicate
and the uncertainties must be taken into account (see for example Refs. \refcite{RNC} where the case of {\it WIMPs}
is discussed at some extent).

Moreover, it is worth noting that -- in every case -- in experiments using discrimination procedures 
the result will not be the identification of the presence of DM particle elastic 
scatterings because of the known existing nuclear recoil-like indistinguishable background
which can hardly be estimated at the needed level of precision.
Finally, the electromagnetic component of the counting rate, rejected in this approach,
can contain the signal or part of it and will be lost by them.

As examples of the above discussion, let us comment few aspects about some activities that recently released results.

The XENON project realized so far two set-ups: XENON10
and XENON100, using dual phase liquid/gas detectors.
Experiments exploiting such technique, like WARP and ZEPLIN (both ended), and CTF-DARK-SIDE,  
perform statistical discrimination 
between nuclear recoil-like candidates and electromagnetic component of the measured 
counting rate through the ratio of the 
prompt scintillation signal ($S1$) and the delayed signal ($S2$) 
due to drifted electrons in the gaseous phase, but after many subtraction procedures
and definition of a ``fiducial volume'' although e.g. the non-uniform far UV light response. 
The XENON100 experiment has released so far data for a total exposure of
224.6 days, using a fiducial volume of only 34 kg of Xenon target mass \cite{xenon}.
The experiment starts from a relevant counting rate and,
in order to try to lower it, needs to apply many data selections, subtractions
and handling. Each selection step can introduce systematic errors 
which can also be variable along the data taking period.
After these selections procedures, an analysis based on
some discrimination between the electromagnetic radiation
and nuclear recoiling candidates is applied, although the two populations of events are mostly overlapped
in the few available calibration data. 
Concerns are discussed in literature about 
the real response of such devices, as e.g. in \cite{paperliq,collar,collarA,collarB,collarC,collarD,collarE,collarF,taupnoz}.
The technical performance of the apparata, confirmed also by similar experiments, has shown that:
i)    the detectors suffer from non-uniformity; some corrections are applied and 
      systematics must be accounted for;
ii)   the response of these detectors is not linear, i.e. the number of photoelectrons/keV 
      depends on the energy scale and also depends on the applied electric field;
iii)  the physical energy threshold is not proved by source calibrations in the 
      energy interval of interest; the calibrations are done with external 
      sources and the lowest energy calibration point is 122 keV of $^{57}$Co; 
      no calibration is possible at the claimed energy threshold (as well as e.g. the experimental 
      determination of efficiency);
iv)   the use of energy calibration lines from Xe activated by neutrons cannot be applied 
      as routine and the studies on a possible calibration with internal sources have not been realized so far;
v)    despite of the small light response (2.28 photoelectron/keVee), absence of calibration in the region of interest
      and non-uniform light response,
      an energy threshold at 1.3 keVee is claimed;
vi)   the energy resolution is poor;
vii)  in the scale-up of the detectors the performances deteriorate;
viii) the behaviour of the light yield for nuclear recoils at low energy is uncertain.
Two events survive after the applied corrections and the cuts procedures, corresponding to 224.6 kg$\times$day exposure;
the quoted estimated background is $(1.0\pm0.2)$ events \cite{xenon}. 

In the double read-out bolometric technique, the heat signal
and the ionization signal are used in order to 
discriminate between electromagnetic events
and nuclear recoil-like events. 
This technique is used by CDMS and EDELWEISS
collaborations. In particular, CDMS-II detector consists of 19 Ge bolometer of about 230 g
each one and 11 Si bolometer of about 100 g each one.
The experiment released data for an exposure of about 190 kg $\times$ day \cite{cdms}
using only 10 Ge detectors in the analysis (discarding all the data collected 
with the other ones) and considering selected time periods for each detector.
EDELWEISS employs a target fiducial mass of about 2 kg of Ge and has released 
data for an exposure of 384 kg $\times$ day
collected in two different periods (July-Nov 08 and  April 09-May 10) \cite{edelweiss}
with a 17\% reduction of exposure due to run selection.
These two experiments claim to have an ``event by event'' discrimination between
{\it noise + electromagnetic background} and {\it nuclear recoil + recoil-like} (neutrons, end-range alphas, fission fragments,...) 
events by comparing the bolometer and the ionizing signals for each event, 
but their results are, actually, largely based on several kinds of huge data selections and on the application of other 
preliminary rejection procedures which are generally poorly described and
often not completely quantified. An example is the time cut analysis used 
to remove the so-called surface electrons that are distributed in the electromagnetic band and 
in the nuclear recoiling one, spanning from low to high energy. No detailed discussion 
about the stability and the robustness of the
reconstruction procedure has been given; a look-up table to identify such event is used but systematical errors
on the stability in time of such table are not discussed. 
In these experiments few nuclear recoil-like events survive the cuts and selection procedures 
applied in the data analysis; these events
are generally assumed in terms of background.
In particular, for the CDMS-II case two events survive after the cuts and selection procedures,
while the quoted estimated background is $(0.8\pm0.1)$ events \cite{cdms}. 
Moreover, most efficiencies and physical quantities entering in the interpretation of
the claimed selected events have never been suitably addressed.
In addition, further uncertainties are present when, as done in
some cases, a neutron background modeling 
is pursued.
As regards their application to the search for time dependence of the data 
(such as the annual modulation signature), it would require -- among others -- 
to face the objective difficulty to control 
all the operating conditions -- at the needed level ($<$ 1\%) -- 
despite the required periodical procedures e.g. for
cooling and for calibration source introduction.
Moreover, the many applied cuts and the discrimination procedures exclude the possibility to obtain the 
needed stability for a reliable annual modulation analysis.
The attempt performed by CDMS-II to search for annual modulation 
in Ge target has been done by using only 8 detectors over 30, by
selecting events in the nuclear recoil band below the 10 keV energy threshold (actually between 5 and 12 keV) 
and by using data that are not continuous over the whole annual periods 
considered in the analysis \cite{cdmsannmod}.
The use of non-overlapping time periods collected with detectors having 
background rate within the signal box that differ 
orders of magnitude cannot allow one to get safely reliable result
(see e.g. arguments in Ref. \refcite{collarcdms}).

Recently, the results of CDMS-II with the Si detectors were published in two 
close-in-time data releases \cite{cdms_si1,cdms_si2}; while 
no events in six detectors (corresponding exposure of 55.9 kg$\times$day before analysis cuts) were reported in the former 
\cite{cdms_si1}, three events in eight detectors (corresponding raw exposure of 140.2 kg$\times$day) were reported 
over the residual background, estimated after subtraction: $\simeq 0.4$ in the second one \cite{cdms_si2}.
The latter result may be interpreted -- under certain assumptions -- in terms of a DM candidate with spin-independent interaction
and a mass around 10 GeV. This is compatible with
interpretations of the DM annual modulation result already reported by DAMA in terms
of this kind of DM candidate and with the other hints reported by CRESST and CoGeNT (see later).

The CRESST experiment exploits the double read-out bolometric technique,
using the heat signal due to an interacting particle 
in the CaWO$_{4}$ crystals and the scintillation light produced.
A statistical discrimination of nuclear recoil-like events from electromagnetic 
radiation is performed. The detector is placed in the Gran Sasso laboratory.
The last data, released by the experiment,
have been collected with 8 detectors of 300 g each one, for an exposure of
about 730 kg $\times$ day \cite{cresst}. As regards the cuts and selection procedures applied,
most of the above discussion also holds. After selections, 67 nuclear recoil-like events
have been observed in the Oxygen band. The background contribution estimated by authors 
ranges from 37 to 43 events, and does not account for all the observed events.
The remaining number of events and their energy distribution may be
interpreted -- under certain assumptions -- in terms of a DM candidate with spin-independent interaction
and a mass in the range of 10-30 GeV. This is compatible with
interpretations of the annual modulation result already reported by DAMA in terms
of this kind of DM candidate and with the possible hints reported by CoGeNT (see later) and CDMS-II.
Improvements in the radiopurity of the set-up 
are planned, in order to reduce known source of background. Future results 
are foreseen.

Other possible positive hints of a possible signal of light Dark Matter candidates inducing nuclear elastic scatterings 
have been reported by the CoGeNT experiment \cite{CoGeNT,CoGeNTA}. The set-up is composed by a 440 g, p-type point contact (PPC) Ge
diode, with a very low energy threshold at 0.4 keVee. It is located in the
Soudan Underground Laboratory. In the data analysis no discrimination 
between electromagnetic radiation and nuclear recoils is applied;
only noise events are rejected. The experiment observes more events than they expect from 
an estimate of the background in the energy range 0.4-3.2 keVee. The energy spectrum of these events is
compatible -- under certain assumptions -- with a signal produced by the interaction of a DM particle 
with a mass around 10 GeV. In addition, considering an exposure of 146 kg $\times$ days CoGeNT experiment
also reports an evidence at about 2.8$\sigma$ C.L. of an annual modulation of the counting rate
in (0.5-0.9) keVee with phase and period compatible -- although the small confidence level --
with a Dark Matter signal.

In conclusion, the comparison of the results achieved by different experiments
must be handled with very cautious attitude without neglecting the many sources of 
uncertainties. This holds both for the model dependent allowed regions and
exclusion plots. In particular, both the negative results and the possible positive hints above--mentioned 
are largely compatible with the model independent annual modulation results of DAMA in many scenarios,
also considering the large uncertainties in theoretical and experimental aspects.
The same consideration can be done for the possible positive hints and the null results 
from the indirect approaches.
Anyhow, let us remind that -- as already stated in Sect. \ref{compatibility} -- 
the obtained model independent evidence of DAMA is compatible with a wide 
set of existing scenarios regarding the nature of the DM candidate
and related astrophysical, nuclear and particle Physics.

\section{DAMA/LIBRA--phase2 and perspectives}

\begin{figure}[!ht]
\centering
\resizebox{0.9\columnwidth}{!}{%
\includegraphics{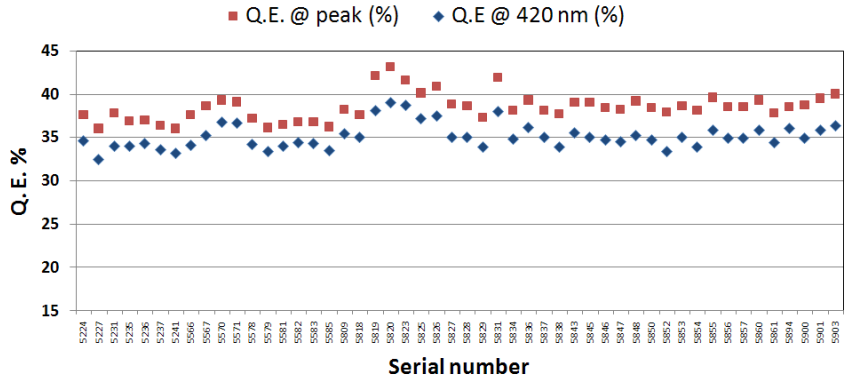} }
\caption{Quantum Efficiency (Q.E.) at peak and at 420 nm of each one of the 50 high Q.E. PMTs, installed in DAMA/LIBRA--phase2.
The averages are 38.5\% and 35.1\%, respectively; the RMS around 1.5\% show that the Q.E. spread
in the PMTs production is well limited. \cite{pmts}}
\label{fig_pmts}
\end{figure}

As shown in Fig. \ref{fg:tem}, in order to increase the experimental
sensitivity of DAMA/LIBRA and to disentangle -- in the corollary investigation on the candidate particle(s) -- at
least some of the many possible astrophysical, nuclear and particle Physics scenarios and related
experimental and theoretical uncertainties, the decreasing of the software energy threshold
is necessary. For this purpose an upgrade at the end of 2010 has been performed, replacing 
all the PMTs with new ones having higher quantum efficiency (see Fig. \ref{fig_pmts}).
After this upgrade, DAMA/LIBRA started its data taking
and is continuously running in this new configuration, 
named DAMA/LIBRA--phase2, since January 2011. Details on the reached performances are reported in Ref. \refcite{pmts}.

Among the perspectives of DAMA/LIBRA--phase2 we recall:
1) the increase of the experimental sensitivity by lowering the software
energy threshold of the experiment; 
2) the improvement of the corollary investigation on the
nature of the Dark Matter particle and related astrophysical, nuclear and 
particle physics arguments; 
3) the investigation of other signal features and second order effects;
4) the improvement of the sensitivity in the investigation of several rare processes other than 
Dark Matter (as done by the DAMA/NaI \cite{allRare} and 
by DAMA/LIBRA--phase1 \cite{papep,cncn,am241}). 
This requires long and heavy full time dedicated 
work for reliable collection and analysis of very large exposures, 
as DAMA collaboration has always done.

The strictly quality control allows DAMA/LIBRA to be
still the highest radiopure set-up in the field with the largest exposed 
full sensitive mass, the full control of running conditions, the largest duty-cycle 
and an exposure orders of magnitude
larger than any other activity in the field.

Some of the main topics, that will be addressed by DAMA/LIBRA--phase2, are:

\begin{itemize} 

\item {\it the velocity and spatial distribution of the Dark Matter particles in the galactic halo}. 

\item {\it the effects induced on the Dark Matter particles distribution in the galactic halo by contributions 
from satellite galaxies tidal streams}. It has been pointed out \cite{Fre04,Fre04A} 
that contributions to the Dark Matter particles  
in the galactic halo should be expected from tidal streams from the Sagittarius Dwarf elliptical galaxy. Considering that this
galaxy was undiscovered until 1994 and considering galaxy formation theories, one has to expect that also other satellite galaxies
do exist and contribute as well. In the recent years other streams of stars and gases have been pointed out both in the Milky Way 
and in nearby galaxies as M31. At present the data available for the kinematics of these 
streams are few, but -- as previously mentioned -- in next years important advancements could be expected from the data of 
the GAIA satellite \cite{GAIA}, whose aim is the measurement of distances and photometric data of millions 
of stars in the Milky Way.
At present, the best way to investigate the presence of a 
stream contribution is to determine in accurate way the phase of the annual modulation, $t_0$, 
as a function of the energy; in fact, for a given halo model $t_0$ would be expected  
to be (slightly) different from 152.5 d and to vary with energy \cite{sag06,ijmd}.

\item {\it  the effects induced on the Dark Matter particles distribution in the galactic halo by the possible existence of caustics}.
It has been shown that the continuous infall of Dark Matter particles in the galactic gravitational field can form caustic
surfaces and discrete streams in the Dark Matter particles halo \cite{lin04}. The phenomenology to point out a similar scenario 
is analogous to that in the previous item; thus, DAMA/LIBRA can as well test this possibility.

\item {\it the detection of possible ``solar wakes''}.
As an additional verification of the possible presence of contributions from streams of Dark Matter particles 
in our galactic halo,
DAMA/LIBRA can investigate also the gravitational focusing effect of the Sun on the Dark Matter particle of a stream.
In fact, one should expect two kinds of enhancements in the Dark Matter particles flow: 
one named ``spike'', which gives an enhancement of Dark Matter particle density along a line collinear 
with the direction of the incoming stream and of the Sun, and another, named ``skirt'', which gives
a larger Dark Matter particle density on a surface of cone whose
opening angle depends on the stream velocity. 
Thus, DAMA/LIBRA will investigate such a possibility
with high sensitivity through second-order time-energy correlations.

\item {\it the study of possible structures as clumpiness with small scale size}. Possible structures as clumpiness with small
scale size could, in principle, be pointed out by exploiting a large exposure which can be collected
by DAMA/LIBRA.

\item {\it the coupling(s) of the Dark Matter particle with the $^{23}$Na, $^{127}$I and electrons and its nature}.
As mentioned, several uncertainties are linked to the coupling(s) between the Dark Matter particle and the ordinary matter.
The exploiting of a new large exposure will allow to better constrain the related aspects.

\item {\it The investigation of possible diurnal effects}. The possible sidereal diurnal effects
can be due to the Earth rotation velocity, to the shadow of the Earth (sensitive just to DM candidates with 
high cross sections and tiny density), and to channeling effects in case of DM particle inducing nuclear recoils.
DAMA/LIBRA can also put more stringent constraints on these aspects. 

\item {\it Etc.}

\end{itemize}

\section{Conclusions}

DAMA/NaI has been pioneer experiment running at LNGS for several years and
investigating as first the model independent Dark Matter annual modulation signature
with suitable sensitivity and control of the running parameters. 
The second generation experiment DAMA/LIBRA having a larger exposed mass and a
higher overall radiopurity, has already significantly increased the sensitivity for the 
investigation of the Dark Matter particle component(s) in the galactic halo.
The cumulative exposure released so far by the two experiments 
is now 1.17 ton $\times$ yr (many orders of magnitude larger than any other activity
in the field), corresponding to 13
annual cycles. The data point out the presence of 
a signal satisfying all the requirements of the model
independent DM annual modulation signature with a
confidence level of 8.9$\sigma$ C.L..
Neither systematic effect nor side reaction able to account for the observed
modulation amplitude and to contemporaneously satisfy all the requirements of
the signature is available
\cite{perflibra,modlibra,modlibra2,scineghe09,taupnoz,vulca010,canj11,tipp11,muons,replica,replicaA}. 

The 8.9$\sigma$ C.L. positive DM model independent result, already achieved so far by DAMA/NaI
and DAMA/LIBRA--phase1, is a reference point. It is compatible with many DM scenarios and, in particular, 
with the possible positive hints recently put forward by activities exploiting other target-materials and/or other approaches;
on the other hand, negative results achieved by applying many subtractions and corrections procedures
are not in robust conflict. The same is for possible model dependent positive hints and null results from indirect approaches.

The data of the last annual cycle of DAMA/LIBRA--phase1
will soon be released. At the end of DAMA/LIBRA--phase1 (end of 2010) 
an important upgrade has been performed, replacing all the 
PMTs with new ones having higher quantum efficiency
to increase the experimental sensitivity lowering the software
energy threshold of the experiment. Another
upgrade at the end of 2012 was successfully concluded, while further 
improvements are planned.
Since January 2011 the DAMA/LIBRA experiment is in data taking 
with the new configuration, named DAMA/LIBRA--phase2. 

A large work will be faced by DAMA/LIBRA--phase2, which is by the fact the intrinsically most sensitive experiment in the field 
of Dark Matter because of its radiopurity, full sensitive exposed mass, full control of running conditions and high duty cycle. 
These qualities will also allow DAMA/LIBRA to further investigate with higher sensitivity several DM features, second order 
effects, and other rare processes.

\section{Acknowledgments}

This paper is dedicated to the memory of Prof. D. Prosperi,
one of the proponents of DAMA and a great friend, mentor and
colleague until his death in 2010. His outstanding work has
left to us and to everyone who met him the passion and the
enthusiasm for physics and a model of high cultural and
human values.

\end{document}